\documentclass[plaindraft,letter]{jpp-AAS}

\usepackage{graphicx,amsmath}
\usepackage{natbib}
\usepackage[pdftex,colorlinks,citecolor=blue]{hyperref}
\usepackage{cleveref}
\usepackage{float}
\usepackage{subcaption}
\usepackage{enumitem}
\usepackage[percent]{overpic}

\def\apjs{Astrophys.~J.~Supp.~Ser.}
\def\apj{Astrophys.~J.}
\def\apjl{Astrophys.~J.~Lett.}

\def\mnras{Mon.~Not.~Roy.~Astron.~Soc.}

\def\prl{Phys.~Rev.~Lett.}
\def\jgr{J.~Geophys.~Res.}

\def\prx{Phys.~Rev.~X}

\def\grl{Geophys.~Res.~Lett.}
\def\pop{Phys.~Plasmas}
\def\pof{Phys.~Fluids}
\def\pofb{Phys.~Fluids~B}

\def\jcp{J.~Comput.~Phys.}
\def\jpp{J.~Plasma Phys.}

\def\ropp{Rev.~Plasma Phys.}

\def\jetp{Sov.~J.~Exp.~Theor.~Phys.}
\def\njp{New J.~Phys.}

\ifCUPmtlplainloaded \else
  \checkfont{eurm10}
  \iffontfound
    \IfFileExists{upmath.sty}
      {\typeout{^^JFound AMS Euler Roman fonts on the system,
                   using the 'upmath' package.^^J}%
       \usepackage{upmath}}
      {\typeout{^^JFound AMS Euler Roman fonts on the system, but you
                   dont seem to have the}%
       \typeout{'upmath' package installed. JPP.cls can take advantage
                 of these fonts, if you use 'upmath' package.^^J}%
       \providecommand\upi{\pi}%
      }
  \else
    \providecommand\upi{\pi}%
  \fi
\fi

\ifCUPmtlplainloaded \else
  \checkfont{msam10}
  \iffontfound
    \IfFileExists{amssymb.sty}
      {\typeout{^^JFound AMS Symbol fonts on the system, using the
                'amssymb' package.^^J}%
       \usepackage{amssymb}%
       \let\le=\leqslant  \let\leq=\leqslant
       \let\ge=\geqslant  \let\geq=\geqslant
      }{}
  \fi
\fi

\ifCUPmtlplainloaded \else
  \IfFileExists{amsbsy.sty}
    {\typeout{^^JFound the 'amsbsy' package on the system, using it.^^J}%
     \usepackage{amsbsy}}
    {\providecommand\boldsymbol[1]{\mbox{\boldmath $##1$}}}
\fi

\newcommand{\pD}[2]{\frac{\partial #2}{\partial #1}}

\newcommand{\D}[2]{\frac{{\rm d} #2}{{\rm d} #1}}
\newcommand{\DD}[2]{\frac{{\rm d}^2 #2}{{\rm d} {#1}^2}}
\newcommand{\bigD}[2]{\frac{{\rm D} #2}{{\rm D} #1}}
\newcommand\bb[1]{\mbox{\boldmath{$#1$}}}

\newcommand{\msb}[1]{\mathsfbi{#1}}

\newcommand{\imag}{{\rm i}}
\newcommand{\rmd}{{\rm d}}
\newcommand{\rme}{{\rm e}}

\renewcommand\bcdot{\,\bb{\cdot}\,}
\newcommand\btimes{\,\bb{\times}\,}
\newcommand\grad{\bb{\nabla}}
\newcommand{\ez}{\hat{\bb{z}}}
\newcommand{\ey}{\hat{\bb{y}}}
\newcommand{\ex}{\hat{\bb{x}}}
\newcommand{\eb}{\hat{\bb{b}}}

\newcommand\edit[1]{#1}

\title[High-$\beta$ collisionless magnetosonic modes]{Microphysically modified magnetosonic modes in collisionless, high-$\bb\beta$ plasmas}

\author[S.~Majeski, M.~W.~Kunz, and J.~Squire]%
{S.~Majeski\ls$^{1}$%
  \thanks{Email address for correspondence: smajeski@princeton.edu}, 
M.~W.~Kunz$^{1,2}$, and J.~Squire$^{3}$}

\affiliation{$^1$Department of Astrophysical Sciences, Princeton University, Peyton Hall, Princeton, NJ 08544, USA\\[\affilskip]
$^2$Princeton Plasma Physics Laboratory, PO Box 451, Princeton, NJ 08543, USA\\[\affilskip]
$^3$Department of Physics, University of Otago, 730 Cumberland St, North Dunedin, Dunedin 9016, New Zealand\\[\affilskip]}

\pubyear{2022}
\volume{}
\pagerange{}
\date{\today}

\begin{document}

\maketitle

\begin{abstract}
With the support of hybrid-kinetic simulations and analytic theory, we describe the nonlinear behaviour of long-wavelength non-propagating (NP) modes and fast magnetosonic waves in high-$\beta$ collisionless plasmas, with particular attention to their excitation of and reaction to kinetic micro-instabilities. The perpendicularly pressure balanced polarization of NP modes produces an excess of perpendicular pressure over parallel pressure in regions where the plasma $\beta$ is increased. For mode amplitudes $|\delta B/B_0| \gtrsim 0.3$, this excess excites the mirror instability. Particle scattering off these micro-scale mirrors frustrates the nonlinear saturation of transit-time damping, ensuring that large-amplitude NP modes continue their decay to small amplitudes. At asymptotically large wavelengths, we predict that the mirror-induced scattering will be large enough to interrupt transit-time damping entirely, isotropizing the pressure perturbations and morphing the collisionless NP mode into the magnetohydrodynamic (MHD) entropy mode. In fast waves, a fluctuating pressure anisotropy drives both mirror and firehose instabilities when the wave amplitude satisfies $|\delta B/B_0| \gtrsim 2\beta^{-1}$. The induced particle scattering leads to delayed shock formation and MHD-like wave dynamics. Taken alongside prior work on self-interrupting Alfv\'en waves and self-sustaining ion-acoustic waves, our results establish a foundation for new theories of electromagnetic turbulence in low-collisionality, high-$\beta$ plasmas such as the intracluster medium, radiatively inefficient accretion flows, and the near-Earth solar wind.
\end{abstract}

%
%
\section{Introduction}

\subsection{Context and motivation}

Nearly half of all the baryonic matter in the Universe resides in a hot and dilute plasma state, in which Coulomb collisions are relatively rare and cosmic magnetic fields greatly influence the trajectories of the constituent particles. Examples include the warm-hot intergalactic medium, having number densities $n\gtrsim 10^{-6}~{\rm cm}^{-3}$ and temperatures $T\sim 10^5$--$10^7~{\rm K}$, and the intracluster medium of galaxy clusters, with $n\gtrsim 10^{-3}~{\rm cm}^{-3}$ and $T\sim 10^7$--$10^8~{\rm K}$. Radiatively inefficient accretion flows such as that onto the supermassive black hole at the Galactic centre, as well as the Solar wind that pervades interplanetary space, provide smaller-scale examples of systems characterized by large collisional mean free paths and small particle gyro-radii. A key feature of these systems is that the transport of momentum and heat are anisotropic with respect to the magnetic-field direction, even when the magnetic energy is much less than the thermal pressure, {\em viz.}~$\beta\doteq 8\upi nT/B^2\gg 1$. This spatial anisotropy is a direct result of the velocity-space anisotropy in the particle distribution function, which is allowed by the rarity of particle-particle collisions and shaped by the particles' primary allegiance to the local magnetic-field direction. In high-$\beta$ plasmas, such field-biased deviations from local thermodynamic equilibrium can have important dynamical consequences on both the large `fluid' scales and the small plasma-kinetic `micro' scales. It is this multi-scale connection between a high-$\beta$ plasma's thermodynamics and its fluid dynamics that is the focus of this paper. In particular, by elucidating the non-linear behaviour of long-wavelength magnetosonic modes, and placing our findings in the company of complementary work on Alfv\'enic and acoustic fluctuations, we demonstrate that even textbook examples of plasma dynamics such as basic waves are fundamentally different in weakly collisional, high-$\beta$ plasmas.

%
%
\subsection{Pressure anisotropy, micro-instabilities, and collisionless damping}

Collisionless and weakly collisional plasmas possess particles whose motions are bound by adiabatic invariants that are otherwise broken in highly collisional MHD plasmas. While there are three adiabatic invariants most commonly considered in plasma physics, two of them -- the magnetic moment $\mu$ for cross-field gyro-motion and the bounce invariant $\mathcal{J}$ for field-parallel bounce motion -- are associated with frequencies that are generally large enough for these invariants to be approximately conserved even when some collisions are present. For describing collective behaviour, these invariants are often adapted into the form of the double adiabats $p_{\perp}/nB$ and $p_{\parallel}B^2/n^3$, which are conserved in time along the flow of the plasma if the density $n$ and magnetic-field strength $B$ change slowly relative to the periodic (gyro- or bounce) motion. In this case, the thermal pressure $p$ is split up into components along and across the magnetic-field direction, $p_{\parallel}$ and $p_{\perp}$ respectively, a result of the invariants each being associated with different components of the particles' motions. In essence, the random thermal motions of a collisionless or weakly collisional plasma are restricted differently depending on whether they are along or across the magnetic field. Their dynamical importance with respect to the magnetic field can also be defined separately, as $\beta_\perp \doteq 8\pi p_\perp/B^2$ and $\beta_\parallel \doteq 8\pi p_\parallel/B^2$. In numerous space and astrophysical environments, the natural variations in the plasma density and magnetic-field strength that are present, coupled with approximate double-adiabatic invariance, lead to the development of pressure anisotropy $\Delta \doteq p_{\perp}/p_{\parallel}-1 \ne 0$. 
In high-$\beta$ plasmas where the thermal pressure is much larger than the magnetic energy, even small deviations from thermal isotropy ($|\Delta|\ll 1$) may be significant enough to grant the pressure anisotropy a role comparable to that of the magnetic energy (i.e., $\beta|\Delta|\sim 1$).

Two mechanisms by which the pressure anisotropy plays this elevated role are the modification of magnetic-field-line tension and the triggering of rapidly growing, kinetic micro-instabilities. An illustration of the former mechanism is a process named `Alfv\'en wave interruption'~\citep{squire16,squire17num,squire17}, in which a linearly polarized Alfv\'en wave whose amplitude satisfies $(\delta B_\perp/B)^2 \gtrsim 2/\beta$ adiabatically generates a pressure anisotropy large enough to nullify the restoring magnetic tension and prevent the wave's propagation. In this paper, we are focused primarily on large-scale compressive fluctuations, for which magnetic tension ends up being of little importance at high~$\beta$. Our focus is therefore primarily on the connection that pressure anisotropy has with ion-Larmor-scale kinetic instabilities, specifically the firehose and mirror instabilities. 

The firehose instability is triggered in pressure-anisotropic plasmas satisfying  $\beta_\parallel\Delta\lesssim-2$. This threshold is commonly referred to as the `fluid firehose' threshold, and corresponds to an exact balance between the restoring magnetic tension force and the destabilizing viscous stress from the negative pressure anisotropy.\footnote{Certain conditions can lead to the dominance of a resonant oblique firehose instability having a less stringent threshold of $\beta_\parallel\Delta\lesssim -1.4$ (\citealt{hm00}; A.F.A.~Bott {\it et al.}, in preparation). These conditions are, in fact, realized in our simulations of long-wavelength fast waves having $|\delta B/B_0|\gtrsim 2/\beta$; see \S\ref{sec:Fastnumerics} and figure~\ref{fig:fastdb1d} in particular.} However, as none of the magnetosonic fluctuations investigated in this paper are subject to self-interruption, the difference between $-2$ and $-1.4$ is of little consequence dynamically, and we generically refer to the `firehose threshold' as being at $-2$. In this case, when small perpendicular fluctuations in the magnetic field are present, the excess parallel pressure leads to a centrifugal force that acts in the bends of the magnetic-field lines. When the pressure anisotropy is sufficiently negative, this force cannot be stably balanced by the magnetic tension and the bends grow very rapidly~\citep{parker58,vs58}, increasingly so on smaller lengthscales (down to the ion-Larmor scale, where they are stabilized by finite-Larmor-radius effects; \citealt{ks67,dv68,yoon93,hm00}). In a driven system, the unstable pressure anisotropy is regulated through a combination of the particles pitch-angle scattering off of these bends and the compensating positive pressure anisotropy associated with the growing magnetic perturbations~\citep{schekochihin08,rosin11,kss14}. Conversely, the mirror instability is triggered when an excessively positive pressure anisotropy satisfies $\beta_\perp \Delta\gtrsim 1$~\citep{barnes66,hasegawa69}. In this case, the enhanced perpendicular pressure is able to push out against local decrements in the magnetic-field strength, causing ion-Larmor-scale `magnetic mirrors' to form. These mirrors resonantly confine particles with large pitch angles ($v_{\perp} > v_{\parallel}$) through their conservation of $\mu$ \citep[e.g.,][]{sk93}. The anisotropic thermal energy of these resonant particles reinforces the outward push against the field lines, further growing the fluctuations (and thus the confining mirror force) until the ends of the mirrors become so kinked that the particles can pitch-angle scatter off of their sharp edges and regulate the pressure anisotropy \citep{kss14,riquelme15,rincon15}.

\citet{kunz20} demonstrated that these kinetic instabilities interfere with the collisionless damping of long-wavelength, parallel-propagating ion-acoustic waves (IAWs). Namely, IAW amplitudes satisfying $|\delta n/n| \gtrsim 2/\beta$ generate a pressure anisotropy large enough to drive firehose and mirror instabilities, whose associated scattering and trapping impede the maintenance of Landau resonances that enable such waves' otherwise potent decay. The result is self-sustaining wave dynamics that evince a weakly collisional plasma: the ion distribution function is near-Maxwellian, the field-parallel flow of heat resembles its Braginskii form (except in regions where large-amplitude magnetic mirrors strongly suppress particle transport), and the relations between various thermodynamic quantities are more ‘fluid-like’ than kinetic.

\subsection{Non-propagating modes, fast waves, and oblique IAWs}

In this work, a combination of elements from both Alfv\'en waves and IAWs is investigated in the study of collisionless magnetosonic modes -- namely, non-propagating (NP) modes (in \S\ref{sec:slow}), fast waves (in \S\ref{sec:fast}), and to a more limited extent oblique IAWs (in Appendix~\ref{app:oiaw}). We investigate fast waves in the limit of perpendicular propagation, in which magnetic tension and collisionless damping play no role, but the associated fluctuations in $B$ and $n$ drive destabilizing pressure anisotropy. The NP modes, on the other hand, are highly oblique, perpendicular-pressure-balanced structures, in which collisionless transit-time damping (or `Barnes damping'; \citealt{barnes66}) is responsible for the entirety of the modes' dynamics. Barnes damping is a form of \citet{landau46} damping in which sinusoidal fluctuations in magnetic-field strength caused by an oblique perturbation (magnetic `mirrors') resonantly confine $\mu$-conserving particles and perform work on their guiding centres, thereby transferring free energy from the electromagnetic perturbations to the particles. For large values of $\beta$, the damping rate of the NP mode is relatively slow, and nonlinear saturation of the damping process can occur before the mode decays by a significant fraction. In this case, trapped particles in near resonance with the mode are rearranged in phase space, flattening the velocity distribution function of the particles $f(v_\parallel)$ in the vicinity of the phase velocity ($v_\parallel \sim 0$) \citep[e.g.,][]{zk63}. Once  $(\partial f/\partial v_\parallel)|_0 \sim 0$, there is no more free energy left to be gained by the distribution from rearranging particles, and the damping process stalls. This swapping of phase-space positions occurs on the order of a bounce time, ${\sim}\Omega^{-1}_{\rm b}$, which is the time it takes for a (just barely) trapped particle to make a full orbit of its confining magnetic mirror. The larger amplitude a mode, the shorter its bounce time, so the nonlinear saturation ensures that large-amplitude NP modes are longer lived than their small-amplitude counterparts. The principal question here is to what extent the pressure anisotropy associated with these modes affects their character and longevity.

%
%
\section{Non-propagating modes: Suppression of nonlinear saturation}\label{sec:slow}

\subsection{Theory}\label{sec:slowtheory}

\subsubsection{Model equations and assumptions}

The linear evolution of the NP mode at long wavelengths can be treated analytically in the drift-kinetic approximation, in which all relevant time- and lengthscales are much larger than those associated with the particles' gyromotion and the velocity distribution function of the particles is gyrotropic. We adopt this framework, and further simplify the calculation by treating the electrons as a massless, neutralizing, isothermal fluid having constant temperature $T_{\rm e}$.\footnote{The choice of isothermal electrons is for consistency with the simulations performed using the {\tt Pegasus++} hybrid-kinetic particle-in-cell code (see~\S\ref{sec:hybrid}), though it can be justified physically in some weakly collisional plasmas such as the ICM, where the electrons are collisional enough to remain near-Maxwellian and fast enough to be approximately isothermal along perturbed magnetic-field lines~\citep[e.g.,][]{kunz11}. This assumption is also consistent with the gyrokinetic theory of collisionless compressive fluctuations in the subsidiary limit $(m_e/m_i)^{1/2}\ll 1$, which predicts that electrons are pressure-isotropic and isothermal along field lines due to rapid conduction if their equilibrium distribution function is isotropic \citep[see \S 2.5.2 of][]{kunz15}.} In this model the velocity of magnetic-field lines, and equivalently the perpendicular fluid flow, is captured by the $\bb{E}{\btimes}\bb{B}$ drift velocity $\bb{u}_\perp$. The perpendicular velocity peculiar to this drift, denoted by $\bb{w}_{\perp}$, then describes the perpendicular particle motion relative to the field lines and the fluid flow, under the constraint that the magnetic moment $\mu\doteq m_{\mathrm{i}}w_{\perp}^2/2B$ is conserved. The component of the particle velocity directed along the local magnetic-field direction is denoted by $v_\parallel$. 

In what follows, we solve for the evolution of small perturbations $\delta f(t,\bb{r},v_\parallel,w_\perp)$ to a spatially uniform `background' ion velocity distribution function $F_0(v_\parallel,w_\perp)$. The parallel ($\parallel$) and perpendicular ($\perp$) coordinate directions are fixed with respect to a uniform background magnetic field, $\bb{B}_0$. Assuming that spatial variations in the plasma are due only to a sinusoidal perturbation having wavenumbers $k_{\parallel}$ and $k_{\perp}$, the relevant equations in their linearized forms are the drift-kinetic Vlasov equation,
\begin{subequations}\label{eqn:linzd}
\begin{equation}\label{eqn:vlasov}
    \left( \pD{t}{} + \imag k_\parallel v_\parallel \right) \left( \delta f +  \frac{\delta B_\parallel}{B_0} \frac{w_\perp}{2} \pD{w_\perp}{F_0} \right) + \frac{e}{m_{\textrm{i}}} \delta E_\parallel \pD{v_\parallel}{F_0} - \imag k_{\parallel}\frac{\delta B_{\parallel}}{B_0}\frac{w^2_{\perp}}{2} \pD{v_\parallel}{F_0} = 0 ;
\end{equation}
the force equation for the evolution of the drift velocity,
\begin{equation}\label{eqn:mom}
    \D{t}{u_{\perp}} = -\frac{\imag k_{\perp}}{m_{\mathrm{i}}n_0} \bigl(\delta p_{\perp{\rm i}} + T_{\textrm{e}} \delta n \bigr)  - \imag k_{\perp}v_\mathrm{A}^2\frac{\delta B_{\parallel}}{B_0} + \imag k_{\parallel}v^2_{\rm A} \frac{\delta B_{\perp}}{B_0};
\end{equation}
the ideal induction equation governing the parallel and perpendicular components of the perturbed magnetic field $\delta\bb{B}$,
\begin{equation}\label{eqn:ind}
    \D{t}{} \frac{\delta B_{\parallel}}{B_0} = -\imag k_{\perp} u_{\perp} \quad{\rm and}\quad \D{t}{} \frac{\delta B_\perp}{B_0} = \imag k_\parallel u_\perp ;
\end{equation}
and a generalized Ohm's law for the parallel electric field,
\begin{equation}\label{eqn:eprl}
    \delta E_{\parallel} = -\imag k_\parallel \frac{T_{\textrm{e}}}{e} \frac{\delta n}{n_0} .
\end{equation}
\end{subequations}
The perturbed number density and perpendicular ion pressure are given by
\begin{equation}
    \delta n \doteq \int\rmd^3\bb{v} \, \delta f \quad{\rm and}\quad \delta p_{\perp{\rm i}} \doteq \int\rmd^3\bb{v} \, \frac{1}{2} m_{\rm i} w^2_\perp \delta f ,
\end{equation}
respectively, with $\rmd^3\bb{v} = 2\upi w_\perp \rmd w_\perp \rmd v_\parallel$. The other symbols have their usual meanings: $e$ is the elementary charge, $m_{\rm i}$ is the ion mass, and $v_{\rm A} \doteq B_0/(4\upi m_{\rm i} n_0)^{1/2}$ is the Alfv\'{e}n speed given $B_0$ and a uniform background density $n_0$ (the zeroth moment of $F_0$). Note that $u_\perp$ is not an explicit moment of the perturbed distribution function, and must be evolved independently using \eqref{eqn:mom}. This combination of the drift-kinetic equation with a fluid equation for the drift velocity and a frozen-in magnetic field is commonly referred to as `kinetic MHD' \citep{kulsrud64,kulsrud83}. 

At this point we take $F_0$ to be a stationary, isotropic, Maxwell--Boltzmann distribution, $F_0=F_{\rm M}(v)$, with $\int\rmd^3\bb{v} \, F_{\rm M}(v)=n_0$ and $\int\rmd^3\bb{v} \, m_{\rm i}v^2 F_{\rm M}(v) = 3n_0 T_{\rm i0} \doteq 3 p_{\rm i0}$. This not only simplifies the analysis, but also ensures that the background distribution function itself is not kinetically unstable. Equation~\eqref{eqn:vlasov} can then be readily integrated in time to obtain
\begin{align}\label{eqn:df}
    \delta f(t,w_\perp,v_\parallel) &= \delta f(0,w_\perp,v_\parallel)\, \rme^{-\imag k_\parallel v_\parallel t} \nonumber\\*
    \mbox{} &- \int^t_0\rmd t' \, F_{\rm M}(v)\, \rme^{-\imag k_\parallel v_\parallel (t-t')} \biggl[ \imag k_\parallel v_\parallel \frac{T_{\rm e}}{T_{\rm i0}} \frac{\delta n(t')}{n_0} - \frac{w^2_\perp}{v^2_{\rm th,i}} \D{t'}{} \frac{\delta B_\parallel(t')}{B_0} \biggr] ,
\end{align}
where $v_{\rm th,i} \doteq (2T_{\rm i0}/m_{\rm i})^{1/2}$ is the ion thermal speed. The first term on the right-hand side of \eqref{eqn:df} represents the parallel phase mixing of the initial perturbation by the free streaming of particles along the (unperturbed) magnetic field. If $\delta f(0,w_\perp,v_\parallel) \propto F_{\rm M}(v)$, integrating this term and then completing the square shows that any moment of the initially perturbed distribution function will decay as $\exp[-(k_\parallel v_{\rm th,i} t/2)^2]$. The second term in \eqref{eqn:df} captures the self-consistent response of the plasma to the induced parallel electric field (${\propto}\delta n/n_0$) and the magnetic mirror force (${\propto}\delta B_\parallel/B_0$). It is this eigenmode response that we first calculate and discuss, before moving on to take the second moments of \eqref{eqn:df} and compute the time-dependent pressure anisotropy in \S\ref{sec:paniso}.

\subsubsection{Eigenmode response for the NP mode}

If we take the fluctuation amplitudes to be proportional to $\exp(-\imag\omega t)$ with complex frequency $\omega$, the dispersion relation that results after combining \eqref{eqn:linzd} may be written as
\begin{equation}\label{eqn:disprel}
    D(\zeta) \doteq \bigl( \omega^2 - k^2 v^2_{\rm A} \bigr) \biggl[ 1 + \frac{T_{\rm i0}}{T_{\rm e}} + \zeta Z(\zeta) \biggl]  + k^2_\perp v^2_{\rm th,i} \, \zeta Z(\zeta) \biggl[ 1 + \frac{T_{\rm i0}}{T_{\rm e}} + \frac{1}{2} \zeta Z(\zeta) \biggr] = 0 ,
\end{equation}
where $\zeta \doteq \omega/|k_\parallel|v_{\rm th,i}$ is the dimensionless phase speed and $Z(\zeta)$ is the plasma dispersion function. The first term in parentheses captures the combined restoring force of the magnetic pressure and tension, and indicates that we are examining magnetosonic modes. Indeed, setting the accompanying multiplicative term in square brackets to zero provides the dispersion relation for a Landau-damped IAW in the limit $(m_{\rm e}/m_{\rm i})^{1/2}\ll 1$. The final term in \eqref{eqn:disprel}, proportional to $k^2_\perp v^2_{\rm th,i}$, couples these Alfv\'enic and acoustic responses; its presence can be traced back to the final term in \eqref{eqn:df} representing the mirror force, and thus introduces collisionless damping of the mode through transit-time damping.

In order to isolate the NP mode, we focus specifically on highly oblique wavenumbers ($k_{\perp} \gg k_{\parallel}$) and low frequencies ($\zeta \ll 1$). In this limit, the plasma dispersion function in~\eqref{eqn:disprel} can be approximated as $Z(\zeta) \approx \imag\sqrt{\upi}$, and we may simplify the dispersion relation further by neglecting terms of order $\zeta^2$. The result is an approximate expression for the decay rate of the NP mode:
\begin{equation}\label{eqn:npdisp}
    \zeta \simeq -\frac{\imag}{\sqrt{\upi}\beta_{\rm i0}} \frac{k^2}{k_{\perp}^2}, \quad{\rm where}\quad \beta_{\rm i0} = \frac{8\upi p_{\rm i0}}{B^2_0} = \frac{v^2_{\rm th,i}}{v^2_{\rm A}} .
\end{equation}
\edit{For $\zeta \ll 1$ to be satisfied by \eqref{eqn:npdisp}, we require that $\beta_{\rm i0} \gg 1$, which  aligns well with our interest in high-$\beta$ plasmas.} Further properties of the NP mode can be found by taking moments of the kinetic equation \eqref{eqn:vlasov}, such as the proportionalities between $\delta n$, $\delta p_{\perp,\mathrm{i}}$, and $\delta B_\parallel$:
\begin{subequations}\label{eqn:eigenmode}
\begin{equation}\label{eqn:dn}
    \frac{\delta n}{n_0} = - \zeta Z(\zeta) \biggl[ 1 + \frac{T_{\rm e}}{T_{\rm i0}}\bigl( 1 + \zeta Z(\zeta)\bigr)\biggr]^{-1} \frac{\delta B_\parallel}{B_0} \simeq -\frac{1}{\beta_0}\frac{k^2}{k^2_\perp} \frac{\delta B_\parallel}{B_0} ,
\end{equation}
\begin{equation}\label{eqn:pprp}
    \frac{\delta p_{\perp\rm i}}{p_{\rm i0}} = -\frac{T_{\rm e}}{T_{\rm i0}} \frac{\delta n}{n_0} + 2\, \frac{\omega^2 - k^2 v^2_{\rm A}}{k^2_\perp v^2_{\rm th,i}}\frac{\delta B_\parallel}{B_0} \simeq \biggl(2+\frac{T_{\rm e}}{T_{\rm i0}}\biggr) \frac{\delta n}{n_0},
\end{equation}
where $\beta_0 \doteq \beta_{\rm i0} ( 1 + T_{\rm e}/T_{\rm i0})$. The latter equation implies approximate perpendicular pressure balance when $k_\parallel \ll k_\perp$, since then
\begin{equation}\label{eqn:pbalance}
    \delta p_{\perp\rm i} + \delta p_{\rm e} + \frac{\delta B^2}{8\upi} \simeq -\frac{k^2_\parallel}{k^2_\perp}\frac{\delta B^2}{4\upi} \ll \frac{\delta B^2}{4\upi}.
\end{equation}
Additionally, the parallel ion pressure perturbation is given by
\begin{equation}\label{eqn:pprl}
    \frac{\delta p_{\parallel\rm i}}{p_{\rm i0}} = -\frac{T_{\rm e}}{T_{\rm i0}} \frac{\delta n}{n_0} - 2 \zeta^2 \bigl( 1+\zeta Z(\zeta) \bigr) \biggl( \frac{\delta B_\parallel}{B_0} + \frac{T_{\rm e}}{T_{\rm i0}}\frac{\delta n}{n_0}\biggr)  \simeq -\frac{T_{\rm e}}{T_{\rm i0}} \frac{\delta n}{n_0} ,
\end{equation}
\end{subequations}
so that $\delta p_{\parallel\rm i} + \delta p_{\rm e} \simeq 0$. Equations~\eqref{eqn:npdisp} and \eqref{eqn:eigenmode} highlight some of the essential properties of the NP mode, namely, \edit{that it does not oscillate, that it decays slowly at high~$\beta$, and that its perturbations to the magnetic-field strength and the density are anti-correlated}.

The physical mechanism behind the damping rate is primarily transit-time magnetic pumping, in which Landau-resonant particles (technically, their guiding centres) that are trapped between large-scale magnetic mirrors formed by an oblique perturbation in the magnetic field extract energy from the mirror force. They experience net heating by betatron acceleration because the number of particles heated in regions where $|B|$ increases (lower $v_\parallel$ particles) is greater than the number of particles cooled where $|B|$ decreases (higher $v_\parallel$ particles). At higher plasma $\beta$ this difference is smaller, hence the $\beta^{-1}$ dependence of the damping rate.\footnote{\edit{Background pressure anisotropy with $\Delta_0 >0$, associated for example with a bi-Maxwellian $F_0=F_{\textrm{bi-M}}(v_\parallel,w_\perp)$, decreases the decay rate of the linear NP mode further by increasing the number of large-pitch-angle particles in the magnetic troughs. Mathematically, as the background pressure anisotropy gets closer to the mirror threshold, the decay rate of the NP mode decreases towards zero, with~\eqref{eqn:npdisp} acquiring a multiplicative factor of $(p_{\parallel\rm i0}/p_{\perp\rm i0})^2(1 - \beta_{\perp\rm i0}\Delta_{\rm 0})$. Such background pressure anisotropy makes it energetically `cheaper' to inflate the magnetic-field lines (in order to maintain perpendicular pressure balance), thereby (partially) offsetting the damping of the field-strength fluctuations. If the concentration of these large-pitch-angle particles leads to more perpendicular pressure than can be stably balanced by the magnetic pressure ({\em viz.}~$\beta_{\perp\rm i0}\Delta_0 > 1$), the troughs must grow deeper to compensate. This process runs away as the resonant particles in the deepening troughs lose energy via betatron cooling, resulting in the mirror instability \citep{sk93,kunz15}. In this paper, we focus solely on the impact of {\em fluctuation-driven} pressure anisotropy on the stability and evolution of magnetosonic modes, and exclude the possibility that the background plasma itself is already kinetically unstable by setting $\Delta_0 = 0$.}}

This type of collisionless damping is susceptible to nonlinear saturation, whereby the particles in the well explore the phase space available to them by $\mu$ conservation, phase-mixing out the original Maxwellian according to their differing bounce times and flattening the distribution function in the magnetic well to create a plateau around $v_\parallel \sim 0$. This effectively increases the plasma $\beta$ of the resonant particles, and the damping rate weakens dramatically. Because of the slow nature of the NP mode's decay rate at high~$\beta$, nonlinear saturation occurs comparatively rapidly, at a rate comparable to the bounce frequency of a thermal particle,
\begin{equation}\label{eqn:bounce}
   \Omega_{\textrm{b}} \doteq \frac{1}{2} k_{\parallel} v_{\textrm{th,i}} \left|\frac{\delta B_\parallel}{B_0}\right|^{1/2} .
\end{equation}
\edit{While most particles bounce at approximately this frequency, particles that are barely trapped bounce much slower due to their prolonged time spent traversing the edge of the magnetic well. As a result, the plateau forms inside-out, reaching a pitch-angle dependent maximum extent set by $|v_\parallel|/w_\perp < \sqrt{|B_{\rm max}|/|B_{\rm min}|-1}$.} For $|\delta B_\parallel/B_0|\gtrsim\beta^{-2}_{\rm i0}$, the bounce frequency~\eqref{eqn:bounce} will be larger than the decay rate~\eqref{eqn:npdisp}, and thus nonlinear saturation will be important. Because of our interest in plasmas having $\beta\gg 1$, even modes that may often be considered `linear' in amplitude will thus decay by only a small amount before experiencing nonlinear saturation, the implication being that these structures should be long lived. That is, unless some process is able to erode the resonant plateau in the perturbed distribution function on a timescale ${\lesssim}\Omega^{-1}_{\rm b}$.

\subsubsection{Generation of pressure anisotropy and triggering of the mirror instability}\label{sec:paniso}

The eigenmode \eqref{eqn:eigenmode} implies a dimensionless pressure anisotropy in the ions given by
\begin{equation}\label{eqn:lindelta}
    \Delta_{\rm NP} \simeq 2 \biggl(1+\frac{T_{\rm e}}{T_{\rm i0}}\biggr) \frac{\delta n}{n_0} \simeq -\frac{2}{\beta_{\rm i0}}\frac{k^2}{k^2_\perp}\frac{\delta B_\parallel}{B_0} .
\end{equation}
This suggests that, for $\delta B_\parallel/B_0 \sim 1$, the pressure anisotropy associated with the NP mode is sufficient to excite both the firehose and mirror instabilities, the former occurring in regions where $\delta B_\parallel > 0$, the latter occurring in regions where $\delta B_\parallel < 0$. There are two considerations that complicate this conclusion.

The first complication concerns the additional pressure anisotropy that is generated when the initial perturbation to the distribution function is anisotropically phase mixed by particles streaming freely along, but not across, the field lines. To see this effect, let us return to the time-dependent solution for the perturbed distribution function, equation~\eqref{eqn:df}, and suppose that, at $t=0$, the plasma hosts an isothermal, pressure-balanced perturbation with
\begin{equation}\label{eqn:IC}
    \delta f(0,w_\perp,v_\parallel) = \frac{\delta n(0)}{n_0} F_{\rm M}(v) = - \frac{2}{\beta_0} \frac{\delta B_\parallel(0)}{B_0} F_{\rm M}(v) .
\end{equation}
This initial condition guarantees that the pressure anisotropy that develops as the particles free stream and the plasma settles into the NP eigenmode is generated self-consistently and not put in by hand. Calculating the difference of the $(1/2) m_{\rm i} w^2_\perp$ and $m_{\rm i}v^2_\parallel$ moments of \eqref{eqn:df} with the initial condition~\eqref{eqn:IC} yields the following expression for the time-dependent pressure anisotropy:
\begin{align}\label{eqn:Delta}
    \Delta_{\rm NP}(t) &= 2 \biggl(\frac{k_\parallel v_{\rm th,i}t}{2}\biggr)^2 \rme^{-(k_\parallel v_{\rm th,i}t/2)^2} \biggl(1 + \frac{T_{\rm e}}{T_{\rm i0}}\biggr) \frac{\delta n(0)}{n_0} \nonumber\\*
    \mbox{} &+ \int^t_0\rmd t' \, \rme^{-[k_\parallel v_{\rm th,i}(t-t')/2]^2} \D{t'}{} \frac{\delta B_\parallel(t')}{B_0} \nonumber\\*
    \mbox{} &+ 2 \int^t_0\rmd t' \, \biggl[ \frac{k_\parallel v_{\rm th,i}(t-t')}{2}\biggr]^2 \rme^{-[k_\parallel v_{\rm th,i}(t-t')/2]^2} \D{t'}{} \biggl[ \frac{T_{\rm e}}{T_{\rm i0}} \frac{\delta n(t')}{n_0} + \frac{\delta B_\parallel(t')}{B_0} \biggr] .
\end{align}
All terms involving the combination $k_\parallel v_{\rm th,i}t/2$ describe the damping effect of phase mixing on the moments of the perturbed distribution function due to the production of fine-scale structure along $v_\parallel$. As discussed by \citet[][their equation~(3.7)]{kunz20}, the first term on the right-hand side of \eqref{eqn:Delta} captures a transiently produced pressure anisotropy resulting from the anisotropy of particle motion: as the magnetized particles free stream along, but not across, the field, the $w^2_\perp$ and $v^2_\parallel$ moments of $\delta f(0)$ phase mix differently. The integral terms in \eqref{eqn:Delta} capture the pressure anisotropy driven by adiabatic invariance as the mode is excited and then decays in time. It is this contribution to $\Delta_{\rm NP}(t)$ that includes the pressure anisotropy of the eigenmode, equation~\eqref{eqn:lindelta}.

\begin{figure}
    \centering
    \mbox{\hspace{4em}$(a)$\hspace{0.485\textwidth}$(b)$\hspace{0.44\textwidth}}\\
    \includegraphics[width=0.48\textwidth]{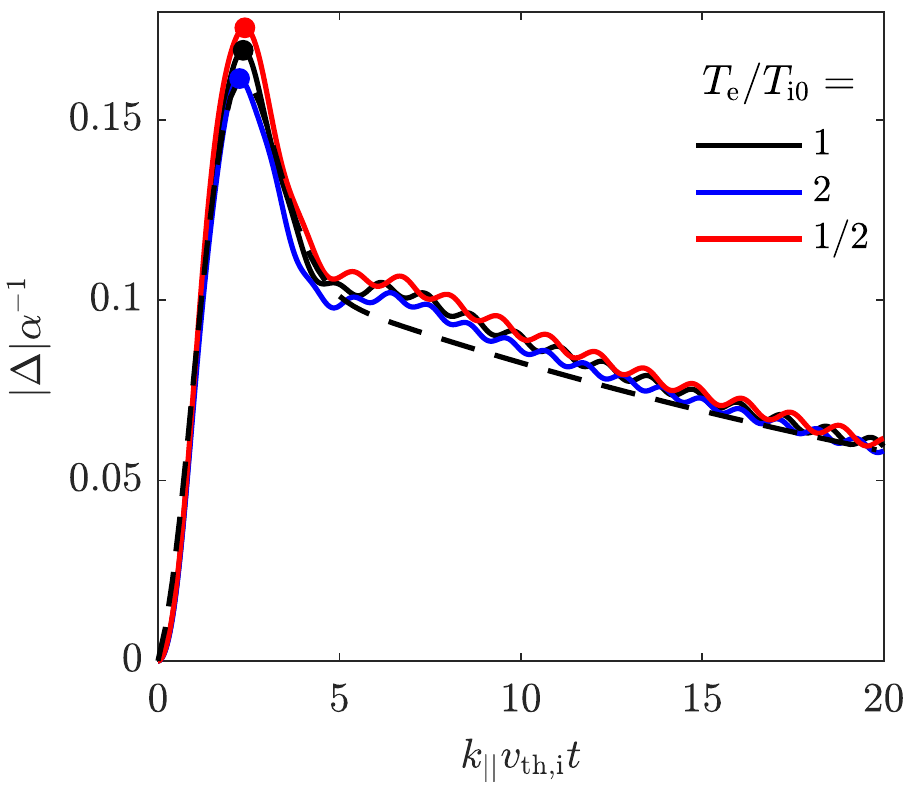}\quad
    \includegraphics[width=0.48\textwidth]{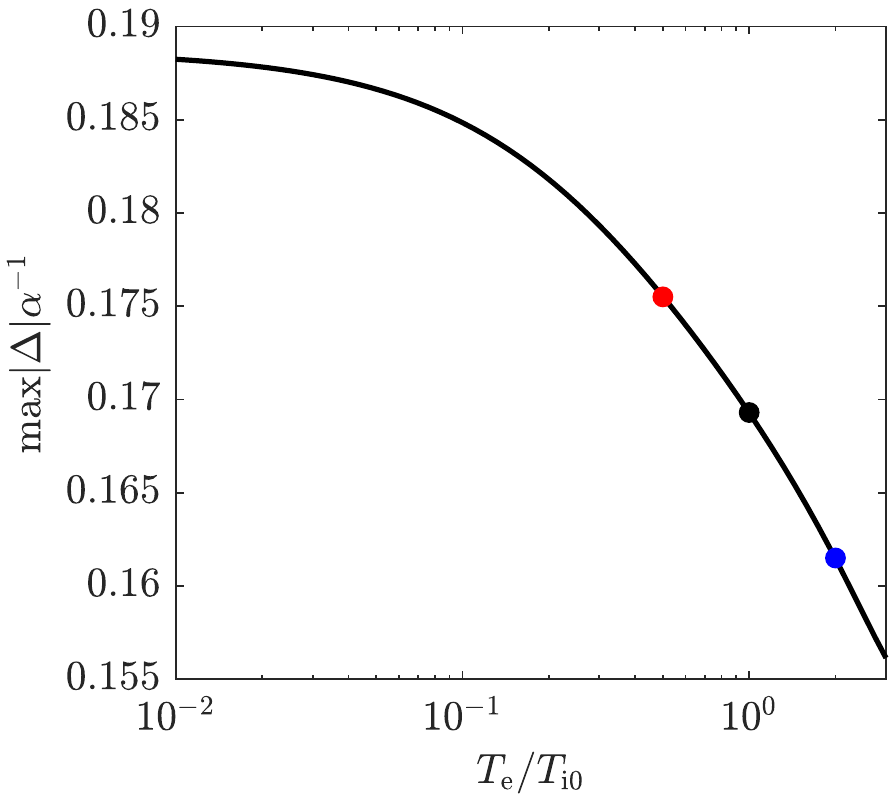}
    \caption{($a$) Solution of~\eqref{eqn:Delta} using the method presented in Appendix~\ref{app:Delta} for the time-dependent root-mean-square pressure anisotropy of a linear NP mode with wavenumber $k_\parallel$ and dimensionless initial amplitude $\alpha\doteq \delta B_\parallel(0)/B_0$ for $\beta_{\rm i0}=16$ and various $T_{\rm e}/T_{\rm i0}$. The small oscillations present after the initial adjustment are due to fast waves generated as the isothermal, pressure-balanced initial condition settles into the NP eigenmode. The approximate analytic solution \eqref{eqn:Delta_approx} is shown with the dashed line. ($b$) Maximum pressure anisotropy (divided by $\alpha$) vs.~$T_{\rm e}/T_{\rm i0}$; its values at $T_{\rm e}/T_{\rm i0}=1/2$, $1$, and $2$ are indicated.}
    \label{fig:linevo}
\end{figure}

The integrals in \eqref{eqn:Delta} can be computed numerically (see appendix~\ref{app:Delta}) and the pressure anisotropy $\Delta_{\rm NP}(t)$ determined for a given initial mode amplitude
\begin{equation}
    \alpha \doteq \left|\frac{\delta B_\parallel(0)}{B_0}\right| .
\end{equation}
The result is shown in figure~\ref{fig:linevo}($a$) at a selection of values of $T_{\rm e}/T_{\rm i0}$. The initial rise in $\Delta_{\rm NP}$ is due to a combination of the anisotropic phase mixing of the initially perturbed density and the pressure anisotropy adiabatically produced as the system settles into the NP eigenmode. After approximately one thermal-crossing time of the mode's parallel wavelength, the eigenmode is established and the slow exponential decay of $\Delta_{\rm NP}$ seen in the figure reflects the Barnes damping of the mode. (The higher-frequency oscillations seen on top of this slow decay are caused by fast modes excited by the initial conditions and represent rapid oscillations about perpendicular pressure balance.)  An approximate analytic solution for $\Delta_{\rm NP}(t)$ may be obtained in the limit of $\beta_{\rm i0}\gg 1$, $(k_\parallel/k_\perp)^2\ll 1$, and $T_{\rm e}/T_{\rm i0} \sim 1$ upon substituting the damped eigenmode~\eqref{eqn:dn} into the time integrals in equation~\eqref{eqn:Delta}. The result is that
\begin{subequations}\label{eqn:Delta_approx}
\begin{align}
    \Delta_{\rm NP}(t) &\simeq 2 \tau^2 \rme^{-\tau^2} \biggl( 1 + \frac{T_{\rm e}}{T_{\rm i0}} \biggr) \frac{\delta n(0)}{n_0} -  \biggl( {\rm erf}(\tau) - \frac{\tau}{\sqrt{\upi}} \rme^{-\tau^2} \biggr) \frac{2}{\beta_{\rm i0}}\frac{\delta B_\parallel(t)}{B_0} \\*
    \mbox{} &= - \left[ 2 \tau^2 \rme^{-\tau^2} +  \rme^{-2\imag\zeta\tau}\biggl( {\rm erf}(\tau) -  \frac{\tau}{\sqrt{\upi}} \rme^{-\tau^2} \biggr)  \right] \frac{2}{\beta_{\rm i0}} \frac{\delta B_\parallel(0)}{B_0} ,\label{eqn:Delta_approx2}
\end{align}
\end{subequations}
where $\tau \doteq k_\parallel v_{\rm th,i} t/2$. The term in square brackets goes as ${\sim}2\tau^2+\tau/\sqrt{\upi}$ for early times, suggesting that the plasma would become mirror-unstable at a time $t_{\rm m} \sim (\sqrt{\alpha} k_\parallel v_{\rm th,i})^{-1}$, comparable to the inverse of the bounce frequency \eqref{eqn:bounce}. With the mode then slowly decaying exponentially, the maximum value of the pressure anisotropy may be estimated by setting $\exp(-2\imag\zeta\tau)\simeq 1$ and maximizing~\eqref{eqn:Delta_approx2} with respect to $\tau$. The result is a maximum pressure anisotropy ${\simeq}2.6 \alpha\beta^{-1}_{\rm i0}$ (cf.~\eqref{eqn:lindelta}) occurring at $k_\parallel v_{\rm th,i} t\simeq 2.3$. The approximate solution~\eqref{eqn:Delta_approx} is traced by the dashed line in figure~\ref{fig:linevo}($a$), and is a manifestly good description of the full solution.

The second complication when using~\eqref{eqn:lindelta} to determine the kinetic stability of the NP mode is related to how the mode perturbs the perpendicular and parallel plasma $\beta$ parameters that feature in the firehose and mirror instability thresholds. Using~\eqref{eqn:eigenmode} and that $\delta B_\perp = - (k_\parallel/k_\perp)\delta B_\parallel$, one obtains
\begin{subequations}
\begin{align}
    \beta_{\parallel{\rm i}} &
    \simeq \beta_{\rm i0} \biggl( 1 + 2 \frac{\delta B_\parallel}{B_0} + \frac{k^2}{k^2_\perp} \frac{\delta B^2_\parallel}{B^2_0} \biggr)^{-1} \biggl[ 1 - \frac{k^2}{k^2_\perp} \frac{1}{\beta_{\rm i0}}\frac{T_{\rm e}}{T_{\rm i0}}\biggl(1+\frac{T_{\rm e}}{T_{\rm i0}}\biggr)^{-1}\frac{\delta B_\parallel}{B_0} \biggr] , \label{eqn:bprl}\\*
    \beta_{\perp{\rm i}} &
    \simeq \beta_{\rm i0} \biggl( 1 + 2 \frac{\delta B_\parallel}{B_0} + \frac{k^2}{k^2_\perp} \frac{\delta B^2_\parallel}{B^2_0} \biggr)^{-1} \biggl[ 1 - \frac{k^2}{k^2_\perp}\frac{1}{\beta_{\rm i0}}\biggl(2+\frac{T_{\rm e}}{T_{\rm i0}}\biggr)\biggl(1+\frac{T_{\rm e}}{T_{\rm i0}}\biggr)^{-1} \frac{\delta B_\parallel}{B_0} \biggr] .\label{eqn:bprp}
\end{align}
\end{subequations}
The final terms in both of these expressions may be dropped in the limit of $\beta_{\rm i0}\gg 1$. Combining the result with \eqref{eqn:lindelta} yields 
\begin{equation}\label{eqn:DNPbeta}
    \beta_{\parallel\rm i}\Delta_{\rm NP} \approx \beta_{\perp\rm i}\Delta_{\rm NP}
    \approx -2 \frac{\delta B_\parallel}{B_0} \biggl( 1 + 2\frac{\delta B_\parallel}{B_0} + \frac{k^2}{k^2_\perp} \frac{\delta B^2_\parallel}{B^2_0} \biggr)^{-1} .
\end{equation}
Equation~\eqref{eqn:DNPbeta} indicates that is impossible to produce a pressure anisotropy that is sufficiently negative to destabilize the plasma to the firehose. Regions in which $\Delta_{\rm NP}<0$ also have a reduced plasma $\beta$, and so the more negative the anisotropy becomes (for larger $\delta B_\parallel>0$), the further the firehose threshold (${\approx}-2/\beta_{\parallel\rm i}$) moves away. \edit{Indeed, minimizing the right-hand side of \eqref{eqn:DNPbeta} for $\delta B_\parallel > 0$, the most negative value of $\beta_{\rm i}\Delta_{\rm NP}$ is found to be ${\approx} -(1+|k/k_\perp|)^{-1} > -1/2$.} In contrast, the plasma in regions where $\delta B_\parallel <0$ that acquire a positive pressure anisotropy have an easier time of reaching the reduced mirror threshold (${\approx}1/\beta_{\perp\rm i}$). Setting the right-hand side of~\eqref{eqn:DNPbeta} to unity and solving for $\delta B_\parallel = -|\delta B_\parallel|$ then provides the following amplitude threshold for the NP mode to trigger the mirror instability:
\begin{equation}\label{eqn:ampthreshold}
    \left|\frac{\delta B_\parallel}{B_0}\right| \gtrsim 0.3 \quad \textrm{(NP mode amplitude threshold)}
\end{equation}
When this criterion is satisfied, we anticipate regions of kinetically unstable plasma to be localized to where $\delta B_\parallel<0$ and to host ion-Larmor-scale mirrors.

With these predictions borne in mind, we now determine the spatial extent of these mirror-unstable regions and discuss how the mirrors growing within them evolve to regulate the pressure anisotropy. 

\subsubsection{Regulation of pressure anisotropy by the mirror instability}\label{sec:slowtheory_inst}

\begin{figure}
    \centering
    \mbox{\hspace{1em}$(a)$\hspace{0.49\textwidth}$(b)$\hspace{0.4\textwidth}}\\
    \includegraphics[width=\linewidth]{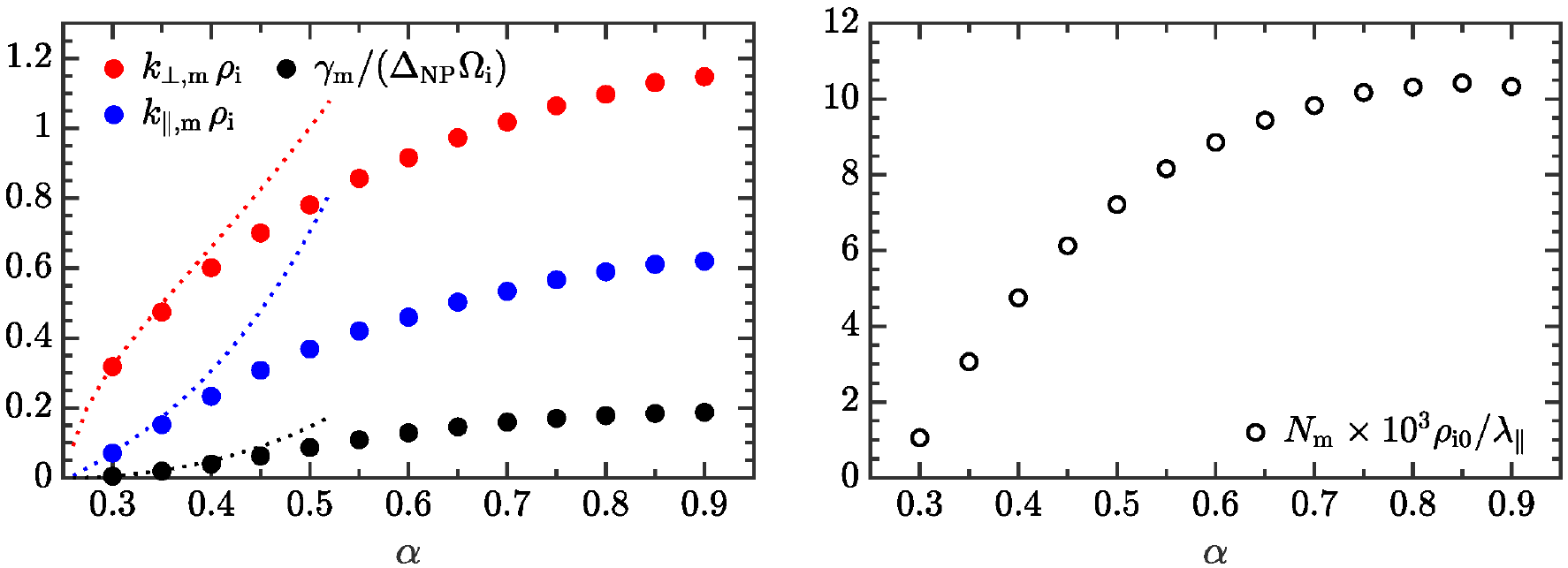}
    \caption{($a$) Perpendicular ($k_{\perp,\rm m}$) and parallel ($k_{\parallel,\rm m}$) wavenumbers of the fastest-growing mirror mode having growth rate $\gamma_{\rm m}$, all computed from linear Vlasov--Maxwell theory using the instability parameter $\Lambda_{\rm m}$ corresponding to a NP mode with $\alpha\doteq |\delta B_\parallel/B_0|$ and $k_\perp/k_\parallel=4$ in a $\beta_{\rm i0}=16$ plasma (see~\eqref{eqn:quad}; these values are weakly dependent upon $\beta_{\rm i0}$ and $k_\perp/k_\parallel$ so long as $\beta_{\rm i0}\gtrsim 10$ and $(k/k_\perp)^2\approx 1$). \edit{Dotted lines correspond to the asymptotic expressions~\eqref{eqn:mirror1}, valid for $\beta_{\perp\rm i}\Lambda_{\rm m}\ll 1$.} ($b$) The predicted number of mirrors $N_{\rm m}$ within the $\delta B_\parallel < 0$ region of a NP mode having wavelength $\lambda_\parallel$ and amplitude $\alpha$ (see~\eqref{eqn:Nm}).}
    \label{fig:mirror}
\end{figure}

In \S\ref{sec:paniso}, we showed that the plasma where $\delta B_\parallel < 0$ becomes mirror-unstable at $t_{\rm m} \sim (\sqrt{\alpha}k_\parallel v_{\rm th,i})^{-1}$ if initialized from isothermal pressure balance. With $\alpha\gtrsim 0.3$ (i.e., when instability is possible), this time is comparable to the timescale over which the NP mode's pressure anisotropy is set up (see figure~\ref{fig:linevo}). We may then view the mirror instability as growing on top of an otherwise weakly decaying positive pressure anisotropy satisfying \eqref{eqn:DNPbeta} with $\delta B_\parallel <0$. The maximum growth rate of the instability depends on how far the local pressure anisotropy ventures beyond the instability threshold, $\Lambda_{\rm m} \doteq \Delta - \beta^{-1}_{\perp\rm i} > 0$. In the asymptotic limit $\beta_{\perp\rm i}\Lambda_{\rm m}\ll 1$, the maximum mirror growth rate and associated wavenumber are given by (\citealt{hellinger07};  A.F.A.~Bott {\it et al.}, in preparation)
\begin{equation}\label{eqn:mirror1}
     \gamma_{\rm m}/\Omega_{\rm i}\approx 0.07 \beta_{\perp\rm i}\Lambda_{\rm m}^2 , \quad k_{\parallel,\rm m}\rho_{\rm i} \approx 0.2 \beta_{\perp\rm i}\Lambda_{\rm m} , \quad k_{\perp,\rm m}\rho_{\rm i}\approx 0.6(\beta_{\perp\rm i}\Lambda_{\rm m})^{1/2}.
\end{equation}
However, because of the sensitive dependence of the instability parameter $\beta_{\perp\rm i}\Lambda_{\rm m}$ on the NP mode amplitude (see~\eqref{eqn:DNPbeta}), with its value ranging from ${\sim}1$ to ${\sim}100$ for $\alpha\in[0.4,0.9]$, only very marginally unstable NP modes ({\em viz.}, $\alpha \approx 0.3$) satisfy the ordering used to derive~\eqref{eqn:mirror1}. \edit{The growth rate and wavenumber when $\beta_{\perp\rm i}\Lambda_{\rm m} \gtrsim 1$ can be obtained by a direct numerical solution of the linearized Vlasov--Maxwell equations for a bi-Maxwellian plasma, with~\eqref{eqn:DNPbeta} specifying the pressure anisotropy for a given NP mode amplitude~$\alpha$ (A.F.A.~Bott, private communication). The resulting growth rates and wavenumbers are shown versus $\alpha$ in figure~\ref{fig:mirror}($a$). (For this figure we used $\beta_{\rm i0}=16$ and $k_\perp/k_\parallel=4$, although the values shown are insensitive to either parameter as long as $\beta_{\rm i0}\gtrsim 10$ and $(k/k_\perp)^2\approx 1$.) As $\alpha$ increases, these quantities approach the empirical values}
\begin{equation}\label{eqn:mirror2}
    \gamma_{\rm m}/\Omega_{\rm i} \approx 0.2\Lambda_{\rm m}, \quad k_{\parallel,\rm m}\rho_{\rm i} \approx 0.6 , \quad k_{\perp,\rm m} \rho_{\rm i} \approx 1.2 .
\end{equation}

In order for the mirror instability to be relevant to the linear evolution of the NP mode, two criteria must be satisfied. First, the mirror growth rate must be much larger than the rate at which the NP mode decays \eqref{eqn:npdisp}, i.e., $\gamma_{\rm m} \sqrt{\upi}\beta_{\rm i} \gg k_\parallel v_{\rm th,i}$. This condition appears to be trivially satisfied in high-$\beta$ plasmas for unstable NP modes with parallel wavelengths $\lambda_\parallel \gtrsim 10^3\rho_{\rm i}$. The second criterion is that the mirror modes must actually fit inside the length of the region that is mirror unstable, {\it viz.}~$2\upi/k_{\parallel,\rm m}\lesssim\ell_{\rm mirror}$. We estimate $\ell_{\rm mirror}$ by asking where in the NP mode the quantity \eqref{eqn:DNPbeta} is larger than unity:
\begin{equation}\label{eqn:quad}
    \Lambda_{\rm m} \simeq \frac{1}{\beta_{\rm i0}} \left( -1 - 4\frac{\delta B_\parallel}{B_0} - \frac{k^2}{k^2_\perp}\frac{\delta B^2_\parallel}{B^2_0} \right) \gtrsim 0.
\end{equation}
Because the leading-order eigenvector components are all real, we can take $\delta B_\parallel = -\alpha B_0 \cos(k_{\parallel}x + k_{\perp}y)$ (as used in our simulations; see \S\ref{sec:hybrid}). Courtesy of our assumption that $k_\perp \gg k_\parallel$, we have that $\delta B_\perp \ll \delta B_\parallel$, so the field lines are approximately straight everywhere and the paths taken by the trapped particles as they bounce are approximately parallel to $\bb{B}_0$. Then, taking the appropriate root of \eqref{eqn:quad} to ensure that the inverse cosine is defined for mirror-unstable amplitudes, we find that the length of the mirror-unstable portion of the wave satisfies
\begin{equation}\label{eqn:mwidth}
    \ell_{\textrm{mirror}} \approx \frac{\lambda_{\parallel}}{\upi}\cos^{-1}\biggl( \frac{2-\sqrt{4-k^2/k^2_\perp}}{\alpha} \biggr) \doteq f_{\rm m}\lambda_\parallel .
\end{equation}
For $\alpha\approx 0.3$--$0.9$ and $k_\parallel\ll k_\perp$, $f_{\rm m}\approx 0.1$--$0.4$. The number of maximally growing mirrors that can fit within $\ell_{\rm mirror}$ is then
\begin{equation}\label{eqn:Nm}
    N_{\rm m} \approx \frac{f_{\rm m}}{4} \biggl(\frac{k_{\parallel,\rm m}\rho_{\rm i}}{2\upi}\biggr)\biggl( \frac{\lambda_\parallel}{\rho_{\rm i}}\biggr) .
\end{equation}
In writing~\eqref{eqn:Nm}, we have included an additional factor of ${\approx}1/4$ to account for the fact that the pressure anisotropy is not expected to be uniform within the mirror-unstable region and so the full extent of $\ell_{\rm mirror}$ is unlikely to be filled with mirrors of identical wavelengths; the bespoke factor of ${\approx}1/4$ was obtained empirically from examining the spatial extent of scattering mirrors formed in the hybrid-kinetic simulations of unstable NP modes presented in~\S\ref{sec:hybrid}. A further, and final, adjustment to $N_{\rm m}$ accounts for the fact that the ion-Larmor radius $\rho_{\rm i} \propto \sqrt{T_{\perp\rm i}}/B$ in the mirror-unstable region is larger than $\rho_{\rm i0}$, primarily because of the decrease in the local magnetic-field strength. Using \eqref{eqn:eigenmode} to express $\delta T_{\perp\rm i}$ in terms of $\delta B_\parallel$, \edit{and appending a multiplicative factor of ${\approx}1/2$ to $\alpha$ to account, if only approximately, for the effective reduction in $\alpha$ due to the non-uniformity of $\delta B_\parallel$ within the mirror-unstable region,} we find that
\begin{equation}\label{eqn:rhoi}
    \frac{\rho_{\rm i}}{\rho_{\rm i0}} \approx \biggl( 1 - \frac{\alpha}{2\beta_{\rm i0}} \frac{k^2}{k^2_\perp} \biggr)^{1/2} \biggl( 1-\alpha + \frac{k^2}{4k^2_\perp} \alpha^2 \biggr)^{-1/2} .
\end{equation}
With $k_{\parallel,\rm m}\rho_{\rm i}$ taken from figure~\ref{fig:mirror}($a$), we can assemble \eqref{eqn:quad}--\eqref{eqn:rhoi} to predict $N_{\rm m}$ for a given $\lambda_\parallel/\rho_{\rm i0}$, $\alpha$, and $k_\parallel/k_\perp$ of the NP mode at $\beta_{\rm i0}\gg 1$.

The result of this procedure is shown in figure~\ref{fig:mirror}($b$) as the open circles. Note that the number of mirrors $N_{\rm m}$ is fairly independent of the NP mode amplitude for $\alpha\gtrsim 0.4$, with the consequence that several mirrors can fit within the mirror-unstable region of a NP mode with $\lambda_\parallel\sim 10^3\rho_{\rm i0}$. However, at the critical amplitude $\alpha \approx 0.3$, only one or perhaps two mirrors are predicted to fit if $\lambda_\parallel \sim 10^3\rho_{\rm i0}$. In this case, the mirror instability might be ineffective at regulating the pressure anisotropy. 

In summary, we predict that a NP mode with $\alpha\gtrsim 0.4$ and $\lambda_\parallel\gtrsim 10^3\rho_{\rm i0}$ should be able to support a robust collection of mirror-unstable fluctuations.

\subsubsection{Effective collisionality induced by the mirror instability}\label{sec:slowcoll}

We now seek an estimate for the effective collision frequency instigated by these mirror-unstable distortions in the magnetic-field lines. For this, we follow the arguments of \citet{newman20} for the pitch-angle diffusion of charged particles in regions of Larmor-scale magnetic irregularities. First, we conjecture that each encounter of an ion with the edges of a single mirror depletes the plasma's temperature anisotropy $\mathcal{A} \doteq \overline{ w_{\perp}^2}/2 - \overline{v_{\parallel}^2}$ by a fraction $\chi$ (here, the overline indicates an average over the ion distribution function). Following \citet{newman20}, we identify $\chi$ with $(3/2)\sin^2\vartheta$, where $\vartheta$ is the local deflection angle of the perturbed magnetic-field lines. \edit{We estimate $\sin^2\vartheta \approx (\delta B_{\perp,\rm m}/B)^2 \approx (k_{\parallel,\rm m}/k_{\perp,\rm m})^2 (\delta B_{\parallel,\rm m}/B)^2$, and leverage prior results on the nonlinear evolution of the mirror instability showing that mirrors can grow to amplitudes $|\delta B_{\parallel,\rm m}/B| \approx 1/3$ before saturating through strong pitch-angle scattering \citep{kss14,riquelme15,sn15}. The result is that}
%
%
\begin{equation}\label{eqn:chi}
    \edit{\chi \approx 0.2 \, (k_{\parallel,\rm m}/k_{\perp,\rm m})^2} . 
\end{equation}
%

To obtain the effective collision frequency $\nu_{\rm eff}$, we then multiply $\chi$ by the number of Larmor-scale mirrors per unit time encountered by a typical particle. For a NP mode with amplitude $\alpha\gtrsim 0.4$, the criterion for a particle to be able to pass through the NP mode's enhancement in $|B|$ is $|v_\parallel|/w_\perp \gtrsim \sqrt{4/3}$. In other words, for a near-Maxwellian distribution of particle velocities, a majority of the particles will be confined to the trough of the NP mode where ion-Larmor-scale mirrors should be present, passing through this mirror-unstable region twice per bounce time. In this case, $N_{\rm m}$ scattering mirrors are encountered by each trapped particle every transit time $\Delta t \approx \upi \Omega^{-1}_{\rm b}$. The average rate of change of the ion anisotropy is then
\begin{equation}\label{eqn:dAdt}
    \frac{\Delta \mathcal{A}}{\Delta t} \approx -\frac{\chi}{\upi} N_{\textrm{m}} \Omega_{\textrm{b}}  \mathcal{A} \doteq - \nu_{\textrm{eff}} \mathcal{A} ,
\end{equation}
where in the last equality we have introduced the effective collision frequency $\nu_{\rm eff}$. Assembling~\eqref{eqn:bounce} and \eqref{eqn:quad}--\eqref{eqn:dAdt}, \edit{we find that
\begin{subequations}\label{eqn:nueff}
\begin{equation}
    \nu_{\rm eff} \approx 0.003 \,\mathcal{G} \Omega_{\rm i0} ,
\end{equation}
where
\begin{equation}
    \mathcal{G} \doteq k_{\parallel,\rm m}\rho_{\rm i} \,\biggl(\frac{k_{\parallel,\rm m}}{k_{\perp,\rm m}}\biggr)^2 \biggl( \alpha-\alpha^2 + \frac{k^2}{4k^2_\perp} \alpha^3 \biggr)^{1/2} \cos^{-1}\biggl( \frac{2-\sqrt{4-k^2/k^2_\perp}}{\alpha} \biggr)
\end{equation}
\end{subequations}
is a function of only the amplitude and wavenumber obliquity of the NP mode.}

\begin{figure}
    \centering
    \mbox{\hspace{1em}$(a)$\hspace{0.49\textwidth}$(b)$\hspace{0.4\textwidth}}\\
    \includegraphics[width=\linewidth]{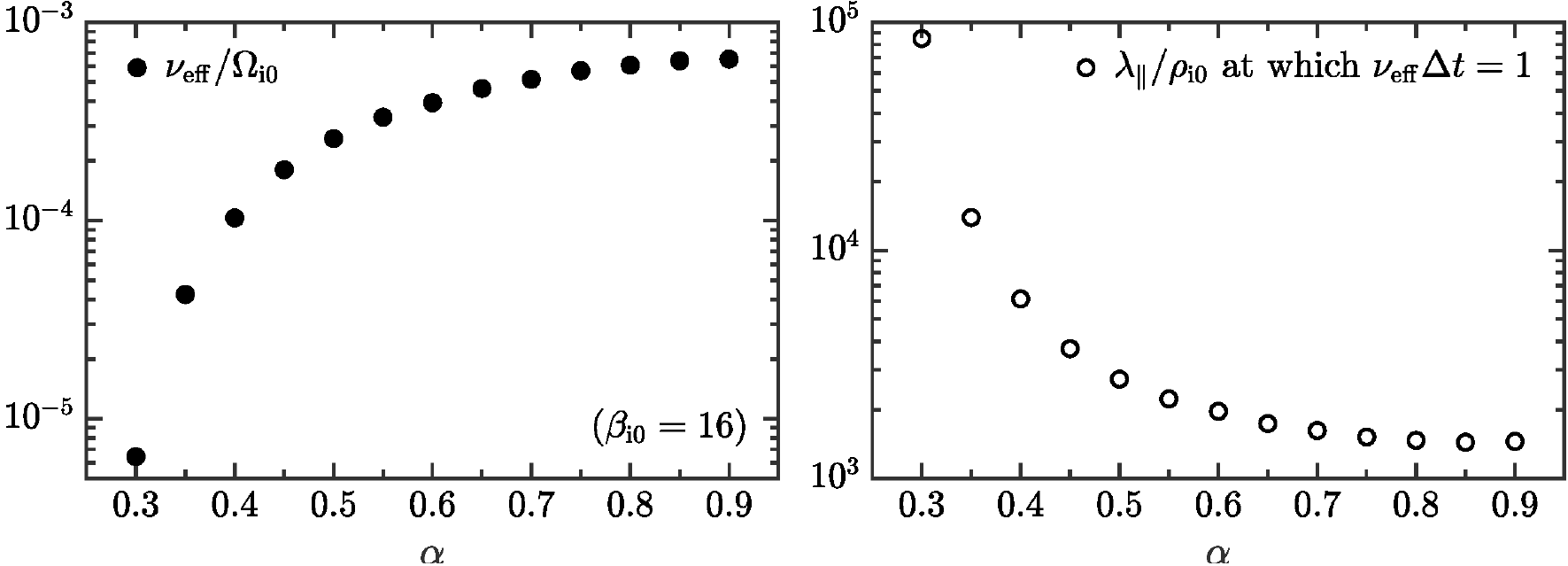}
    \caption{($a$) Predicted scattering frequency $\nu_{\rm eff}$ (see~\eqref{eqn:nueff}) caused by the mirror instability for a NP mode with amplitude $\alpha$, using the values of $k_{\parallel,\rm m}\rho_{\rm i}$ in figure~\ref{fig:mirror}($a$). ($b$)~Minimum parallel wavelength $\lambda_\parallel$ of a NP mode for which $\nu_{\rm eff}\Delta t \ge 1$, where $\Delta t = \upi\Omega^{-1}_{\rm b}$. Such modes should host mirrors whose scattering frequency is comparable to the transit time. \edit{The data in both panels correspond to $\beta_{\rm i0}=16$ and $k_\perp/k_\parallel=4$, although the values shown are insensitive to either parameter as long as $\beta_{\rm i0}\gtrsim 10$ and $(k/k_\perp)^2\simeq 1$.}}
    \label{fig:nueff}
\end{figure}

Equation~\eqref{eqn:nueff} states that {\em the predicted $\nu_{\rm eff}$ is independent of the wavelength of the NP mode and increases with increasing $\alpha$}, key features that are tested (and confirmed) in~\S\ref{sec:scaledependence}. The predicted dependence of $\nu_{\rm eff}$ upon $\alpha$ at $\beta_{\rm i0}=16$ and $k_\perp/k_\parallel=4$ is shown in figure~\ref{fig:nueff}($a$); \edit{the values shown are insensitive to either parameter as long as $\beta_{\rm i0}\gtrsim 10$ and $(k/k_\perp)^2\approx 1$}. The predicted collision frequency drops gradually from  $\alpha=0.9$ to $0.5$, and then falls sharply by more than an order of magnitude to $\nu_{\rm eff}\lesssim 10^{-5}\Omega_{\rm i0}$ at $\alpha=0.3$. In panel~($b$), we plot the minimum parallel wavelength $\lambda_\parallel$ of a NP mode for which $\nu_{\rm eff}\Delta t\ge 1$, where $\Delta t = \upi\Omega^{-1}_{\rm b}$. Such modes should host mirrors whose scattering frequency is comparable to the transit time. Note that, for $\alpha=0.3$, $\lambda_\parallel/\rho_{\rm i0}$ must be ${\gtrsim}10^5$ for the scattering frequency to be larger than the inverse transit time. It is worth bearing these numbers in mind when interpreting the simulation results presented in~\S\S\ref{sec:slowcollnum} and~\ref{sec:scaledependence}.

\subsubsection{Suppression of nonlinear saturation of the NP mode}\label{sec:suppression}

Once $\nu_\mathrm{eff}$ becomes competitive with the bounce frequency, the induced scattering will isotropize the ion distribution function faster than the nonlinear saturation can maintain the plateau in $\delta f(v_\parallel)$ around $v_\parallel \sim 0$. In this case, the nonlinear saturation is suppressed and the NP mode should resume its decay at a rate comparable to \eqref{eqn:npdisp} \citep{johnston71}. At some point during this decay, the mode amplitude will pass below its critical threshold for triggering the mirror instability \eqref{eqn:ampthreshold}, and the mirror modes themselves will become short-wavelength decaying NP modes. Near the mirror-instability threshold, these short-wavelength NP modes decay very slowly, and so the associated magnetic-field-strength fluctuations will remain nonlinear for some time after the large-scale NP mode is no longer formally mirror unstable. We therefore conjecture that the NP mode will continue to decay until the mirror fluctuations (and their induced scattering) have had sufficient time to dissipate. Excepting perhaps the case of asymptotically long NP mode wavelengths, then, there should be some delay between when the NP mode passes below threshold and when its nonlinear saturation is re-established.

The preceding arguments imply that three distinct regimes exist for collisionless NP modes in high-$\beta$ plasmas: (i) When the mode amplitude satisfies $|\delta B_\parallel/B_0| < 0.3$, the associated pressure anisotropy is too small to trigger the mirror instability, and the mode experiences slow Barnes damping until the damping nonlinearly saturates as the distribution function flattens around $v_\parallel \sim 0$. These pressure-balanced structures are thus long-lived. (ii) When $|\delta B_\parallel/B_0|\gtrsim 0.3$, the pressure anisotropy triggers the mirror instability in regions where $\delta B_\parallel < 0$ and eventually introduces an effective collisionality that, for sufficiently large NP mode wavelengths, suppresses the maintenance of a nonlinear plateau. As a result, linear decay resumes until the NP mode decays back well below its amplitude threshold. (iii) Because the induced scattering rate~\eqref{eqn:nueff} does not scale with the wavelength of the NP mode, one might expect a third fluid-like regime results at very long wavelengths when $\nu_\mathrm{eff} \gg k_\parallel v_\mathrm{th,i}$ and the collisionless damping is arrested altogether. We discuss the realizability of this third regime and speculate on its behaviour in~\S\ref{sec:scaledependence}.

%
%
\subsection{Numerical results}\label{sec:hybrid}

\subsubsection{Method of solution and initial conditions}\label{sec:method}

To test the theory presented in \S\ref{sec:slowtheory} and explore the nonlinear evolution of a mirror-infested NP mode, we employ the hybrid-kinetic particle-in-cell code \texttt{Pegasus++} (\citealt{kunz14}; Arzamasskiy {\it et al.}, in preparation). \texttt{Pegasus++} evolves the ion distribution function $f(t,\bb{r},\bb{v})$ using a collection of positively charged macro-particles that interact with the self-consistent electromagnetic fields $\bb{E}(t,\bb{r})$ and $\bb{B}(t,\bb{r})$, which are in turn evolved on a discrete mesh using Faraday's law and a generalized Ohm's law that includes the inductive electric field, the Hall effect, and a thermoelectric field caused by pressure gradients in the (assumed massless) electron fluid. The latter ensures quasi-neutrality. For simplicity, we adopt an isothermal equation of state for the electrons with temperature $T_{\rm e} = T_{\rm i0}$. Both the interpolation of fields to the macro-particle locations, and the deposition of the macro-particles' phase-space information on the mesh, are performed using second-order-accurate triangle-shaped stencils.

All simulations of the NP mode are performed on a two-dimensional mesh that is elongated in the direction of a mean magnetic field $\bb{B}_0 = B_0\ex$ and spans one full NP mode wavelength, $L_x \times L_y = \lambda_\parallel \times \lambda_\perp$. The latter ranges from $\lambda_\parallel = 1000\rho_{\rm i0}$ to $4000\rho_{\rm i0}$, with aspect ratios of either $\lambda_{\parallel}/\lambda_{\perp}=4$ or $8$. When varying these two dimensions, the transverse dimension is never smaller than $250\rho_{\rm i0}$, thereby guaranteeing sufficient scale separation between the NP mode and any ion-Larmor-scale instabilities. In all runs, the spatial resolution is $\Delta x = \Delta y \simeq 0.3\rho_{\rm i0}$ and the number of macro-particles per cell is either $10^4$ or $5\times 10^3$ (the latter used only in our largest simulations); these values are similar to those used in previously published {\tt Pegasus} simulations of collisionless Alfv\'{e}n waves \citep{squire17num} and IAWs \citep{kunz20} in firehose/mirror-susceptible plasmas. \edit{A digital low-pass filter is applied to the computed moments of the ion distribution function in order to reduce the impact of grid-scale, finite-particle-number noise on the evolution of the NP mode and the trajectories of the particles}.

At $t=0$ we perturb the magnetic field using the vector potential
\begin{equation}
    \bb{A}(x,y) = -\frac{\alpha B_0}{|k|} \sin(k_\parallel x + k_\perp y) \ez,
\end{equation}
where $k_\parallel = 2\upi/\lambda_\parallel$, $k_\perp = 2\upi/\lambda_\perp$, and $\alpha$ is a dimensionless number quantifying the mode amplitude. To excite the NP mode, the associated change in the magnetic pressure, 
\begin{equation}\label{eqn:dBsq}
    \frac{\delta B^2}{8\upi} = - \frac{\alpha B^2_0}{8\upi} \cos(k_\parallel x + k_\perp y ) \left[ \frac{2k_\perp}{|k|} - \alpha \cos(k_\parallel x + k_\perp y ) \right] ,
\end{equation}
must be exactly balanced by a perturbation to the perpendicular pressure of the plasma (cf.~\eqref{eqn:pbalance}). In order to keep the initialization of the latter relatively simple, we choose to begin not from an exact NP eigenmode but rather from an isothermal perturbation to the plasma density $\delta n$, in which case the perturbed perpendicular pressure is simply $\delta p_\perp = \delta n (T_{\rm i0} + T_{\rm e})$. Balancing this expression by~\eqref{eqn:dBsq} and solving for $\delta n$ leads to the initial ion distribution function
\begin{equation}
    f(0,x,y,v) = F_{\rm M}(v) \left\{ 1 + \frac{\alpha}{\beta_0} \cos(k_\parallel x + k_\perp y) \left[ \frac{2k_\perp}{|k|} - \alpha \cos(k_\parallel x + k_\perp y) \right] \right\} .
\end{equation}
In this case, the initial total (magnetic plus thermal) pressure in the simulation domain is constant and equal to  $(B^2_0/8\upi)(1+\beta_0)$; recall that $\beta_0 \doteq \beta_{\rm i0}(1+T_{\rm e}/T_{\rm i0})$. Starting from a pressure-isotropic plasma has the advantage that any pressure anisotropy that develops is generated self-consistently and not put in by hand. It is also consistent with the assumptions made to obtain the analytic solution for $\Delta_{\rm NP}(t)$, equation~\eqref{eqn:Delta}.

In all \edit{but two of} our simulations, we set $\beta_{\rm i0} = 16$, \edit{a value large enough to allow comparison} with the asymptotic expressions derived in \S\ref{sec:slowtheory}, but not so large that we cannot capture a full decay time of the linear NP decay rate. We vary $\alpha\in[0.1,0.8]$, spanning the predicted NP amplitude threshold for triggering the mirror instability \eqref{eqn:ampthreshold}. Special attention is paid to the case with $\lambda_\parallel = 2000\rho_{\mathrm{i}0}$, $\lambda_\perp = 500\rho_{\mathrm{i}0}$, and $\alpha=0.8$; we refer to this as our fiducial case. \edit{Two additional runs, one with $\beta_{\rm i0}=4$ and the other with $\beta_{\rm i0}=36$, both having $\alpha=0.8$, $\lambda_\parallel=1000\rho_{\rm i0}$, and $\lambda_\perp=250\rho_{\rm i0}$, were also performed.}

Hereafter, $\langle\,\cdot\,\rangle$ denotes a spatial average taken over the entire domain; \edit{$\langle\,\cdot\,\rangle_{\rm m}$ denotes a spatial average taken over the mirror-unstable region of the NP mode}; and $\langle\,\cdot\,\rangle_k$ denotes a spatial average taken over the $y$-direction while accounting for the changing position of the wavefront (so as to align all of the perturbed and unperturbed regions within the domain). The latter is referred to as a `wavefront average'; note that it leaves the $x$-coordinate (in the direction of $\bb{B}_0$) unchanged.

%
%
\subsubsection{Overall evolution of the fiducial run}\label{sec:NPoverall}

\begin{figure}
\centering
\includegraphics[width=0.97\textwidth]{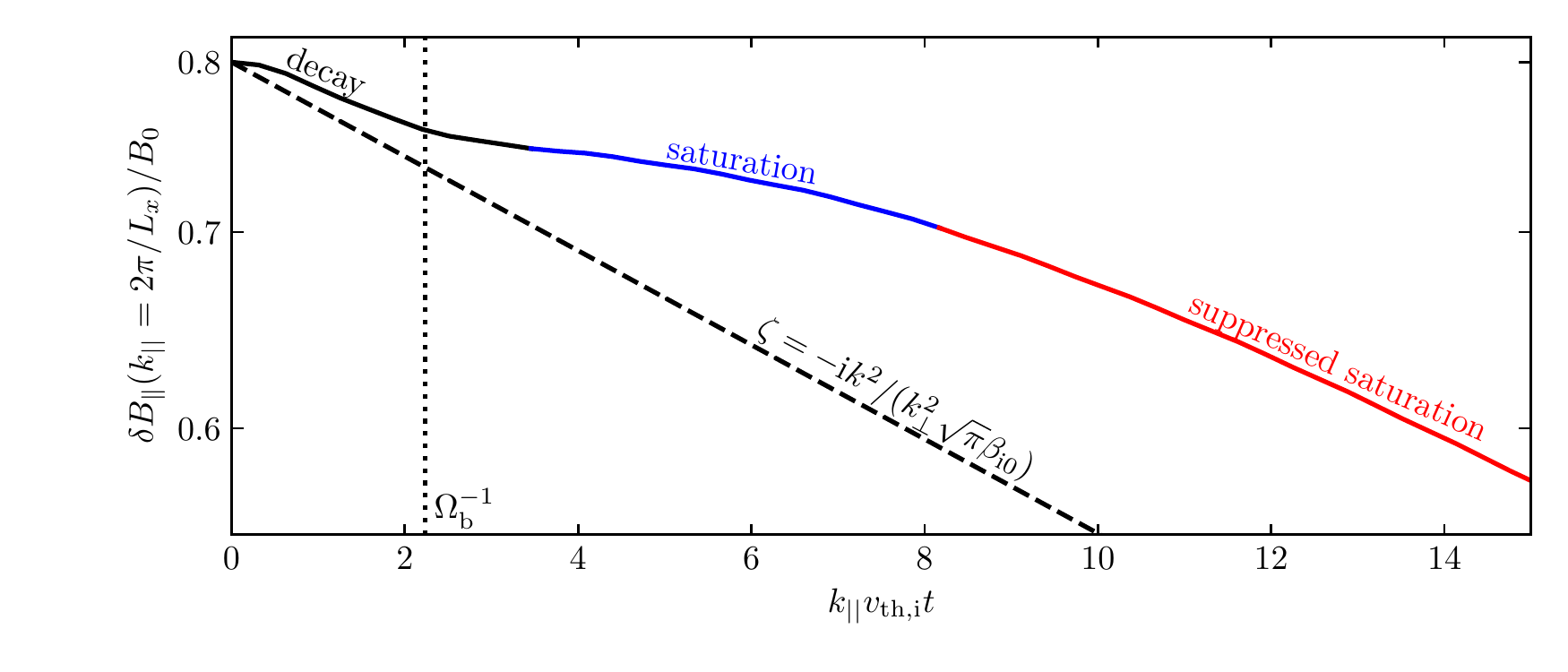}
\caption{Amplitude of the magnetic-field-strength perturbation of the NP mode vs.~time from the fiducial run, with the different phases of the predicted evolution labelled and colour-coded. The dashed line indicates the linear decay rate \eqref{eqn:npdisp} of the NP mode in a pressure-isotropic plasma with $\beta_{\rm i0}\gg 1$. See~\S\ref{sec:NPoverall} for discussion.}
\label{fig:slowmap}
\end{figure}

\begin{figure}
\centering
\includegraphics[width=0.99\textwidth]{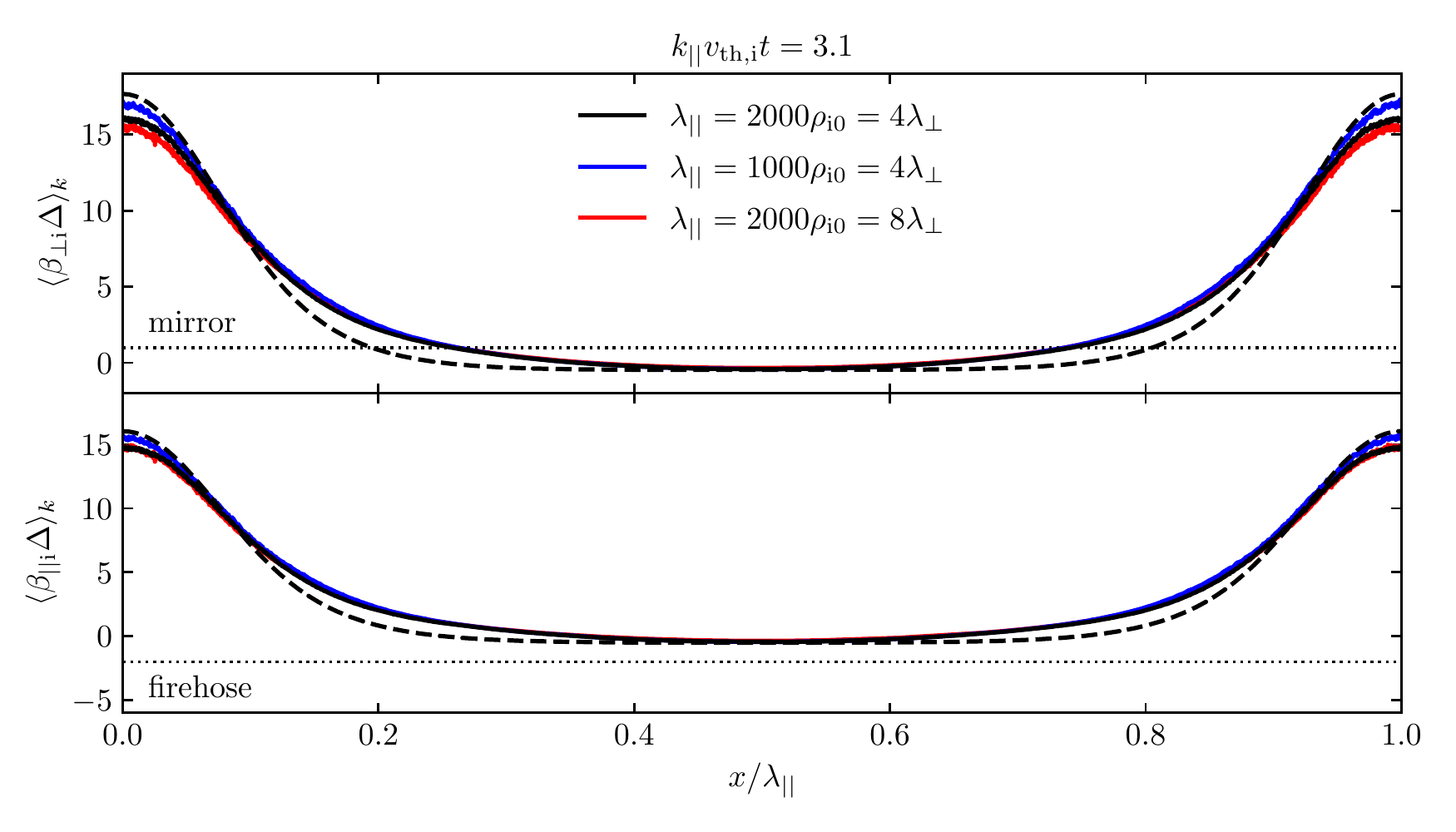}
\caption{Wavefront-averaged profiles of $\beta_{\perp\rm i}\Delta$ and $\beta_{\parallel\rm i}\Delta$ at $k_\parallel v_{\rm th,i}t=3.1$, when the pressure anisotropy is near its maximum value, compared against the theoretical predictions from the linear eigenmode~\eqref{eqn:DNPbeta}, for $\alpha = 0.8$ and different NP mode wavelengths $\lambda_\parallel$ and $\lambda_\perp$. The fiducial run corresponds to the solid black line. Positive values of $\beta_{\rm i}\Delta$ far exceeding the mirror threshold occur in the regions where $\delta B_\parallel < 0$. Elsewhere, negative pressure anisotropy is compensated by a decrease in $\beta_{\rm i}$ to avoid exciting the firehose instability.}\label{fig:dbtheoryprof}
\end{figure}

\begin{figure}
\centering
\includegraphics[width=0.99\textwidth]{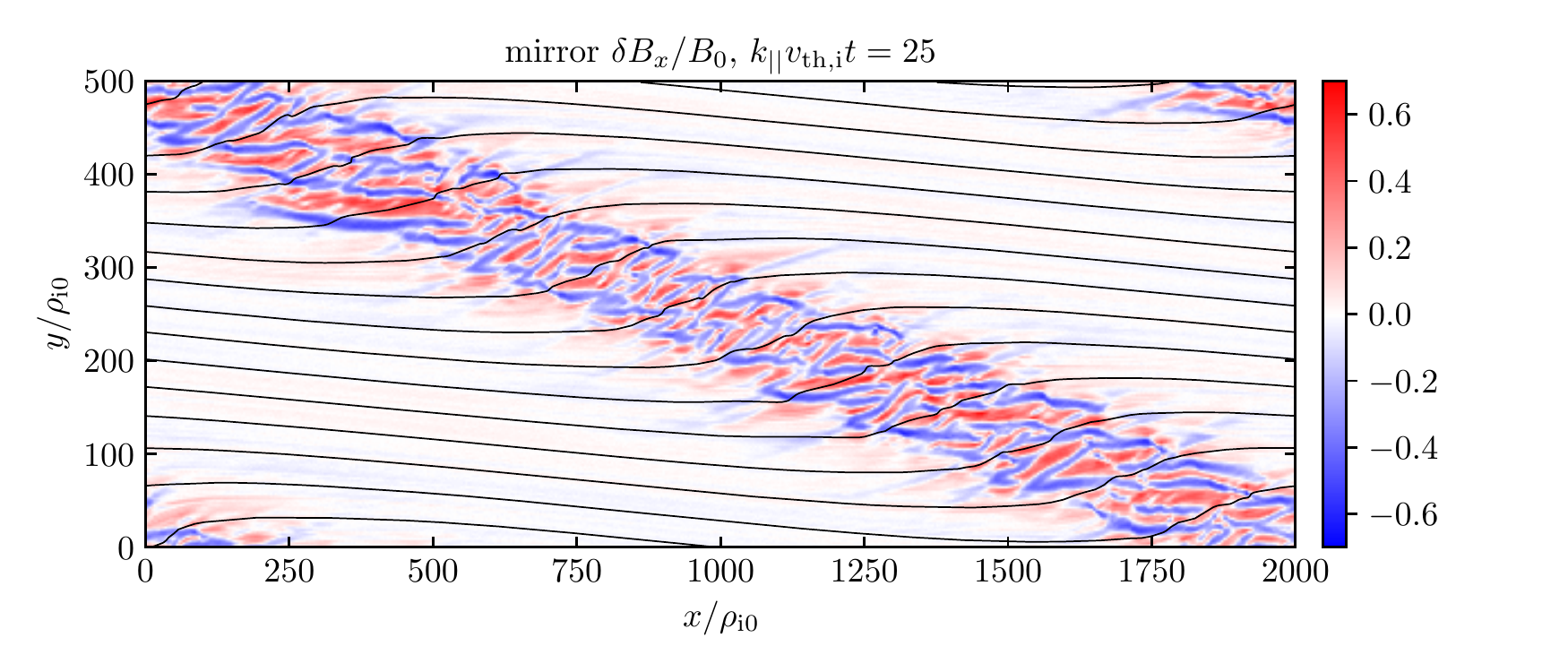}
\caption{The $x$-component of the magnetic-field perturbation, filtered to remove wavenumbers associated with the $\alpha=0.8$ NP mode, at $k_\parallel v_{\rm th,i}t = 25$ in the fiducial run. By this time, the mirror instability is fully nonlinear, causing large-amplitude, small-wavelength deflections in the magnetic-field direction that pitch-angle scatter particles.}\label{fig:mirrorfield}
\end{figure}

We begin our discussion of the \texttt{Pegasus++} results by using the fiducial run to make contact with some of the predictions laid out in \S\ref{sec:slowtheory}. These predictions include the excitation and subsequent linear collisionless damping of the NP mode, its nonlinear saturation, the simultaneous generation of mirror-unstable pressure anisotropy in the regions of the mode where $\delta B_\parallel < 0$, and the resumption of linear damping following the pitch-angle scattering of trapped ions by the saturated Larmor-scale mirrors at a rate larger than the bounce frequency. Figure~\ref{fig:slowmap} illustrates these evolutionary phases by depicting the amplitude of the NP mode versus time. After a rapid adjustment from the isothermal pressure-balanced initial condition, the NP mode emerges and decays at the linear rate (black line) for approximately one bounce time, $\Omega^{-1}_{\rm b}$. Immediately thereafter, the decay stalls (blue line) as nonlinear saturation sets in. Figure~\ref{fig:dbtheoryprof} demonstrates that, meanwhile, the NP mode has produced a large, positive pressure anisotropy in the regions where $\delta B_\parallel < 0$ and almost zero pressure anisotropy elsewhere, consistent with the prediction~\eqref{eqn:DNPbeta} (dashed line). The mirror-unstable region (with $\langle\beta_{\perp\rm i}\Delta\rangle_k$ above the dotted line in figure~\ref{fig:dbtheoryprof}) is seen to occupy ${\sim}40\%$ of the NP mode wavelength, consistent with~\eqref{eqn:mwidth} for $\alpha = 0.8$. It is in this mirror-unstable region that the magnetic field acquires moderate-amplitude, oblique fluctuations in its strength on ion-Larmor scales, which are clearly apparent in figure~\ref{fig:mirrorfield}. The strongest fluctuations occupy roughly a quarter of the box length and acquire amplitudes comparable to that of the mean field. The associated distortions in the field lines ultimately scatter particles at a rate comparable to the bounce frequency (see figure~\ref{fig:nuvxt} and the accompanying discussion in \S\ref{sec:slowcollnum}). As a result, the NP mode amplitude enters a `suppressed saturation' phase (figure~\ref{fig:slowmap}, red line), during which the nonlinear plateau is eroded by the mirror-induced collisionality and the Barnes damping resumes.\footnote{We were not able to discern any fluctuations above the noise floor in the out-of-plane component $B_z$, which would be indicative of the ion-cyclotron instability \citep[e.g.,][]{gl94}. \edit{Such fluctuations appeared, however, in the run at $\beta_{\rm i0}=4$ and $\alpha=0.8$; at this lower value of $\beta_{\rm i0}$, the ion-cyclotron threshold is comparable to the mirror threshold \citep[e.g.,][]{hellinger06}. Nevertheless, no substantive differences in the subsequent evolution of the NP mode were seen.}}


In the remainder of \S\ref{sec:hybrid}, we examine these phases in more detail and their dependence on mode amplitude and scale separation, starting with the mirror-induced scattering and its impact on the NP mode's pressure anisotropy.

%
%
\subsubsection{Effective collisionality: particle scattering and trapping}\label{sec:slowcollnum}
\begin{figure}
\centering
\includegraphics[width=0.96\textwidth]{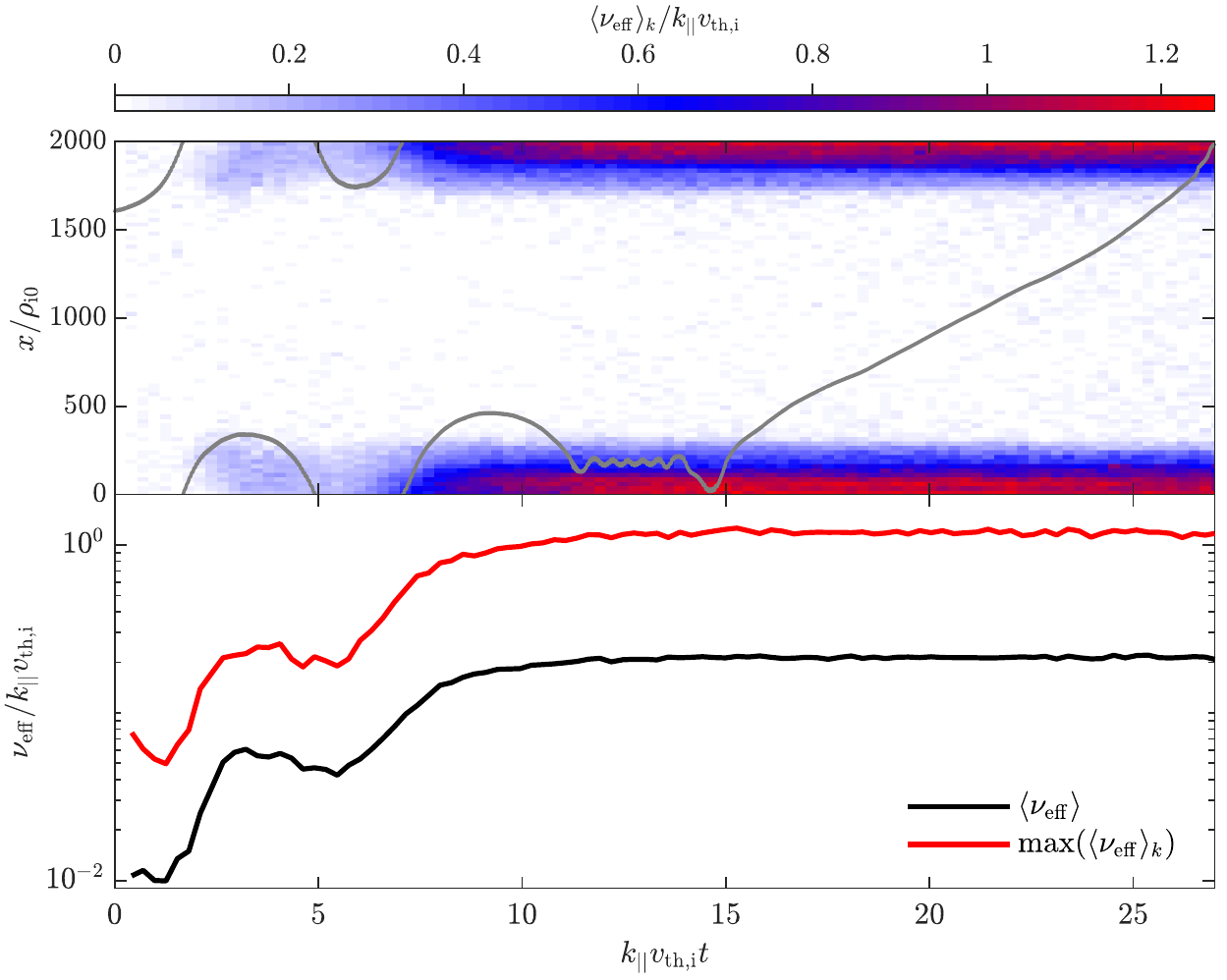}
\caption{Effective collisionality $\nu_{\rm eff}$ caused by the mirror instability in the fiducial run with $\alpha=0.8$ and $\lambda_\parallel=2000\rho_{\rm i0}=4\lambda_\perp$. Top: Space-time diagram of $\langle\nu_{\rm eff}\rangle_k$ (colour). A illustrative particle trajectory is shown with the grey line, exhibiting resonant bouncing, followed by trapping within a mirror fluctuation, and eventual scattering out of resonance with the NP mode. Bottom: Box-averaged (black) and maximum wavefront-averaged (red) collision frequencies vs.~time.}
\label{fig:nuvxt}
\end{figure}

Figure~\ref{fig:nuvxt} displays the evolution of the mirror-induced effective collisionality $\nu_{\rm eff}$ in the fiducial run, calculated following the method used in \citet{kss14,kunz20}, \citet{msk16}, and \citet{squire17num}. Namely, the individual magnetic moments of ${\sim}10^4$ particles are tracked and monitored for (both abrupt and accumulated) changes by at least a factor of $\kappa=1.2$~(as used by \citealt{kunz20} to measure firehose/mirror-induced scattering in unstable IAWs). \edit{The time intervals $\tau$ between which these changes occur are stored, 
along with the locations at which the changes occurred, and a spatially dependent effective collision frequency $\nu_{\rm eff}$ is calculated from the mean scattering time $\langle \tau \rangle$ using $\nu_{\rm eff} \doteq (\ln\kappa)^2/\langle \tau \rangle$.} This calculation was also performed using $\kappa\in[1.1,1.5]$, with no significant differences arising.

In the bottom panel, the box-averaged effective collisionality (black line) and maximum value of the wavefront-averaged effective collisionality (red line) are shown as functions of time. Both exhibit rapid growth during the initial phase of the mirror instability and then reach a quasi-steady state, with ${\rm max}(\langle\nu_{\rm eff}\rangle_k) \approx 0.0035\Omega_{\rm i0} \approx 2.5\Omega_{\rm b}$. We have found the timescale for the scattering rate to reach this steady state to be largely independent of the wavelength of the NP mode, although it increases somewhat with decreasing $\alpha$ because of the slower linear growth rate of the mirror instability. The space-time diagram of the wavefront-averaged collisionality shown in the top panel indicates that the maximum value of $\nu_{\rm eff}$ is localized to the centre of the mirror-unstable region, with slightly smaller values occurring near this region's boundaries where the mirror amplitudes are smaller (cf.~figure~\ref{fig:mirrorfield}). A large fraction of the thermal plasma is subject to this collisionality, because the mode amplitude is large enough that most of the plasma particles are confined in the regions where $\delta B_\parallel < 0$. For example, when $\alpha=0.8$, particles whose pitch angles satisfy $|v_\parallel|/w_\perp \leq \sqrt{{\rm max}(B)/{\rm min}(B)-1} \approx 2.8$ would be mirror-confined in the absence of collisions. Outside of these regions, where the plasma is stable, the collisionality is very low; as a result, the box-averaged collisionality is more than a factor of 5 smaller than the maximum value. The top panel also shows the path of a single tracked particle as a grey line. The initial evolution demonstrates bouncing within the $\delta B_\parallel <0$ region. Once the mirror fluctuations reach nonlinear amplitudes, the particle is temporarily trapped within a growing mirror. Eventually, it scatters enough in pitch angle to become de-trapped and traverses the $\delta B_\parallel > 0$ region, breaking its resonance with the NP mode.

%
%
\subsubsection{Evolution of pressure anisotropy}

\begin{figure}
\centering
\includegraphics[width=0.95\textwidth]{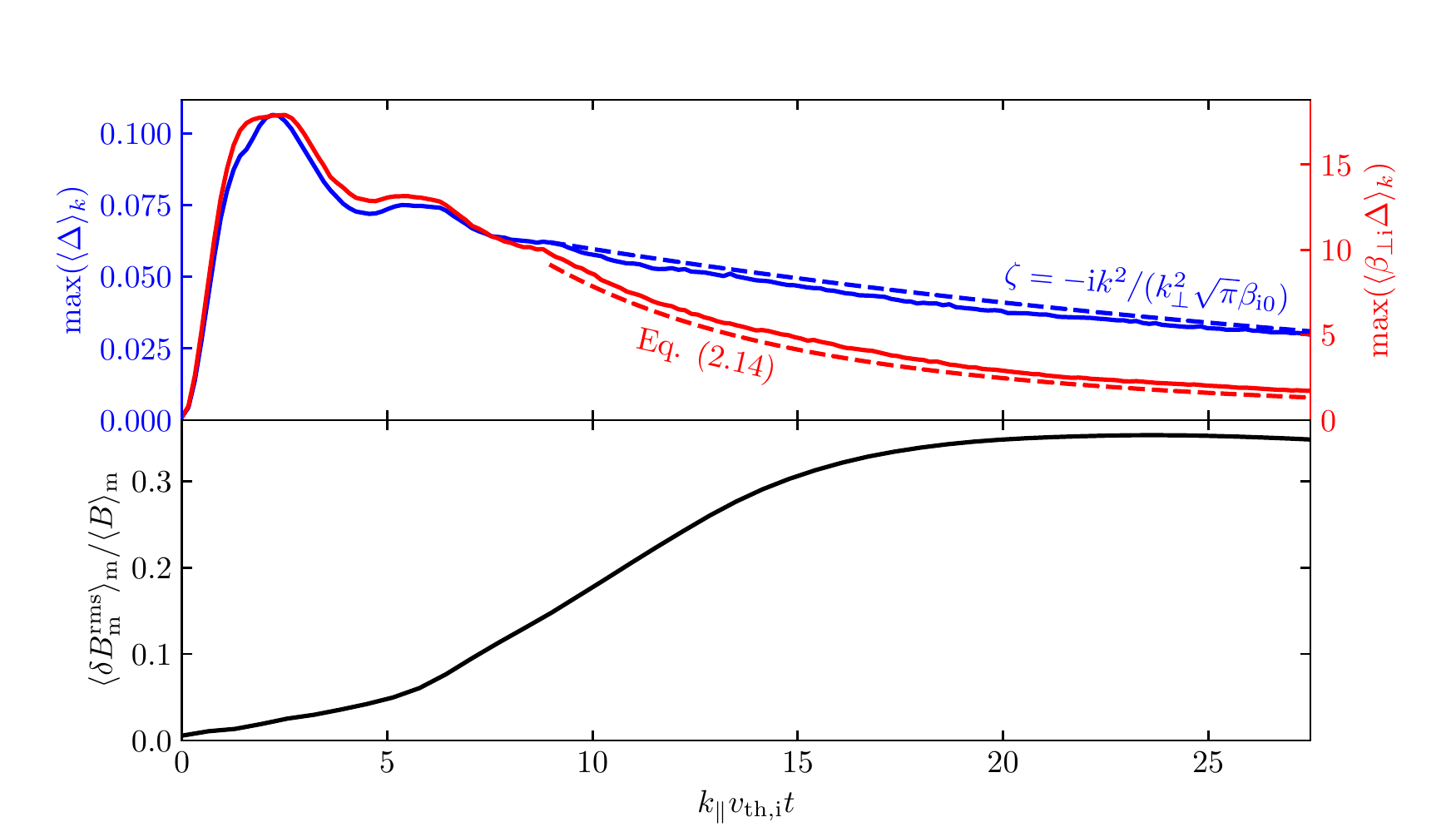}
\caption{Top: Maximum of the wavefront-averaged $\Delta$ (solid blue line) and $\beta_{\rm \perp i}\Delta$ (solid red line) versus time in the fiducial run. The evolution of $\langle\Delta\rangle_k$ matches well the predicted linear evolution (blue dashed line), suggesting that the rapid reduction of $\beta_{\rm \perp i}\Delta$ is due mostly to the resumed decay of the NP mode and the decrease in $\beta_{\perp\rm i}$ caused by the growing mirror fluctuations. Bottom: Root-mean-square amplitude of the mirror fluctuations, averaged over the mirror unstable region \edit{and normalized to the average `background' (i.e., guide-field plus NP-mode) magnetic-field strength in the mirror region}. The growth of the mirror instability coincides with a drop in $\langle\beta_{\rm \perp i}\Delta\rangle_k$.}
\label{fig:dandbvt}
\end{figure}
The top panel of figure~\ref{fig:dandbvt} shows the evolution of the maximum of the wavefront-averaged $\Delta$ and $\beta_{\rm \perp i}\Delta$ in the fiducial run. The bottom panel depicts the growth of the root-mean-square amplitude of the mirror fluctuations, averaged over the mirror-unstable region where $\delta B_\parallel <0$ and \edit{normalized to the average `background' (i.e., guide-field plus NP-mode) magnetic-field strength in this region.} The fluctuations grow large enough to scatter particles and restore the linear decay of the NP mode, through which the pressure anisotropy decays. Indeed, $\langle\Delta\rangle_k$ is similar to the linear prediction~\eqref{eqn:Delta_approx}, denoted here by the blue dashed line. Likewise, $\langle\beta_{\perp\rm i}\Delta\rangle_k$ is modeled well by~\eqref{eqn:DNPbeta} with the substitution $\delta B_\parallel/B_0 = -\alpha\exp(-\imag\zeta k_\parallel v_{\rm th,i}t)$ where $\zeta$ is the linear eigenvalue~\eqref{eqn:npdisp}. This expression is traced by the dashed red line in figure~\ref{fig:dandbvt}, \edit{where we have started the decay at $k_\parallel v_{\rm th,i} t =9$ and set $\alpha=0.68$ in order to account for the delay due to the (temporary) nonlinear saturation}. At larger scale separations, we anticipate that faster pitch-angle scattering induced by the mirrors will be able to regulate the pressure anisotropy more efficiently than its linear decay, at which point the mode will no longer resemble the collisionless linear NP eigenmode (see~\S\ref{sec:scaledependence}). 
\begin{figure}
\centering
\includegraphics[width=0.99\textwidth]{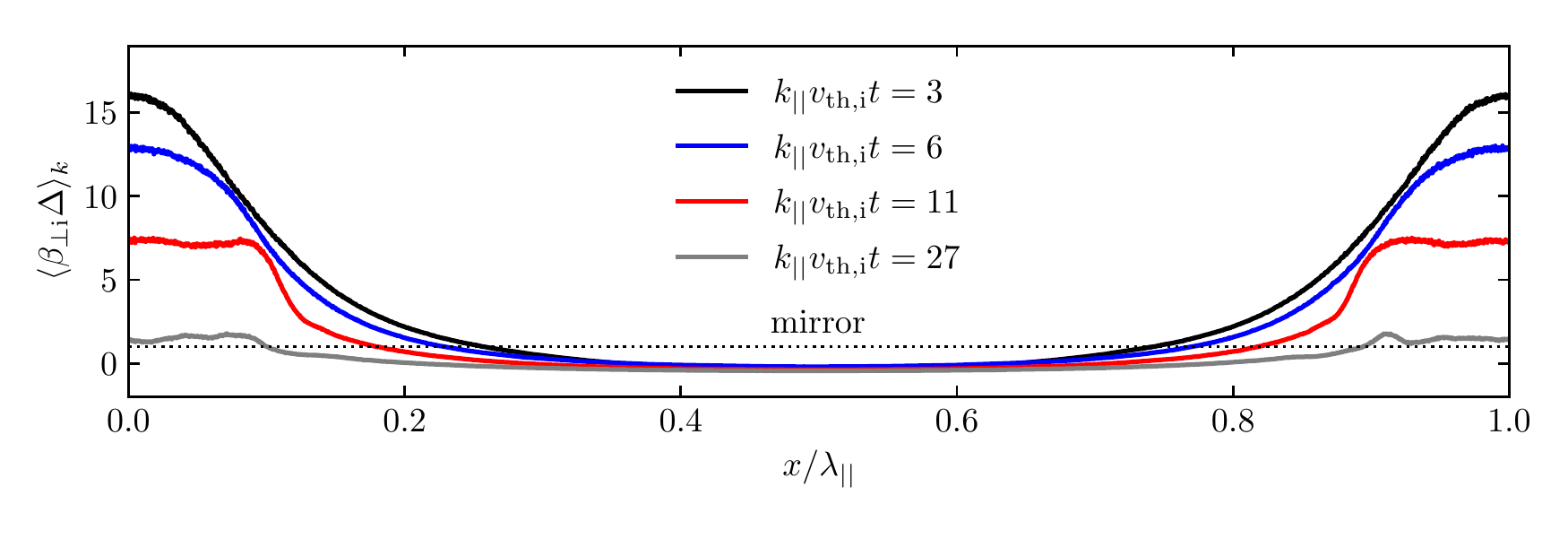}
\caption{Temporal evolution of the wavefront-averaged profile of $\beta_{\rm \perp i}\Delta$. Four times are shown: just after the adjustment into the NP eigenmode during the initial decay phase (black line); an intermediate time during which the NP mode decay is saturated (blue line); after the mirrors become nonlinear and scatter particles fast enough to suppress the NP mode's saturation (red line); and later once $\beta_{\rm \perp i}\Delta$ has been reduced enough that the mirrors are marginally stable (grey line).}
\label{fig:dbvt}
\end{figure}

The growth of mirrors leads to modifications in the shape of the NP mode profile, as shown in figure~\ref{fig:dbvt}. The evolution of the wavefront-averaged profile of $\beta_{\rm \perp i}\Delta$ in the fiducial run at $k_\parallel v_{\rm th,i}t = 3$, 6, 11, and 27 is shown. The profile in the region where the mirror instability is active has flattened, although the mode seems to remain close to the linear eigenmode, as evidenced by figure~\ref{fig:dandbvt}. The reduction in $\beta_{\rm \perp i}\Delta$ occurs considerably faster than the linear decay of $\Delta$ by itself, which highlights the importance of $\beta_{\perp \rm i}$ in achieving marginal stability. This reinforces the idea that the mirror fluctuations do not so much act directly on the anisotropy to achieve $\beta_{\perp \rm i}\Delta = 1$, but rather they enable the NP mode to decay and reduce both $\Delta$ \textit{and} $\beta_{\perp i}$ to achieve marginal stability more rapidly than would otherwise occur.

%
%
\subsubsection{Suppression of nonlinear saturation and resumption of transit-time damping}

\begin{figure}
\centering{
\mbox{\hspace{3em}$(a)$\hspace{0.44\textwidth}$(b)$\hspace{0.45\textwidth}}\\
  \begin{subfigure}{0.475\linewidth}
    \includegraphics[width=\linewidth]{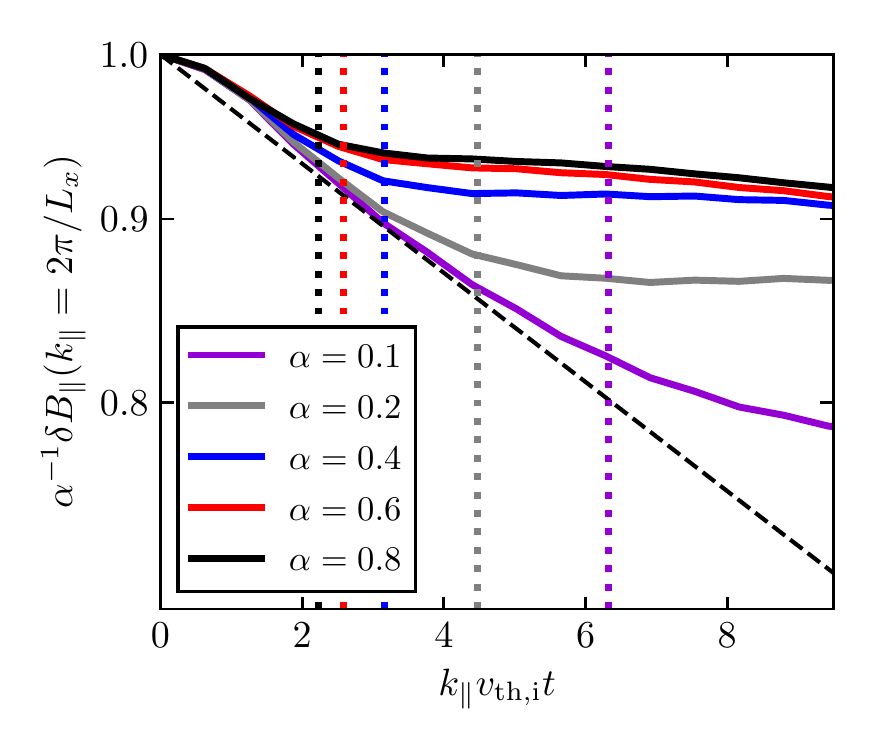}
  \end{subfigure}
  \quad
  \begin{subfigure}{0.47\linewidth}
    \includegraphics[width=\linewidth]{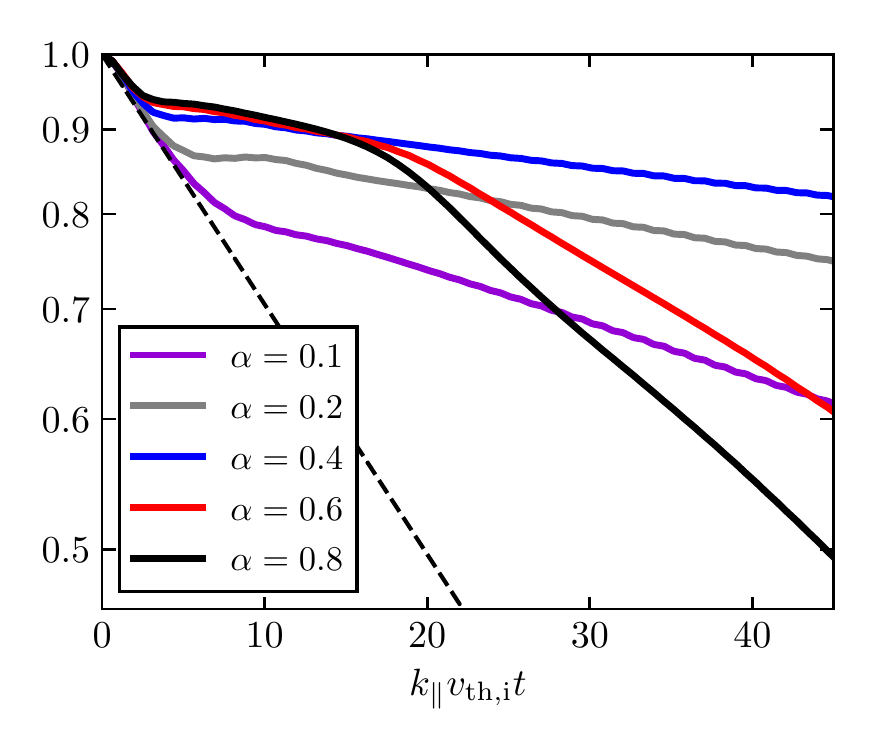}
  \end{subfigure}
 }
\caption{Amplitude of the magnetic-field-strength perturbation of the NP mode, normalized to its initial value, vs.~time for $\lambda_\parallel = 1000\rho_{\rm i0} = 4\lambda_\perp$ and different $\alpha$. ($a$) Early times, during which the NP mode nonlinearly saturates after approximately one bounce time ${\sim}\Omega^{-1}_{\rm b}$ (vertical dotted lines; see~\eqref{eqn:bounce}). The dashed line indicates the linear decay rate~\eqref{eqn:npdisp}. ($b$) Late times, showing suppression of nonlinear saturation for amplitudes $\alpha \ge 0.6$.}\label{fig:satvamp}
\end{figure}

The effects of nonlinear saturation and mirror-induced collisionality across a variety of NP mode amplitudes can be seen in figure~\ref{fig:satvamp}. For reasons of computational cost, for these runs we used $\lambda_\parallel=1000\rho_{\rm i}$ rather than the fiducial $2000\rho_{\rm i}$. A Fourier transform is used to select the magnitude of the box-wavelength perturbation to the background field (i.e., the amplitude of the NP mode); this quantity is plotted as a function of time. In panel ($a$), the initial phase of evolution is featured, at first demonstrating linear decay at a rate similar to the prediction~\eqref{eqn:npdisp} (shown by a black dashed line), approximately independent of $\alpha$. After roughly one bounce time (marked by dotted lines of matching colour), the decay begins to stall and the mode amplitude tends towards a constant value. This nonlinear saturation occurs at earlier times for larger mode amplitudes, trending with the $\alpha^{-1/2}$ scaling of the bounce time (see~\eqref{eqn:bounce}). At amplitudes $\alpha\gtrsim 0.4$, more than 90\% of the original mode amplitude is preserved by the nonlinear saturation, suggesting that large-amplitude collisionless NP modes at high~$\beta$ can be rather long lived. 

Figure~\ref{fig:satvamp}($b$) shows the behaviour of these modes over longer timescales. For amplitudes $\alpha \le 0.4$, nonlinear saturation remains and the linear decay rate is never again realized. \edit{By contrast, the larger values of pressure anisotropy in the $\alpha=0.6$ and $0.8$ NP modes produce mirror fluctuations with amplitudes large enough to interfere with the maintenance of the nonlinear plateau. These modes are then able to decay further and convert magnetic energy into particle energy through a balance between plateau generation and pitch-angle scattering. The linear decay rate is fully re-established at $\alpha = 0.8$. A slightly weaker decay rate is seen in the $\alpha=0.6$ case because of the slower mirror-induced scattering rate relative to $k_\parallel v_{\rm th,i}$; at the larger scale separation of $\lambda_\parallel=2000\rho_{\rm i0}$ (not shown), the full linear decay rate is re-established for $\alpha=0.6$.} With the value of $\lambda_\parallel$ used in these runs being twice smaller than that in the fiducial run, it is notable that the time at which near-linear decay is restored by mirror-induced scattering is the same in units of $\Omega_{\rm i0}$. At scale separations much larger than those we are able to simulate currently, we therefore anticipate the nonlinear plateau to be eroded almost instantly by rapid mirror growth and its associated particle scattering.

\begin{figure}
\centering
\includegraphics[width=0.94\textwidth]{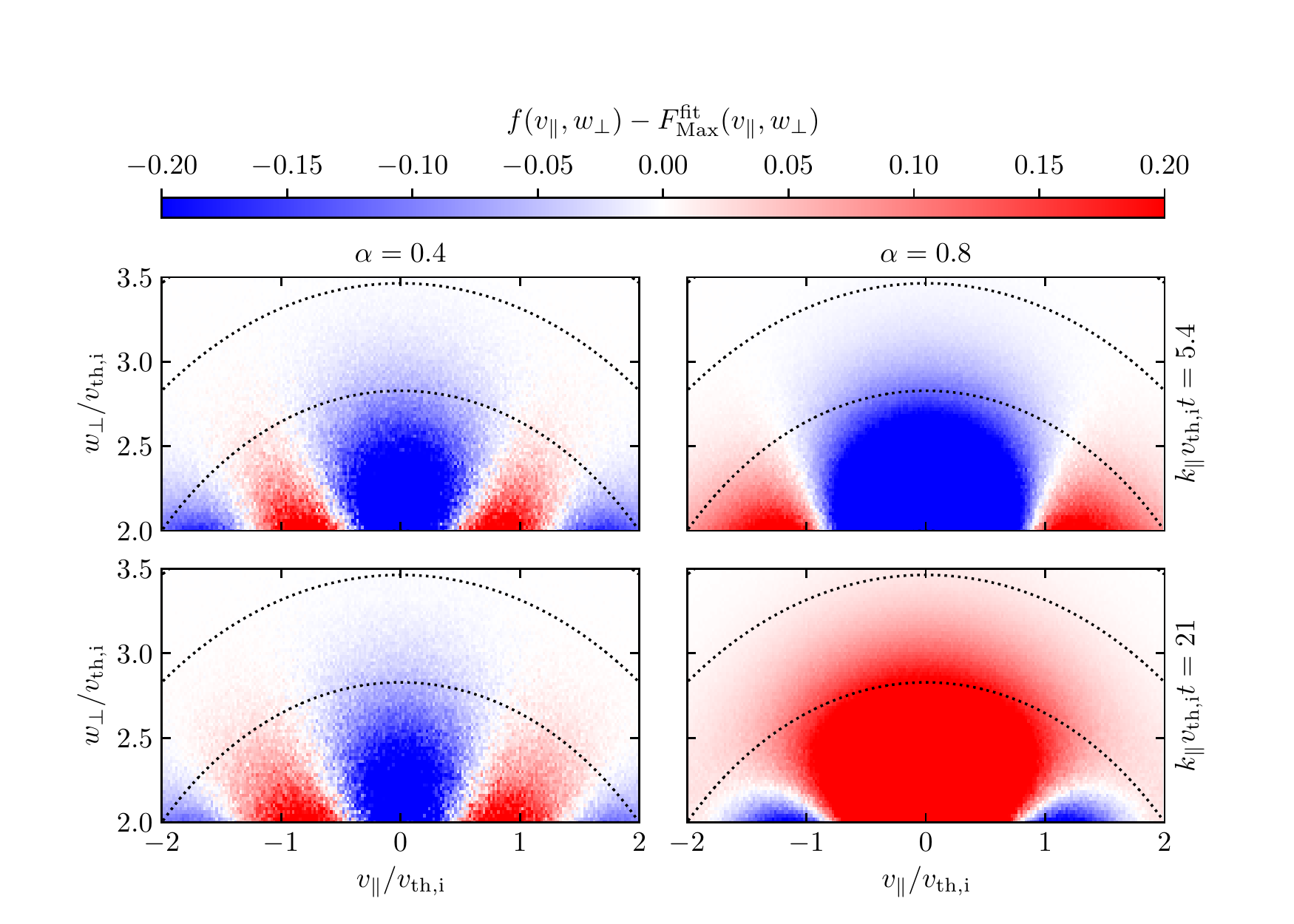}
\caption{\edit{Ion velocity distribution functions $f(v_\parallel,w_\perp)$ measured within the $\delta B_\parallel < 0$ region, with bi-Maxwellian fits subtracted, from two simulations having $\lambda_\parallel=4\lambda_\perp$ and either $\alpha=0.4$ (left column) or $0.8$ (right column). The top (bottom) row of panels corresponds to a time $k_\parallel v_{\rm th,i}t=5.4$ ($=21$)}. The colour bar has been allowed to saturate for the purpose of showing detail. Dotted lines represent isocontours of total energy, $w_\perp^2+v_\parallel^2={\rm const}$.}\label{fig:plateau}
\end{figure}

Our final piece of evidence that the nonlinear plateau is maintained at subcritical NP mode amplitudes and eroded at supercritical amplitudes is also the most direct. \edit{In figure~\ref{fig:plateau} we show ion velocity distribution functions $f(v_\parallel,w_\perp)$ measured within the $\delta B_\parallel < 0$ region, with bi-Maxwellian fits subtracted, from two runs having $\lambda_\parallel=4\lambda_\perp$ and either $\alpha=0.4$ (left column) or $0.8$ (right column). These distribution functions were time-averaged over two intervals of duration $4(k_\parallel v_{\rm th,i})^{-1}$ centred about $k_\parallel v_{\rm th,i}t = 5.4$ (top row) and $k_\parallel v_{\rm th,i}t = 21$ (bottom row). In the $\alpha=0.4$ run, the distribution is reduced with respect to the bi-Maxwellian at high pitch angles where the ions are well trapped, which indicates flattening in the parallel distribution about $v_\parallel \sim 0$. This feature persists beyond $k_\parallel v_{\rm th,i}t = 21$, and is the cause of the stalled decay seen in figure~\ref{fig:satvamp}. In the $\alpha=0.8$ run, the flattening observed in the early-time distribution function is removed later on, allowing transit-time damping to resume (see figure~\ref{fig:slowmap}). In fact, a considerable enhancement in the phase-space density exists at higher $w_\perp$ near $v_\parallel=0$; we suspect that the resumed damping allows for further betatron heating of trapped particles, leading to a small population of nonthermal particles in the $w_\perp$ tail of the distribution. Note that the width of the flattened regions at early times increases dramatically with amplitude, as is expected from the trapping criterion $|v_\parallel|/w_\perp < \sqrt{|B_{\rm max}|/|B_{\rm min}|-1}$.}

%

%
%
\subsubsection{Dependence on scale separation}\label{sec:scaledependence}
\begin{figure}
\centering
\includegraphics[width=0.98\textwidth]{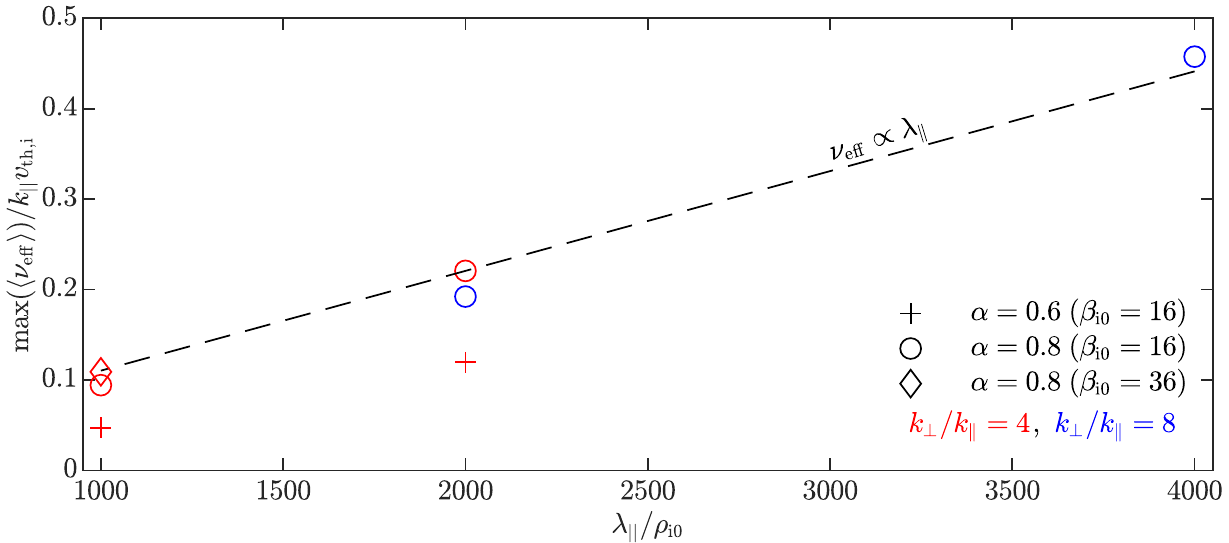}
\caption{Maximum value of the measured mirror-induced effective collision frequency $\nu_{\rm eff,max}$ vs.~NP mode wavelength for $\beta_{\rm i0}=16$ at two different wavenumber obliquities and two different initial amplitudes \edit{(an additional run having $\alpha=0.8$, $\beta_{\rm i0}=36$, and $\lambda_\parallel/\rho_{\rm i0}=1000$ is also included)}. The predicted scaling $\nu/(k_\parallel v_{\rm th,i}) \propto \lambda_{\parallel}$ is shown (dashed black line), normalized to the fiducial case (red circle at $\lambda_\parallel=2000\rho_{\rm i0}$).}
\label{fig:nuscaledep}
\end{figure}
The effective collision frequency predicted by~\eqref{eqn:nueff} suggests that, if the initial NP mode amplitude and wavenumber obliquity were held constant, then increasing the wavelength of the mode should have no effect on the collision frequency. This can be recast as a more illustrative relationship between the thermal crossing time and the collision frequency, $\nu_\mathrm{eff}/(k_\parallel v_{\rm th,i}) \propto \lambda_{\parallel}$. Figure~\ref{fig:nuscaledep} shows the maximum value of the box-averaged effective collision frequency normalized to $k_\parallel v_{\rm th,i}$ for a few different NP mode wavelengths, wavenumber obliquities, and amplitudes. The measured values exhibit good agreement with the proportionality expectation at both wavenumber obliquities. This evidence implies that, at yet longer wavelengths, the collision frequency will continue to increase with respect to the transit time. Note that the measured collisionality for $\alpha=0.6$ is approximately a factor of two smaller than  for $\alpha=0.8$, in qualitative agreement with the prediction featured in figure~\ref{fig:nueff}($a$) that the scattering should decrease with decreasing NP mode amplitude. The fact that the simulated NP mode with $\alpha=0.4$ and $\lambda_\parallel=1000\rho_{\rm i0}$ does not have its nonlinear saturation interrupted by mirrors is also consistent with the prediction in figure~\ref{fig:nueff}($b$). \edit{Finally, the collisionality measured in the run having $\alpha=0.8$, $\lambda_\parallel/\rho_{\rm i0}=1000$, and $\beta_{\rm i0}=36$ (red diamond) is comparable to that in the otherwise-equivalent $\beta_{\rm i0}=16$ run (red circle), consistent with the theoretical expectation \eqref{eqn:nueff} that $\nu_{\rm eff}$ should be independent of $\beta_{\rm i0}$ for $\beta_{\rm i0}\gtrsim 10$.}\footnote{\edit{The effective collisionality measured in the $\beta_{\rm i0}=4$ run satisfies ${\rm max}(\langle\nu_{\rm eff}\rangle)\simeq 0.027 k_\parallel v_{\rm th,i}$, a value that is larger than predicted because of the additional scattering from Larmor-scale magnetic perturbations driven by the ion-cyclotron instability and because the prediction \eqref{eqn:nueff} is accurate only for $\beta_{\rm i0}\gtrsim 10$.}}

As conjectured in~\S\ref{sec:suppression}, the linear scaling of $\nu_{\rm eff}/(k_\parallel v_{\rm th,i})$ with $\lambda_\parallel$ suggests a possible fluid-like regime at sufficiently long NP-mode wavelengths. To investigate this regime, if only approximately, we examine the linear decay rate of NP modes in the presence of a constant pitch-angle scattering rate, shown in figure~\ref{fig:npdisp}. The details of how we determined this decay rate are given in Appendix~\ref{app:mslin}; note that the real part of the frequency is zero for all scattering rates, i.e., the mode remains non-oscillatory. On the left-hand side of the plot, the collision frequency is small and the collisionless NP mode is recovered; on the right-hand side, the collision frequency is large and the mode becomes the MHD entropy mode. The MHD entropy mode is similar to the kinetic NP mode in that it too has no real frequency, but in the fully collisional limit it involves only a density perturbation. For the employed values of $k_\perp/k_\parallel=4$ and $\beta_{\rm i0}=16$, the transition between these two regimes occurs at $\nu \approx 3 k_\parallel v_{\rm th,i}$. Using an asymptotic expansion at high~$\beta$ and $k\simeq k_\perp$, one can show that the transitional collisionality scales approximately as $\nu \sim (3/4)\sqrt{\beta_{\rm i}} k_\parallel v_{\rm th,i}$. With \edit{$\nu_{\rm eff}/\Omega_{\rm i0} \sim (3$--$6)\times 10^{-4}$} for $\alpha\gtrsim 0.6$ (see figures~\ref{fig:nueff} and \ref{fig:nuscaledep}), we estimate that the transition to the collisional regime requires a scale separation \edit{$\lambda_\parallel/\rho_{\rm i0}\gtrsim 10^4\sqrt{\beta_{\rm i0}}$}. Under this condition, the mirror-induced scattering will both isotropize the pressure perturbation and prevent resonant particles from continuously sapping energy from the wave, thereby reducing the decay rate and morphing the collisionless NP mode into the MHD entropy mode \edit{(at least for as long as the mirrors continue to scatter particles faster than ${\sim}\sqrt{\beta_{\rm i}} k_\parallel v_{\rm th,i}$)}. Unfortunately, unless the scale separation is extremely large (e.g., $\lambda_\parallel/\rho_{\rm i0}\gtrsim 10^5$ for our parameters), the decay rate will not be much slower than in the $\nu=0$ case. In the absence of affordable numerical simulations to test this point,\footnote{\edit{To accomplish this at $\beta_{\rm i0}=16$ would require a parallel wavelength roughly $40\times$ larger than used in our largest simulation. With the computational cost being ${\propto}(\lambda_\parallel/\rho_{\rm i0})^2$ (accounting for the proportionally longer run times needed at larger scale separations), such a run would require ${\sim}10^9$~CPU-hours to complete.}} we simply conjecture that at asymptotic wavelengths the reduction in the decay rate would allow these NP structures to become long lived once again, much like their below-threshold, non-linearly saturated counterparts.

\begin{figure}
\centering
\includegraphics[width=\textwidth]{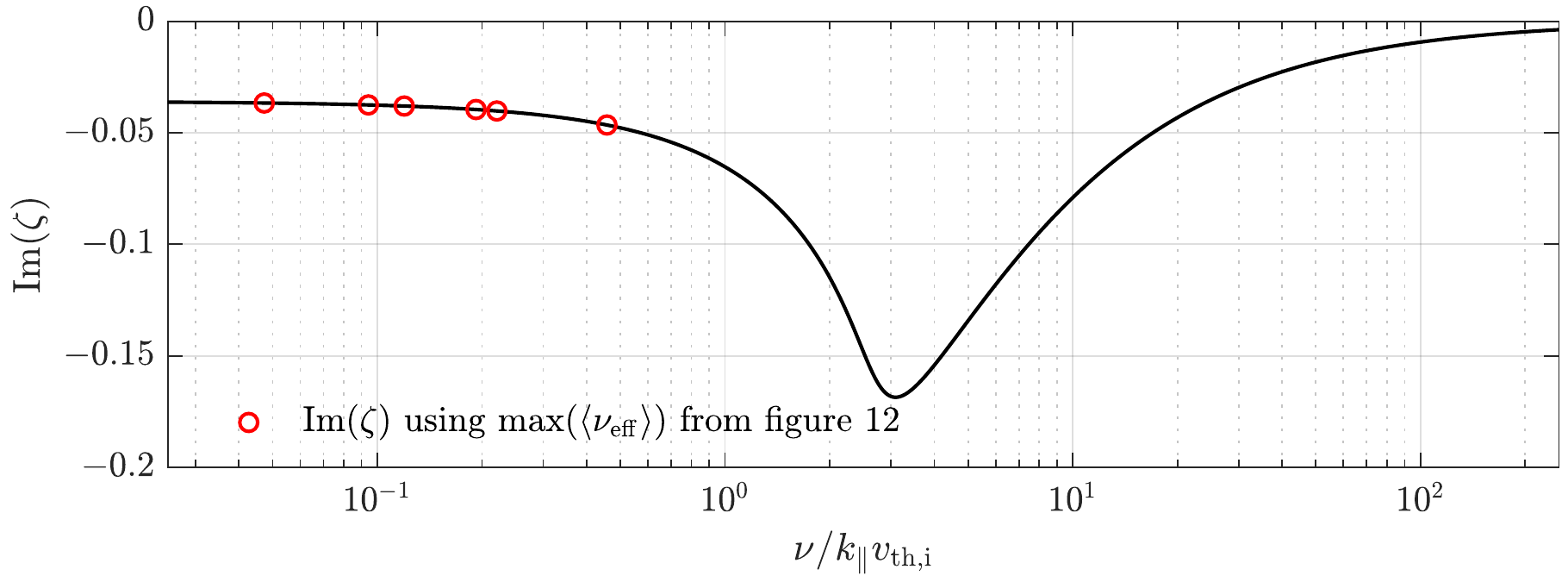}
\caption{Linear decay rate of the NP mode obtained from the Landau-fluid CGL-MHD equations~\eqref{eq:cglmhd} (see Appendix~\ref{app:mslin} for details). The dimensionless (complex) frequency $\zeta\doteq \omega/(|k_\parallel|v_{\rm th,i})$ is computed numerically as a function of collisionality $\nu/(|k_\parallel|v_{\rm th,i})$ for $k_\perp = 4|k_\parallel|$, $\beta_{\rm i0} = 16$, and $T_\mathrm{e} = T_\mathrm{i0}$. Overlaid are red circles marking the maximum box-averaged scattering rates measured in our hybrid-kinetic simulations (see figure~\ref{fig:nuscaledep}).}\label{fig:npdisp}
\end{figure}

\subsection{Summary of key results on the NP mode}

For the reader's benefit, we summarize here the essential findings of our investigation of the NP mode in magnetized, high-$\beta$, collisionless plasmas:
\vspace{0.5em}
\begin{itemize}[align=left,leftmargin=2em,itemindent=0pt,labelsep=0pt,labelwidth=1.5em]
  \setlength\itemsep{0.5em}
  \item Transit-time (Barnes) damping of NP modes nonlinearly saturates before substantial collisionless decay when the mode amplitude $|\delta B_\parallel/B_0| \gtrsim \beta_{\rm i0}^{-2}$.
  \item The near-perpendicular pressure balance of the NP eigenmode polarization ensures the production of large positive $\beta_{\rm i}\Delta$ and only weakly negative $\beta_{\rm i}\Delta$.
  \item Above a threshold amplitude of $|\delta B_\parallel/B_0| \approx 0.3$, the pressure anisotropy affiliated with the NP eigenmode becomes unstable to the mirror instability; at no point is the plasma firehose unstable.
  \item Once the growing mirror fluctuations become nonlinear, they pitch-angle scatter particles according to~\eqref{eqn:nueff}, a rate which is independent of the NP mode wavelength.
  \item At wavelengths satisfying $\sqrt{\beta_{\rm i}} \gtrsim \nu/(k_\parallel v_{\rm th,i}) \gtrsim |\delta B_\parallel/B_0|^{1/2}$, the induced scattering is only fast enough to erode the nonlinear plateau, causing the mode to resume its decay close to the linear (collisionless) rate.
  \item At longer wavelengths satisfying $\nu/(k_\parallel v_{\rm th,i}) \gg \sqrt{\beta_{\rm i}}$, transit-time damping will be interrupted entirely. We predict that in this limit the mode will behave more like the MHD entropy mode.
\end{itemize}

%
%
\section{Fast modes: Wave steepening and viscous damping}\label{sec:fast}

%
%
\subsection{Theory}\label{sec:fasttheory}

\subsubsection{Model equations and assumptions}

Collisionless fast magnetosonic waves are in many ways simpler than their non-propagating counterparts, particularly so if their wavevectors are nearly perpendicular to the background magnetic field, \textit{viz.}~$k_\perp \gg k_\parallel$. In this limit, collisionless damping is extremely weak, and magnetic tension plays virtually no role in the mode's propagation. In fact, for exactly perpendicular propagation ($k_\parallel=0$), Landau and Barnes damping are entirely absent at long wavelengths due to the limited cross-field transport of magnetized particles. In this case, no kinetic information about these modes other than their pressure anisotropy is needed, and they can be described entirely within double-adiabatic MHD -- a model that results from taking the first three fluid moments of the drift-kinetic system (see Appendix~\ref{app:mslin}) and dropping the heat fluxes. Setting $\bb{B} = B\ey$ and $\grad=\ex \,\partial/\partial x$, these equations are
\begin{subequations}\label{eq:dmhd}
\begin{gather}
    \bigD{t}{n} = -n \pD{x}{u_\perp}, \label{eqn:dmhd_n} \\*
    m_{\rm i} n \bigD{t}{u_\perp} = - \pD{x}{} \biggl( p_{\perp\rm i} + p_{\rm e} + \frac{B^2}{8\upi} \biggr), \label{eqn:dmhd_force} \\*
    \bigD{t}{B} = -B \pD{x}{u_\perp} , \label{eqn:dmhd_B}\\*
    \bigD{t}{}\biggl(\frac{p_{\perp\rm i}}{nB}\biggr) = 0 , \label{eqn:dmhd_pprp} \\*
    \bigD{t}{}\biggl(\frac{p_{\parallel\rm i} B^2}{n^3}\biggr) = 0,\label{eqn:dmhd_pprl}
\end{gather}
\end{subequations}
where ${\rm D}/{\rm D}t \doteq \partial / \partial t + u_\perp \partial/\partial x$. Although the right-hand side of~\eqref{eqn:dmhd_force} is independent of the parallel pressure, and so~\eqref{eqn:dmhd_pprl} is not needed to close this set of equations, equation~\eqref{eqn:dmhd_pprl} is nevertheless useful for calculating the fast-wave pressure anisotropy. As in~\S\ref{sec:slow}, we adopt a simple equation of state for the electrons, $p_\mathrm{e} = n T_\mathrm{e}$, with $T_{\rm e}$ being constant.\footnote{Having the electrons respond double-adiabatically would simply double the pressure anisotropy associated with the fast wave and send $T_{\rm e}/2T_{\rm i0}\rightarrow T_{\rm e}/T_{\rm i0}$ in \eqref{eqn:omegafast}.}

In what follows, we investigate analytically two features of fast-wave propagation in a collisionless, magnetized plasma, adopting the simple but illustrative case of $k_\parallel = 0$. First, we demonstrate that such waves nonlinearly steepen quicker in double-adiabatic MHD than they do in standard (pressure-isotropic) MHD, a direct consequence of the proportional relationship between $T_\perp$ and $B$ associated with $\mu$ conservation, equation~\eqref{eqn:dmhd_pprp}. Second, we show how the resulting pressure anisotropy can destabilize the plasma to both firehose and mirror instabilities. We then estimate the effective scattering frequency introduced into the plasma by these instabilities and discuss how the consequent regulation of the pressure anisotropy affects the characteristics of the fast wave.

Before proceeding, it is useful to linearize \eqref{eq:dmhd} to obtain the fast-wave dispersion relation and eigenmode. Perturbing the plasma about a uniform background having density $n_0$, isotropic ion pressure $p_{\rm i0}$, and magnetic-field strength $B_0$, we find that
\begin{equation}\label{eq:fasteigvec}
    \frac{\delta p_{\perp,\mathrm{i}}}{p_{\rm i0}} = 2\frac{\delta B}{B_0}  \qquad \text{and} \qquad \frac{\delta p_{\parallel\rm i}}{p_{\rm i0}} = \frac{\delta n}{n_0} = \frac{\delta B}{B_0} .
\end{equation}
These equations state that the density and pressure anisotropy are positively correlated with the magnetic-field strength, with the parallel ion temperature remaining constant. The dispersion relation of this double-adiabatic (`da') fast wave is
\begin{equation}\label{eqn:omegafast}
    \omega = k_\perp v_\mathrm{A} \sqrt{1 + \beta_{\rm i0} \biggl(1+\frac{T_\mathrm{e}}{2T_\mathrm{i0}} \biggr)} \doteq k_\perp v_\mathrm{ms,da},
\end{equation}
so that the bulk velocity $u_\perp = v_{\rm ms,da}(\delta B/B_0)$. For comparison, the dispersion relation of a fast wave in single-adiabatic (`sa') MHD is
\begin{equation}
    \omega = k_\perp v_\mathrm{A} \sqrt{1 + \beta_{\rm i0}\biggl(\frac{\Gamma}{2} +\frac{T_\mathrm{e}}{2T_\mathrm{i0}} \biggr)} \doteq k_\perp v_\mathrm{ms,sa} ,
\end{equation}
where $\Gamma$ is the adiabatic index of the ions. The proportional relation between the magnetic-field strength and the density in the double-adiabatic model means that $v_{\rm ms,da}>v_{\rm ms,sa}$. This increase will play a role in allowing double-adiabatic fast waves to form shocks faster than single-adiabatic fast waves, especially so at high~$\beta$.

%
%
\subsubsection{Wave steepening in double- versus single-adiabatic MHD}

For waves in which the perturbed quantities determine the wave propagation speed, steepening may result. Large-amplitude waves in particular generate significant differences in the propagation speed between the peaks and the troughs, a situation expected to occur in both double- and single-adiabatic MHD fast waves. In this section, we perform a series of manipulations to the system \eqref{eq:dmhd} in order to quantify this effect. Before proceeding, it is convenient to renormalize quantities using the Alfv\'en speed $v_{\rm A} = B_0/(4\upi m_{\rm i} n_0)^{1/2}$ and the wavelength $\lambda$ as follows: $u_{\perp} = \widetilde{u}_{\perp} v_\mathrm{A}$, $B = \widetilde{B}{B_0}$, $n = \widetilde{n} n_0$, $x = \widetilde{x} \lambda$, $t = \widetilde{t} \lambda/v_\mathrm{A}$, and $p_{\perp,\mathrm{i}} = \widetilde{p}_{\perp\rm i} m_\mathrm{i}n_0 v_\mathrm{A}^2$. We also note that, if the perturbations satisfy $\delta \widetilde n = \delta \widetilde B$ at $t=0$, then these two quantities will remain equal for all times (see equations~\eqref{eqn:dmhd_n} and \eqref{eqn:dmhd_B}); we can then eliminate $\delta \widetilde n$ in favour of $\delta \widetilde B$.\footnote{This reduction is equivalent to assuming an adiabatic index of $\Gamma=2$. In fact, when comparing the results of this analysis to an MHD treatment with isothermal electrons, the substitution $\Gamma=2$ recovers the double-adiabatic result (see~\eqref{eq:dabreak} and~\eqref{eq:sabreak}).} Meanwhile, if $\delta \widetilde{B}$ is small and its associated perturbations in $\widetilde{p}_{\perp,\mathrm{i}}$ and $\widetilde{n}$ are given by \eqref{eq:fasteigvec}, equation~\eqref{eqn:dmhd_pprp} becomes
\begin{equation}
    \pD{\widetilde{t}}{} \biggl( \frac{\widetilde p_{\rm \perp,i}}{\widetilde n \widetilde B} \biggr) \approx - \widetilde{u}_{\perp} \frac{\beta_{\rm i0}}{2} \pD{\widetilde{x}}{(\delta\widetilde{B})^2} \sim \mathcal{O}\bigl[(\delta \widetilde{B})^3\bigr].
\end{equation}
Hence, to second order in $\delta \widetilde{B}$, we may treat $\widetilde{p}_{\rm\perp,i} = (\beta_{\rm i0}/2)\widetilde{B}^2$ as the equation of state if the initial condition is an eigenmode. 

Under these conditions, equations~\eqref{eq:dmhd} may be combined to obtain the following system:
\begin{equation}\label{eqn:matlin}
    \pD{\widetilde{t}}{}
    \left[
    \begin{array}{c}
    \widetilde{u}_{\perp} \vphantom{\bigg(}\\*
    \delta \widetilde{B} 
    \end{array}
    \right]
    +
    \left[
    \begin{array}{cc}
    \widetilde{u}_{\perp} & 1+\beta_{\rm i0} \biggl(1 + \dfrac{T_{\rm e}/2T_{\rm i0}}{1+\delta \widetilde{B}} \biggr) \\*
    1+\delta\widetilde{B} & \widetilde{u}_{\perp}
    \end{array}
    \right]
    \pD{\widetilde{x}}{}
    \negthickspace
    \left[
    \begin{array}{c}
    \widetilde{u}_{\perp} \vphantom{\bigg(} \\*
    \delta\widetilde{B} 
    \end{array}
    \right] = 0.
\end{equation}
Defining $\bb{W} = [\widetilde u_\perp, \delta \widetilde B]^\mathrm{T}$, equation~\eqref{eqn:matlin} can be rewritten as $\partial_{\widetilde{t}} \bb{W} + \msb{A}(\bb{W}) \partial_{\widetilde{x}} \bb{W} = 0$, with $\msb{A}(\bb{W})$ being the evolution matrix. By first finding the eigenvalues $l^{(i)}$ and left eigenvectors $\bb{L}^{(i)}$ of $\msb{A}(\bb{W})$, this system can be solved via its characteristic equations, which are given by $\bb{L}^{(i)} \bcdot \rmd\bb{W} = 0$.  These characteristic equations are obeyed along space-time trajectories following $\rmd \widetilde x/\rmd \widetilde t = l^{(i)}$. However, because our equation of state is only valid up to second order in the wave amplitude, we need only to retain those terms of first order in the evolution matrix, and hence in its eigenvalues. Therefore, we expand the characteristic equations to first order and integrate them to find that the combinations
\begin{equation}\label{eqn:charappx}
    \eta_{\pm} = \widetilde u_\perp \pm \widetilde v_\mathrm{ms,da} \delta \widetilde B
\end{equation}
are approximately constant along
\begin{equation}
 \biggl( \D{\widetilde{t}}{\widetilde{x}} \biggr)^\pm = \frac{\eta_+ + \eta_-}{2} \pm \widetilde v_\mathrm{ms,da} \biggl[ 1 + \frac{1+\beta_\mathrm{i0}}{4 \widetilde v_\mathrm{ms,da}^3}(\eta_+ -\eta_-) \biggr].
\end{equation}
These can be reformulated as two nonlinear advection equations,\footnote{This process is analogous to that used in the derivation of approximate Riemann solvers for numerical solutions of the MHD equations \citep[e.g.,][]{stone08}. Commonly, the left eigenvector is assumed to be constant when integrating the characteristic equations. Here, we keep terms up to first order in $\delta B$ within $\bb{L}$ to more accurately resolve the wave steepening. The careful reader will note that these expressions do not  transform directly back to an approximate form of~\eqref{eqn:matlin}. This approach focuses on the characteristics of $\msb{A}$, so the leading-order behaviour of~\eqref{eqn:matlin} and the eigenvalues/vectors of $\msb{A}$ are approximated accurately; this is in contrast to expanding $\msb{A}$ itself and truncating past the first correction in $\delta\widetilde{B}$.}
\begin{equation}
    \frac{\partial \eta_\pm}{\partial \widetilde t} + \biggl( \D{\widetilde{t}}{\widetilde{x}} \biggr)^\pm \frac{\partial \eta_\pm}{\partial \widetilde x} = 0.
\end{equation}
Note that, if the initial conditions are those of the fast eigenmode (as previously assumed in the assertion that $\delta \widetilde B = \delta \widetilde n$ for all time), then $\eta_-=0$ for all time. We are then left with
\begin{equation}
    \frac{\partial\eta_+}{\partial\widetilde{t}} + \biggl[ \frac{\eta_+}{2} + \widetilde{v}_\mathrm{ms,da} \biggl( 1 + \frac{1+\beta_{\rm i0}}{4\widetilde{v}_\mathrm{ms,da}^3}\eta_+\biggr)\biggr]\frac{\partial\eta_+}{\partial\widetilde{x}} = 0,
\end{equation}
the solution of which for $\widetilde u_{\perp}$ is given by the method of characteristics as
\begin{equation}\label{eq:usln}
    \widetilde{u}_{\perp}(\widetilde{t},\widetilde{x}) = \delta\widetilde{u}_{\perp 0} \biggl(\widetilde{t},\widetilde{x} - \widetilde{v}_\mathrm{ms,da}\widetilde{t} \biggl[ 1 + \delta\widetilde{B}_0(\widetilde{x}_\mathrm{i}) + \frac{1+\beta_{\rm i0}}{2\widetilde{v}_\mathrm{ms,da}^2} \delta\widetilde{B}_0(\widetilde{x}_\mathrm{i})  \biggr] \biggr),
\end{equation}
where the subscript `0' denotes an initial value, and $x_\mathrm{i}$ is the $x$-position of the source of a given characteristic.
\begin{figure}
\centering
\includegraphics[width=0.95\textwidth]{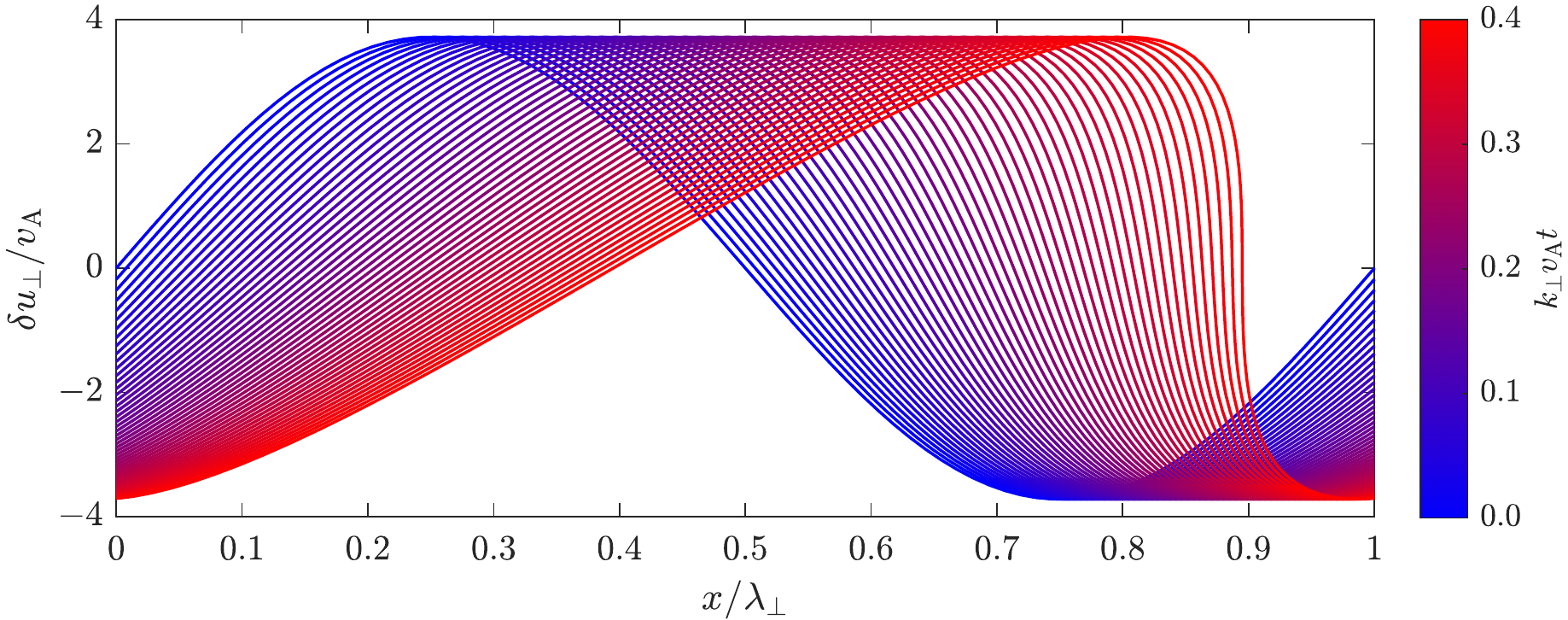}
\caption{Approximate solution~\eqref{eq:usln} to the fast-wave steepening problem with initial amplitude $\alpha = 0.3$ and $\beta_{\rm i0}=25$. The solution has just begun to form a shock, indicating a shock-formation time of $k_\perp v_{\rm A} t_\mathrm{s} \sim 0.4$. }\label{fig:faststeep}
\end{figure}

The time-dependent solution for an example large-amplitude, double-adiabatic fast wave is shown in figure~\ref{fig:faststeep}. This solution is strictly valid only until a shock has formed, at a time that may be determined by evaluating the eigenvalue $l^+$ at the location $x_0$ where its derivative achieves its largest negative value:
\begin{equation}\label{eq:dabreak}
    t^{\rm da}_\mathrm{s} = \bigl[l^+(x_0)\bigr]^{-1} \approx \biggl[\alpha k_\perp v_\mathrm{ms,da} \biggl(1 + \frac{v_{\rm A}^2}{v_\mathrm{ms,da}^2} \frac{1+\beta_{\rm i0}}{2} \biggr)\biggr]^{-1} ,
\end{equation}
where $\alpha \doteq \delta\widetilde{B}(0)$ is the initial fast-wave amplitude. This double-adiabatic (`da') shock-formation time is to be compared to the corresponding time in a single-adiabatic MHD plasma, in which $pn^{-\Gamma}=p_0n^{-\Gamma}_0$. The general problem of fast-wave steepening in MHD plasmas has been studied thoroughly under many conditions~\citep{hada85,odblom98,sujith05}. Following an analogous process to that used for the double-adiabatic fast wave, we find the single-adiabatic (`sa') shock-formation time
\begin{equation}\label{eq:sabreak}
    t_\mathrm{s}^{\rm sa} \approx \biggl[\alpha k_\perp v_\mathrm{ms,sa} \biggl(1 + \frac{v_{\rm A}^2}{v_\mathrm{ms,sa}^2} \frac{1   +\Gamma(\Gamma-1)\beta_{\rm i0}/2}{2} \biggr)\biggr]^{-1}.
\end{equation}
Simplifying \eqref{eq:dabreak} and \eqref{eq:sabreak} at high~$\beta$, and setting $T_{\rm e} = T_{\rm i0}$ and $\Gamma = 5/3$, yields
\begin{equation}
    k_\perp v_{\rm A}t_\mathrm{s}^\mathrm{da} \approx \frac{\sqrt{6}}{4\alpha \sqrt{\beta_\mathrm{i0}}} \qquad \text{and} \qquad  k_\perp v_{\rm A}t_\mathrm{s}^\mathrm{sa} \approx \frac{12\sqrt{3}}{29\alpha \sqrt{\beta_\mathrm{i0}}} \simeq 1.17 k_\perp v_{\rm A}t^{\rm da}_{\rm s}.
\end{equation}    
The single-adiabatic shock-formation time is thus larger than the double-adiabatic shock-formation time. When $T_{\rm e}/T_{\rm i0} = 0$, their ratio reaches a maximum of ${\simeq}1.23$; for $T_{\rm e} \gg T_{\rm i0}$, it approaches unity. This increase is a consequence of the direct correlation between the magnetic-field strength and the perpendicular (ion) pressure in double-adiabatic MHD, which amplifies local changes in the mode propagation speed.

%
%
\subsubsection{Pressure anisotropy and its regulation by kinetic instabilities}\label{sec:fastanis}

By contrast with the NP mode, the fast wave generates a fluctuating pressure anisotropy as the wave propagates. At sufficiently large $\beta$, both firehose and mirror instabilities may therefore be triggered. With $\delta p_{\perp,\mathrm{i}}$ and $\delta p_{\parallel,\mathrm{i}}$ given by~\eqref{eq:fasteigvec}, the amplitude threshold for triggering both firehose and mirror instabilities is
\begin{equation}\label{eq:fastthresh}
    \left|\frac{\delta B}{B_0}\right| \gtrsim \frac{2}{\beta_{\rm i}} \quad\textrm{(fast-wave  amplitude  threshold)} .
\end{equation}
At high~$\beta$, this criterion can be satisfied for even small-amplitude fluctuations, justifying the use of the linear eigenvector and unperturbed $\beta_{\rm i}$ in determining the threshold.

To assess whether these micro-instabilities will be able to grow, we compare their linear growth rates to the linear frequency of the fast wave at high $\beta$, $\omega_{\rm fast}\sim k_\perp v_{\rm th,i}$. We adopt the maximal mirror growth rate from~\eqref{eqn:mirror1}, and use the maximal oblique firehose growth rate $\gamma_{\rm f} \approx 0.3\Omega_{\rm i}\Lambda^{1/2}_{\rm f}$ where $\Lambda_{\rm f} \doteq |\Delta + 2/\beta_{\rm i}|$ (\citealt{yoon93}; A.F.A.~Bott~{\em et al.}, in preparation), both of which are appropriate for the near-threshold conditions we anticipate in our fast-wave simulations. Assuming $|\delta B/B_0|\gtrsim 2\beta^{-1}_{\rm i}$, we find that
\begin{equation}\label{eq:growvprop}
    \frac{\gamma_{\rm m}}{\omega_{\rm fast}} \sim 0.01 \beta^{-1}_{\rm i}\frac{\lambda_\perp}{\rho_{\rm i}} \quad{\rm and}\quad
    \frac{\gamma_{\rm f}}{\omega_{\rm fast}} \sim 0.1 \beta^{-1/2}_{\rm i} \frac{\lambda_\perp}{\rho_{\rm i}} ,
\end{equation}
where $\lambda_\perp = 2\upi/k_\perp$ is the wavelength of the fast wave. It is immediately apparent from \eqref{eq:growvprop} that, at high~$\beta$, very large scale separation between the fast-wave wavelength and the ion-Larmor scale is necessary to allow enough time for mirror fluctuations to grow and become nonlinear. The scaling with $\beta_{\rm i}$ is much weaker for the firehose instability, and so there will exist wavelengths at which mirror regulation of the pressure anisotropy is effectively non-existent but the firehose regulation is rapid. For this reason our \texttt{Pegasus++} simulations, which focus on $\beta_{\mathrm{i}0}=25$, require $\lambda_\perp \gg 10^3 \rho_{\mathrm{i}0}$ to realize both mirror and firehose regulation.

The unstable Larmor-scale fluctuations will ultimately grow to amplitudes at which the particles' rate of pitch-angle scattering is sufficient to hold the pressure anisotropy at marginal stability. This rate may be estimated by calculating the pressure anisotropy driven by a small-amplitude fast wave in a weakly collisional plasma (following \citealt{braginskii65}) and asking what value of effective collisionality $\nu_{\rm eff}$ would be required to keep $|\Delta|\sim 2\beta^{-1}_{\rm i}$. With the former given in the collisional regime by $\Delta \sim -(\grad\bcdot\bb{u})/\nu_{\rm eff} \sim (k_\perp v_{\rm ms}/\nu_{\rm eff}) |\delta B/B_0|$, the limiting collisionality is
\begin{equation}\label{eq:nufast}
    \nu_{\rm eff} \sim k_\perp v_{\rm ms} \frac{\beta_{\rm i}}{2} \biggl|\frac{\delta B}{B_0}\biggr| .
\end{equation}
Note its explicit dependence upon the scale of the fast wave, an indirect consequence of the pressure anisotropy of the fast wave being continuously driven by the fluctuating wave. This is very different from the case with the aperiodic NP mode, in which the pressure anisotropy -- an essential feature of the mode's perpendicular pressure balance -- actually decays in time through transit-time damping.

\subsubsection{Viscous damping and collisional propagation}

\begin{figure}
\centering
\includegraphics[width=0.99\textwidth]{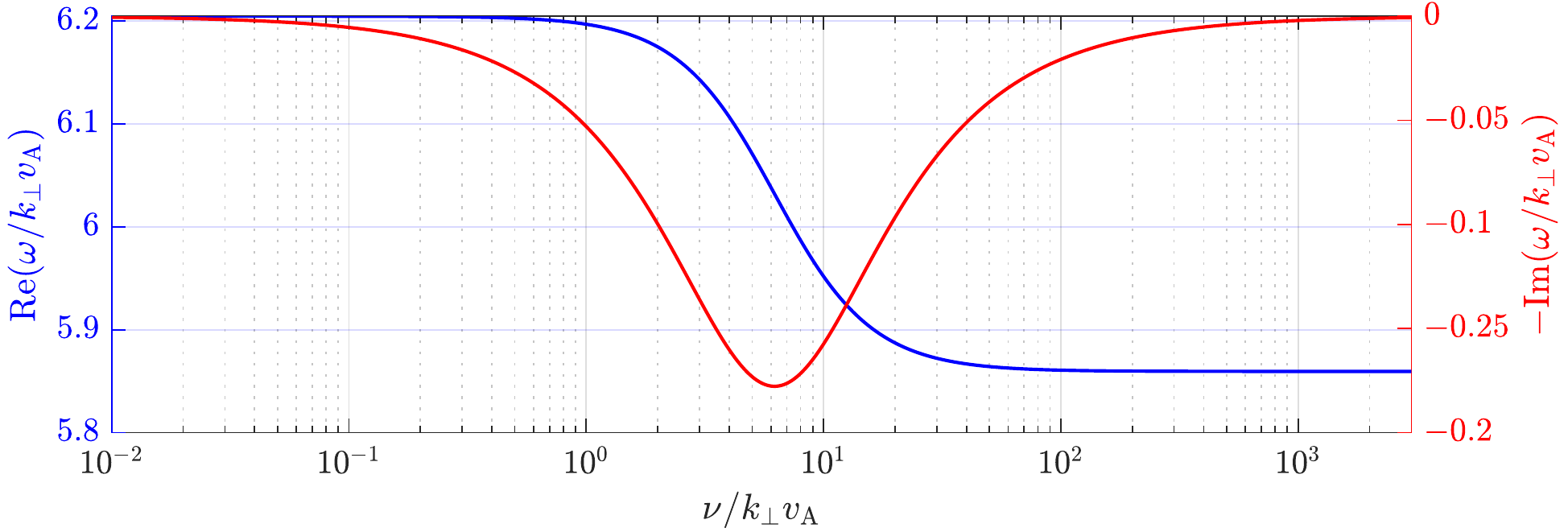}
\caption{Exact solution to the dispersion relation~\eqref{eq:fastdisp} for a $k_\parallel=0$ fast wave in a plasma having collision frequency $\nu$, $\beta_{\rm i0} = 25$, and $T_\mathrm{e}/T_\mathrm{i0}=1$.}\label{fig:fastdisp}
\end{figure}

The estimate of the effective collisionality~\eqref{eq:nufast} suggests that, depending on the wave amplitude, one should see a variety of fast-wave behaviour. For example, if $|\delta B/B_0| \gg 2\beta_{\rm i0}^{-1}$, then the implied collisionality can be large enough to push the fast wave into the collisional Braginskii-MHD regime ($\nu\gg\omega$). Making the presently unjustified yet instructive assumption that this collisionality is distributed uniformly in space, the fast-wave dispersion relation at arbitrary $\nu$ can be obtained after including isotropizing collisional terms $-\nu \Delta p/nB$ and $\nu \Delta p B^2/n^3$ on the right-hand sides of~\eqref{eqn:dmhd_pprp} and ~\eqref{eqn:dmhd_pprl} respectively, then linearizing the resulting system of equations. We find that
\begin{equation}\label{eq:fastdisp}
    \omega^3 - i\nu\omega^2 - \omega k^2_\perp v_{\rm ms,da}^{2} +i\nu k^2_\perp v_{\rm ms,sa}^{2} = 0.
\end{equation}
The numerical solution to~\eqref{eq:fastdisp} is shown in figure~\ref{fig:fastdisp}. In the collisionless limit $\nu \rightarrow 0$, one recovers propagation at the double-adiabatic fast speed; taking $\nu \rightarrow \infty$ returns propagation at the single-adiabatic fast speed. Viscous damping occurs at intermediate values of $\nu \sim \mathrm{Re}(\omega) \sim k_\perp v_{\rm th,i}$ around the transition between the double- and single-adiabatic regimes, where the scattering rate is comparable to the wave's oscillation frequency. The damping rate is always small compared to the wave frequency.

The dispersion relation~\eqref{eq:fastdisp} alongside the amplitude threshold~\eqref{eq:fastthresh} and the predicted effective collision frequency~\eqref{eq:nufast} imply three regimes for the behaviour of perpendicularly propagating fast modes in a high-$\beta$ plasma. For small amplitudes satisfying $|\delta B/B_0| < 2\beta_{\rm i0}^{-1}$, the mode propagates normally as a collisionless fast mode. It will steepen and eventually form a shock on the double-adiabatic shock time $t_{\rm s}^{\rm da}$. In the near-threshold regime where $|\delta B/B_0| \gtrsim 2\beta_{\rm i0}^{-1}$, the scattering rate from triggered mirror and firehose instabilities will not quite reach the value~\eqref{eq:nufast}, though scattering is still expected to occur and result in some viscous damping. The wave will also steepen to form a shock, but only a fraction of the wavelength will be kinetically unstable and therefore the shock will occur on a hybrid of the double- and single-adiabatic shock times. Lastly, at amplitudes well above the threshold, the scattering rate should be given by~\eqref{eq:nufast}. The viscous damping will be very weak,  the wave will host firehose/mirror scattering sites throughout most of its wavelength, and the shock time should be better represented by the single-adiabatic model.

We now test these ideas using numerical simulations.

%
%
\subsection{Numerical results}\label{sec:Fastnumerics}

\subsubsection{Method of solution and initial conditions}\label{sec:Fastmethod}

Due to the large scale separations needed to obtain asymptotic $\nu_\mathrm{eff}$ for both firehose and mirror fluctuations (\S\ref{sec:fastanis}), we use a combination of \texttt{Pegasus++} and (much cheaper) Landau-fluid CGL-MHD simulations. All simulations initialize a $k_\parallel=0$ fast wave in an otherwise Maxwellian plasma using the collisionless eigenmode~\eqref{eq:fasteigvec}, {\em viz.},
\begin{equation}
\begin{gathered}
    \bb{B}(0,x) = B_0\bigl[ 1 + \alpha\sin(k_\perp x) \bigr] \ey , \quad \bb{u}(0,x) = v_{\rm ms,da}\alpha \sin(k_\perp x) \ey , \\* \frac{n(0,x)}{n_0} = \frac{p_{\parallel\rm i}(0,x)}{p_{\rm i0}} = 1 + \alpha\sin(k_\perp x) , \quad \frac{p_{\perp\rm i}(0,x)}{p_{\rm i0}} = \bigl[ 1 + \alpha\sin(k_\perp x) \bigr]^2,
\end{gathered}
\end{equation}
where $k_\perp = 2\upi/\lambda_\perp$ and $\alpha$ is a dimensionless number quantifying the mode amplitude. For the {\tt Pegasus++} runs, the mesh is two-dimensional and elongated in the propagation direction, with size $L_x\times L_y = \lambda_\perp \times 100\rho_{\rm i0}$. The size of the domain in the $y$ direction is large enough to capture all relevant firehose and mirror fluctuations. We set $\beta_{\rm i0}=25$ and $T_{\rm e}=T_{\rm i0}$; the slightly larger value of $\beta_{\rm i0}$, as compared to that used in the simulations of the NP mode ($\beta_{\rm i0}=16$), results in a shorter numerical integration time (and thus computational savings) without changing the physical character of the fast wave. The spatial resolution and the number of macro-particles per cell are the same as in the NP simulations (\S\ref{sec:method}). In the manuscript we only show results from a {\tt Pegasus++} run having $\lambda_\perp = 8000\rho_{\rm i0}$, corresponding to the largest domain size that we simulated. We found that this value of $\lambda_\perp/\rho_{\rm i0}$ was the minimum required for the mirrors to have time to grow and begin scattering particles before the wave oscillates and the sign of the driven pressure anisotropy reverses.

In the accompanying Landau-fluid simulations, the full system of CGL-MHD equations is solved using a new Riemann solver implemented in a version of the finite-volume \texttt{Athena++} simulation code~\citep{stone08} that includes Landau-fluid heat fluxes (J.~Squire {\it et al.}, in preparation). These equations are given in Appendix~\ref{app:mslin}; they reduce to~\eqref{eq:dmhd} in our chosen geometry. For these runs, $\beta_{\rm i0}$ is varied between 1 and 100 to study the variance of the shock time. A `limiter' collisionality $\nu_{\rm lim}$ is set either to $0$ or to $\alpha \beta_{\rm i0} k_\perp v_{\rm ms,da}$, depending on whether the focus is on wave steepening and shock formation ($\nu=0$) or the effects of the instability-induced scattering. This anomalous scattering rate is active only within regions of the domain where the pressure anisotropy would be kinetically unstable, {\em viz.}, where $\beta_{\rm i}\Delta \leq -2$ and $\beta_{\rm i}\Delta\geq 1$; elsewhere it is zero. It serves to isotropize the plasma pressure where mirror or firehose fluctuations would otherwise do so in a kinetic system, by contributing a term proportional to $-\nu_{\rm lim}\Delta p$ to the right-hand sides of the evolution equations for $p_\perp$ and $p_\parallel$.

As in~\S\ref{sec:slow}, $\langle\,\cdot\,\rangle$ denotes a spatial average taken over the entire domain, while $\langle\,\cdot\,\rangle_k$ denotes a spatial average performed along the wavefront (in this case, the $y$ direction).

%
%
\subsubsection{Wave steepening and shock formation}

Our first goal is to test the expression~\eqref{eq:dabreak} for the shock-formation time $t_{\rm s}$. We perform a parameter survey by varying $\beta_{\rm i0}$ and the wave amplitude $\alpha$ using the CGL-MHD code with the micro-instability-limiting scattering turned off. At each time step in the simulation, the local density gradient (using a four-cell average) is calculated throughout the domain and its maximum value is recorded as a measure of the wave steepening. As a fast wave steepens, the growth rate of this maximum gradient increases until eventually the shock forms and the maximum gradient in the domain begins to plateau. We define the numerically calculated shock-formation time to be the time at which the rate of change of this maximum gradient drops below half of its own peak value. The resulting times are compared with~\eqref{eq:dabreak} in figure~\ref{fig:faststeep}. When testing the dependence on $\beta_{\rm i0}$ (blue, left), the perturbation amplitude is set to $\alpha=0.01$; when testing the dependence on amplitude (red, right), $\beta_{\rm i0} = 25$.

\begin{figure}
\centering
\includegraphics[width=0.99\textwidth]{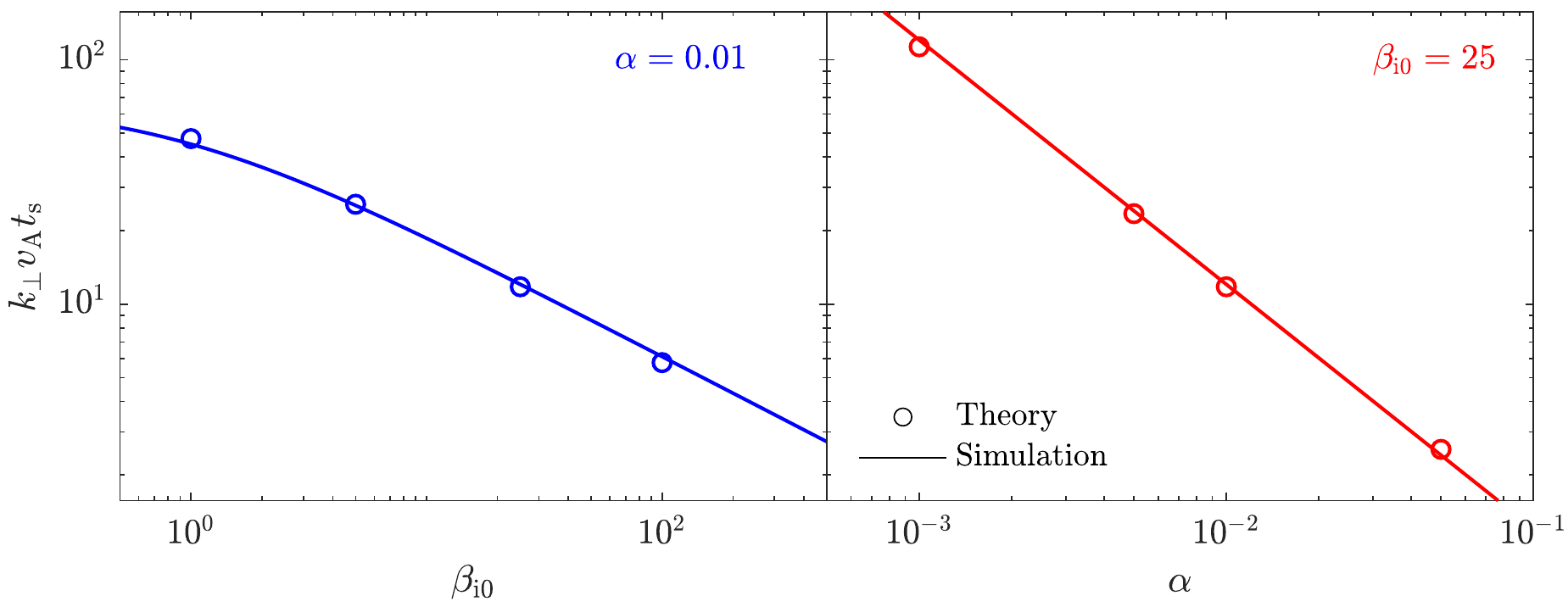}
\caption{Shock-formation time versus $\beta_{\rm i0}$ and $\alpha$ for a double-adiabatic fast wave computed from CGL-MHD simulations (lines) and predicted analytically using~\eqref{eq:dabreak} (circles). The simulated waves are estimated to have formed a shock at the time when the rate of change of the maximum density gradient drops below half of its own peak value.}\label{fig:faststeep2}
\end{figure}

Overall, the agreement between~\eqref{eq:dabreak} and the numerically calculated shock-formation times is quite good. Small variations occur due to differences in the rates at which the maximum gradients plateau and to minute fluctuations in the maximum value of the gradient after the shock is formed (this value does not necessarily reach a perfect steady state). Perhaps unsurprisingly, at high $\beta$ where $v_\mathrm{ms,da} \approx v_\mathrm{A}\sqrt{3\beta_{\rm i0}/2}$, the ratio of the wave-crossing time and the shock-formation time is $t_\mathrm{cross}/t_\mathrm{s,da} \approx 4\alpha/3$. This means the number of wavelengths propagated prior to forming a shock is dependent upon the mode amplitude only.

%
%
\subsubsection{Generation of pressure anisotropy and triggering of kinetic instabilities}

Prior to shock formation, the linearized fluctuations~\eqref{eq:fasteigvec} suggest that pressure anisotropy at a level capable of triggering both mirror and firehose instabilities will exist when the fast-wave amplitude satisfies $|\delta B/B_0| \gtrsim 2/\beta_\mathrm{i}$. For these supercritical amplitudes, the wavefront should carry with it rapidly growing firehose fluctuations and more slowly growing mirror fluctuations, as per~\eqref{eq:growvprop}. To test this idea, we performed a large-scale \texttt{Pegasus++} simulation, the parameters of which are described in~\S\ref{sec:Fastmethod}; the initial wave amplitude $\alpha = 0.1$ and $\beta_{\rm i0}=25$. 
\begin{figure}
\centering{
  \begin{subfigure}{0.99\linewidth}
    \includegraphics[width=\linewidth]{figures/fig17.pdf}
    \caption{Pressure anisotropy times the ion beta from a {\tt Pegasus++} simulation of a collisionless fast wave, showing that the compression and rarefaction of the magnetic-field lines generates oppositely signed anisotropies that move with the wavefront. Some sloshing due to firehose regulation of the negative pressure anisotropy causes an additional reversal of $\Delta$ in the final time frame.}\label{fig:fast_delta_microa}
  \end{subfigure}
  \begin{subfigure}{0.99\linewidth}
    \includegraphics[width=\linewidth]{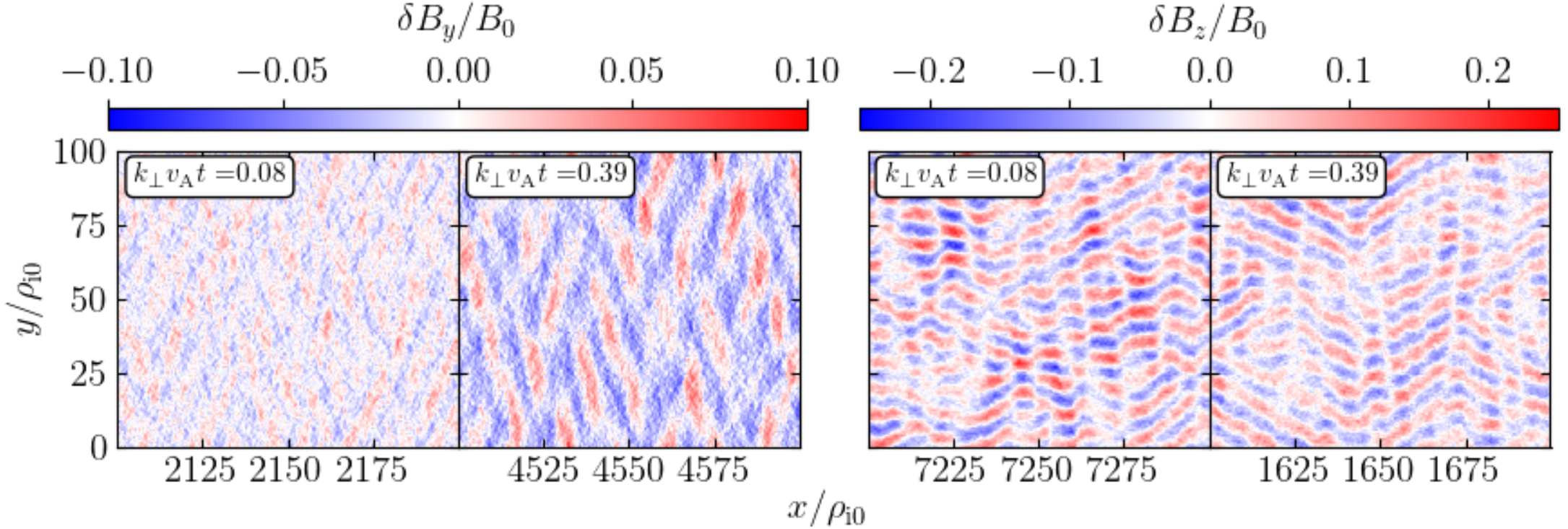}
    \caption{Zoomed-in regions showing $\delta B_y$ and $\delta B_z$, with the contribution from the background fast wave removed. Recall that the mean field is oriented in the $y$ direction. In the left set of panels, the mirror instability, with its oblique orientation and dominance in $\delta B_\parallel=\delta B_y$, grows relatively slowly in the co-moving region of fast-wave compression from $k_\perp v_{\rm A}t = 0.08$ to $0.39$. The firehose instability in the right set of panels is predominantly oblique and exhibits rapid growth and saturation; smaller-amplitude parallel firehoses appear in $\delta B_x$ (not shown). These firehose fluctuations reside downstream of the mirrors, where the fast-wave $\delta B<0$.}\label{fig:fast_delta_microb}
  \end{subfigure}
 }
\caption{}\label{fig:fast_delta_micro}
\end{figure}

Figure~\ref{fig:fast_delta_micro}($a$) depicts the pressure anisotropy generated by the fast wave as it propagates through space at three different times ($k_\perp v_{\rm A}t = 0.0$, 0.08, 0.39; note that the aspect ratio of the plotted domain is far from unity, and that the mean magnetic field is in the $y$ direction). Initially, the positive and negative pressure anisotropies in the wave are equal in magnitude. Shortly thereafter, the (unstable) negative anisotropy is reduced significantly due to the rapid growth of the (primarily oblique) firehose instability. The positive pressure anisotropy does not show a comparable decrease, and in fact increases somewhat from its initial value. This is likely because the rapid change in the negative-anisotropy regions, which perturbs the wave and causes some deviation from the eigenmode, is not matched by a comparable regulation from the positive side because of the relatively slow mirror growth. Figure~\ref{fig:fast_delta_micro}($b$) zooms in on the corresponding magnetic-field fluctuations that emerge in two separate co-moving regions where the plasma is mirror unstable (left) or firehose unstable (right). To accentuate these fluctuations, the large-scale contribution from the fast wave has been removed. At $k_\perp v_\mathrm{A} t = 0.08$, oblique firehose fluctuations are strong and nonlinear; parallel firehose fluctuations are also present, though subdominant, in $\delta B_x$ (not shown). At this time, there is only a hint of mirror fluctuations emerging above the noise level. In the final frame ($k_\perp v_\mathrm{A} t = 0.39$) however, highly oblique mirror modes have grown to large amplitudes in the region encompassed approximately by $x/\rho_{\rm i0} \in [4000,5000]$. The scale separation achieved in this simulation ($L_x/\rho_{\rm i0}=8000$) was the minimum at which we could observe mirror fluctuations with strengths comparable to their firehose counterparts; increasing the scale separation further would come at considerable computational expense.

%
%
\subsubsection{Effective collisionality: particle scattering}

\begin{figure}
\centering
\includegraphics[width=0.95\textwidth]{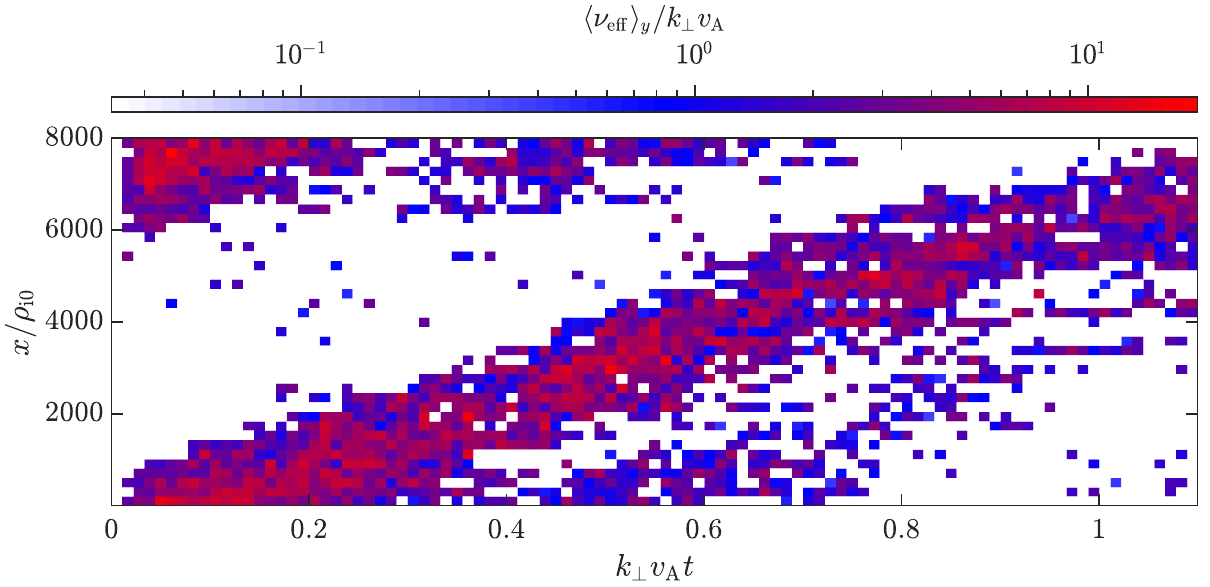}
\caption{Space-time diagram of the effective collision frequency measured in a {\tt Pegasus++} fast wave. The simulation parameters are $\beta_{\rm i0} = 25$, $\alpha = 0.1$, and $T_e/T_{\rm i0}=1$; using these numbers in~\eqref{eq:nufast} predicts $\nu_\mathrm{eff} \approx 16k_\perp v_\mathrm{A}$.}\label{fig:fastnuvxt}
\end{figure}

Following~\S\ref{sec:slowcollnum}, the effective collisionality was determined for the fast wave shown in figure~\ref{fig:fast_delta_micro} by tracking thousands of ion macro-particles and measuring the frequency at which their $\mu$ changes statistically by a factor of 1.2 or more. Figure~\ref{fig:fastnuvxt} depicts this scattering rate as a function of the position along the wave ($x/\rho_\mathrm{i0}$) and the time ($k_\perp v_\mathrm{A}t$). Sites of strong scattering are associated with the firehose modes, which appear more or less instantly and travel along with the trough of the wave. The trail of the scattering sites indicates that the trough of the wave moves at ${\approx}6v_{\rm A}$, as expected for a fast mode with $\beta_\mathrm{i0} = 25$. In this simulation, the rapid regulation of the pressure anisotropy by the firehose instability causes sloshing. The sloshing temporarily drives a higher positive pressure anisotropy, and therefore enhanced mirror growth, for a short period beginning at $kv_{\rm A}t \approx 0.4$. The measured scattering rate in the firehose-unstable regions is comparable to the predicted asymptotic scattering rate for a $\beta_\mathrm{i0}=25$ fast wave with $\alpha=0.1$ and $T_\mathrm{e}=T_\mathrm{i0}$, {\em viz.}~$\nu_\mathrm{eff} \approx 16 k_\perp v_\mathrm{A}$ (see~\eqref{eq:nufast}). The mirror instability in this case also scatters particles at an average rate of a few times $k_\perp v_\mathrm{A}$, but these scattering sites are much less coherent and do not coincide with the peak in the positive pressure anisotropy. This delayed growth is a result of the limited achievable scale separation in our simulations, which only barely allows mirrors to grow to nonlinear levels within a fast-wave crossing time.

\begin{figure}
\centering
\includegraphics[width=0.99\textwidth]{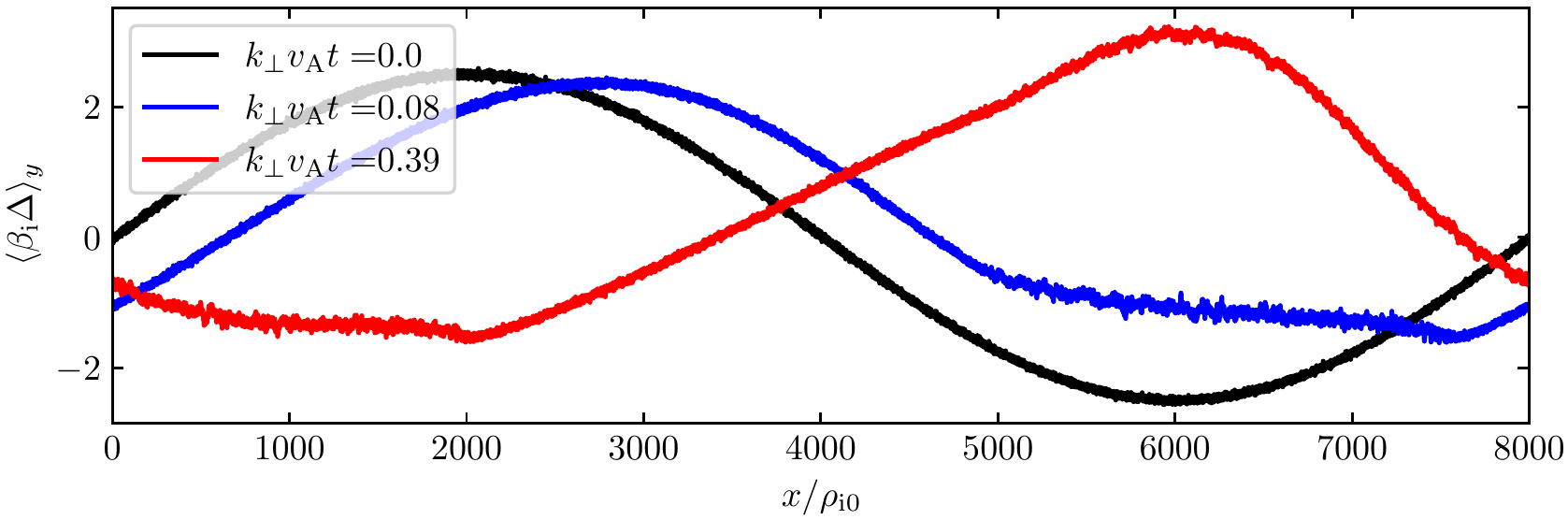}
\caption{Wavefront-averaged $\beta_\mathrm{i}\Delta$ in the fast wave for the same time frames as figure~\ref{fig:fast_delta_micro}. Pressure-anisotropy regulation from the firehose instability \edit{maintains $\beta_\mathrm{i}\Delta \gtrsim -1.4$}, while the mirror fluctuations cause some distortion of the mode above $\beta_\mathrm{i}\Delta \approx 1$ but are unable to regulate fully the positive anisotropy to marginally unstable values. An increase in the rate at which positive pressure anisotropy is generated by the steepened wave and the asymmetry in the anisotropy's regulation by micro-instabilities causes an enhancement of the positive pressure anisotropy in the final time shown.
}\label{fig:fastdb1d}
\end{figure}

The effects of the induced scattering on the fast wave's pressure anisotropy are visible in figure~\ref{fig:fastdb1d}, which shows $\langle\beta_{\rm i}\Delta\rangle_y$ at the same times as in figure~\ref{fig:fast_delta_micro}. \edit{The negative anisotropy is regulated within a very short time by the firehose instability to a value close to the oblique threshold $\beta_\mathrm{i}\Delta \simeq -1.4$}. This regulation persists, but is not matched on the mirror-unstable side. Some steepening has also occurred, as expected, but the positive anisotropy has not been driven down near marginal mirror stability. In order for mirror fluctuations to regulate the positive pressure anisotropy to marginal stability, they would need to grow faster with respect to the fast-wave crossing time; equation~\eqref{eq:growvprop} suggests that this could be achieved by increasing $\lambda_\perp/\rho_{\rm i0}$ even further (beyond $\lambda_\perp/\rho_{\rm i0} = 10^4$), \edit{or perhaps by decreasing $\beta_{\rm i0}$ (though in this case the amplitude threshold~\eqref{eq:fastthresh} would increase, necessitating larger fast-wave amplitudes that would shock almost immediately)}. Unfortunately, such large scale separations become prohibitively expensive to simulate using \texttt{Pegasus++}, and so from this point onward we employ the CGL-MHD code with pressure-anisotropy limiters.\footnote{\edit{As with the NP mode's decay rate, the fast-wave steepening time also increases linearly with the wavelength, here $\lambda_\perp$, and so the overall cost scales ${\propto}(\lambda_\perp/\rho_{\rm i0})^2$. A {\tt Pegasus++} simulation with a scale separation of $\lambda_\perp/\rho_{\rm i0} = 10^4$ would cost ${\gtrsim}10^7$~CPU-hours.}}

%
%
\subsubsection{Viscous damping and collisional steepening}\label{sec:fastcgl}

To study fast-wave behaviour at asymptotically large scale separations, we employ the Landau-fluid CGL-MHD code. These simulations are performed using a larger $\beta$ parameter than used in the \texttt{Pegasus++} run, $\beta_{\rm i0} = 100$ rather than 25, and with $\alpha = 0.2$. These parameters have the advantage that a large portion of the fast wave is initially above the threshold for instability while the wave remains somewhat linear in amplitude. As discussed in~\S\ref{sec:Fastmethod}, this code introduces a user-specified constant scattering rate in (and only in) the kinetically unstable regions of the plasma. We set this scattering rate according to~\eqref{eq:nufast} using the initial mode amplitude. In reality, this scattering rate should decay alongside the amplitude, and so our treatment will not precisely reproduce the results that would be obtained from a more rigorous kinetic calculation.

\begin{figure}
\centering{
  \mbox{\hspace{1em}$(a)$\hspace{0.45\textwidth}$(b)$\hspace{0.45\textwidth}}\\
  \begin{subfigure}{0.49\linewidth}
    \includegraphics[width=\linewidth]{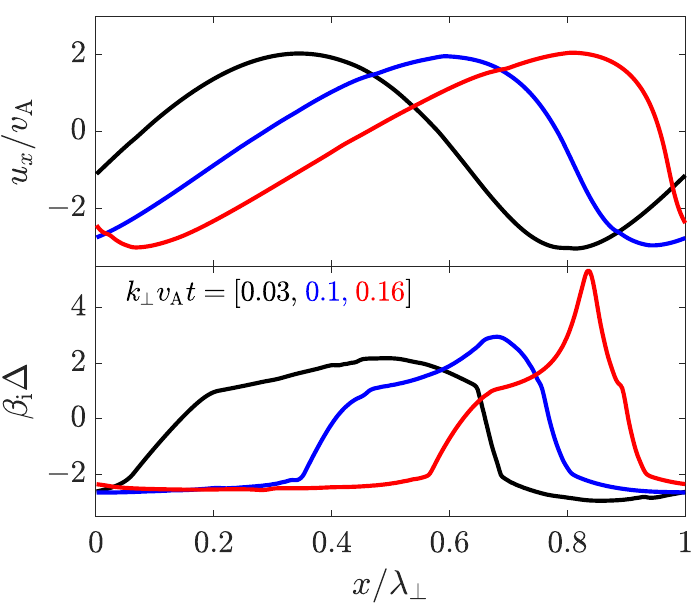}\label{fig:fast_decay}
  \end{subfigure}
  \begin{subfigure}{0.481\linewidth}
    \includegraphics[width=\linewidth]{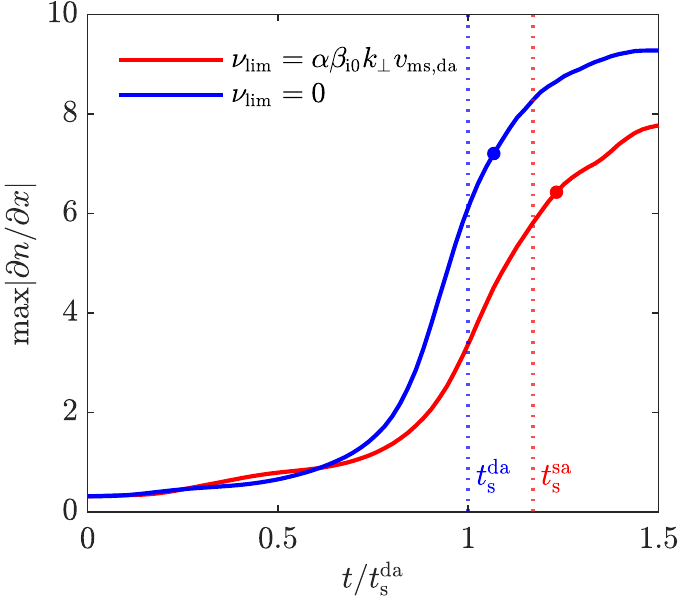}\label{fig:shock_v_nu}
  \end{subfigure}
 }
\caption{($a$) Propagation of an $\alpha=0.2$ fast wave with $\beta_{\rm i0}=100$ and $\nu_{\rm lim}$ set by~\eqref{eq:nufast}. The top panel shows wave steepening in the fluid velocity, with no noticeable viscous decay on the timescale of shock formation. The bottom panel shows regulation of the pressure anisotropy to near the mirror and firehose thresholds. A peak appears in $\beta_{\rm i}\Delta$ due to the rapid generation of positive pressure anisotropy in the steepening wavefront. ($b$) The maximum density gradient found within the domain of the same $\alpha = 0.2$, $\beta_{\rm i}=100$ fast wave, compared against an equivalent run with $\nu_{\rm lim}=0$. The predicted shock times are labelled by $t_{\rm s}^{\rm da}$ and $t_{\rm s}^{\rm sa}$, and the shock times detected by the same method used for figure~\ref{fig:faststeep} are denoted by circular markers. The growth of the maximum gradient continues for a longer time in the single-adiabatic case than in the double-adiabatic case, indicating delayed shock formation.}\label{fig:0p2fastsims}
\end{figure}

In figure~\ref{fig:0p2fastsims}, the propagation and nonlinear steepening of the CGL-MHD fast wave are presented. The top panel in figure~\ref{fig:0p2fastsims}($a$) shows the bulk fluid velocity perpendicular to the background field at three different times, exhibiting steepening without a significant change in wave amplitude. This indicates that no significant viscous dissipation occurs on a timescale comparable to the shock-formation timescale \edit{(as predicted by \eqref{eq:fastdisp})}. The bottom panel shows the pressure anisotropy of the wave at the same times, multiplied by $\beta_{\rm i}$. The anisotropy is substantially reduced below what it would be in the absence of the limiting collisionality, particularly on the firehose-unstable side, although it is not perfectly regulated to the instability thresholds. In particular, a peak in the positive pressure anisotropy becomes prominent starting from $k_\perp v_{\rm A} t \approx 0.1$. This is a result of wave steepening, as the sharp gradient at the wavefront generates positive $\Delta$ much faster than the slow decline in the wake generates negative $\Delta$, as well as faster than our (constant) limiting collisionality is able to regulate. Figure~\ref{fig:0p2fastsims}($b$) displays the evolution of the maximum absolute value of the density gradient from this run, alongside that from a comparable run with $\nu_{\rm lim}=0$. On the abscissa is the simulation time normalized by the double-adiabatic shock time $t_{\rm s}^{\rm da}$ (see~\eqref{eq:dabreak}). We calculated the shock time for each run using the same detection method as in figure~\ref{fig:faststeep2}; these times, marked by filled circles in the figure, agree reasonably well with the predicted values of $t_{\rm s}^{\rm da}$ and $t_{\rm s}^{\rm sa}$ for the collisionless and collisional cases, respectively. The difference in steepening rate between the two runs can be interpreted as $\nu_{\rm lim}$ forcing a more MHD-like, rather than collisionless, evolution in the fast wave. The collisional isotropization at the peaks of the wave (which are also the most rapidly moving regions) effectively changes the local adiabatic index of the ions, slowing down the steepening process and yielding better agreement with $t_{\rm s}^{\rm sa}$ than with $t_{\rm s}^{\rm da}$. In this sense then, all of the essential characteristics of large-amplitude, high-$\beta$, collisionless fast waves approach that of single-adiabatic MHD as a result of induced micro-instabilities.

%
%
\section{Summary and discussion}

This exploration of microphysically unstable magnetosonic modes brings closure to a systematic investigation of isolated waves in collisionless, high-$\beta$ plasmas that started with the discovery of self-interrupting Alfv\'en waves and continued with the demonstration of self-sustaining sound. In summary, through the action of adiabatic invariance, the consequent production of pressure anisotropy, and the excitation of rapidly growing, micro-scale kinetic instabilities\dots
\vspace{0.5em}
\begin{itemize}[align=left,leftmargin=2em,itemindent=0pt,labelsep=0pt,labelwidth=1.5em]
  \setlength\itemsep{0.5em}
  \item collisionless linearly polarized Alfv\'en waves with amplitudes satisfying $(\delta B_\perp/B_0)^2\gtrsim 2/\beta_{\rm i0}$ retard their own propagation and spur their own viscous decay \citep{squire16,squire17num}; 
  \item collisionless IAWs with amplitudes satisfying $|\delta n/n|\gtrsim 2/\beta_{\rm i0}$ avert their otherwise potent Landau damping and propagate in a manner akin to sound waves in a weakly collisional fluid  \citep{kunz20};
  \item collisionless NP modes with amplitudes satisfying $|\delta B_\parallel/B_0| \gtrsim 0.4$ and wavelengths $\lambda_\parallel\gtrsim 10^4\beta^{1/2}_{\rm i0}\rho_{\rm i0}$ \edit{are predicted to} interrupt their transit-time damping and behave similarly to MHD entropy modes (at smaller wavelengths, these large-amplitude NP modes \edit{have been shown to} decay via transit-time damping, which is sustained against its nonlinear saturation by weak mirror-induced collisionality) \edit{(this paper)}; and
  \item collisionless fast waves with amplitudes satisfying $|\delta B/B_0| \gtrsim 2/\beta_{\rm i0}$ and wavelengths $\lambda_\perp \gg 10^2\beta_{\rm i0}\rho_{\rm i0}$ acquire an effective adiabatic index of $5/3$ and therefore propagate and nonlinearly steepen at single-adiabatic rates \edit{(this paper)}.
\end{itemize}
\vspace{0.5em}
Notwithstanding the somewhat narrow focus on the behaviour of isolated eigenmodes, the simple demonstration that micro-scale physics effectively filters out what kinds of macro-scale fluctuations are allowed in a high-$\beta$ plasma is of broad relevance to observed space and astrophysical systems and to theories for electromagnetic turbulence. The most immediate application to the former is the near-Earth solar wind. For example, \citet{verscharen16} used linear theory to conjecture that plasma instabilities could be driven by compressive fluctuations in the $\beta\gtrsim{1}$ solar wind through the adiabatic production of pressure anisotropy, leading to `collisionless isotropization' of solar-wind protons. Our work supports this idea quantitatively from first principles. \citet{verscharen17} then measured the polarization of compressive fluctuations within the solar wind at $1~{\rm au}$ using data from the {\it Wind} spacecraft, finding that the eigenmode relationships detected were best represented by MHD, rather than collisionless, slow modes. \citet{coburn22} approached this same issue from a different angle, measuring the dispersion relation of compressive modes in the solar wind and determining which scattering rates best reproduced them. They concluded that the mean free path predicted by their wave measurements is ${\sim}10^3$ times smaller than that set by Coulomb collisions, finding that the dispersion relation of the measured fluctuations most closely resembles that of Braginskii-MHD slow modes. Both of these observational results find a natural explanation in the context of our paper, at least for those portions of the wind having $\beta\gtrsim 1$ that have been measured to be constrained by the firehose and mirror instability thresholds~\citep{kasper02,hellinger06,bale09,chen16}.

To the extent that nonlinearly interacting fluctuations in strong electromagnetic turbulence retain some characteristics of their linear eigenmodes, the above conclusions cast doubt on whether some well-established pillars of MHD and gyrokinetic turbulence theory \citep{gs95,lg01,schekochihin09,schekochihin22} are applicable to high-$\beta$ plasmas. For example, with each fluctuation generating and responding to pressure anisotropy in an amplitude-, wavelength-, and polarization-dependent way, it is suspect that inertial-range compressive fluctuations are simply passively mixed by the Alfv\'en-wave cascade and, in turn, exert no back-reaction on the Alfv\'enic fluctuations. \edit{Likewise, shorter-wavelength fluctuations would reside within (and be altered by) a patchy, yet locally uniform, pressure anisotropy produced by the ensemble of much longer-wavelength fluctuations, implying a loss of strict locality in the turbulent cascade. While the question of how a background pressure anisotropy affects electromagnetic kinetic turbulence has been addressed using reduced (gyrokinetic) models \citep{kunz15,kunz18}, those studies did not address this potential non-locality, nor did they incorporate the impact of kinetically unstable fluctuations and the associated anomalous scattering. At this point it is unclear how all this additional physics plays out within a turbulent cascade governed by a scale-by-scale `critical balance' between the characteristic linear and nonlinear frequencies, an organizing principle for strong turbulence that appears to hold (albeit in a modified form) even in the presence of strong pressure anisotropies \citep{bott21,squire23}. The mutual interactions between what are conventionally considered to be energetically decoupled cascades, and the impact of this coupling on the constant flux of energy, the locality of interactions, and the universality of critical balance, ought to be investigated. Some progress on this front has recently been made by \citet{arzamasskiy22}, who showed using hybrid-kinetic simulations that strong Alfv\'enic turbulence with $(\delta B_\perp/B_0)^2\gtrsim 2/\beta_{\rm i0}$ self-consistently produces a parallel viscous scale comparable to the driving scale of the cascade and involves non-local energy transfers in $k$ space associated with the excitation of ion-Larmor-scale kinetic instabilities. Incorporating compressive fluctuations into this study would be informative, not only with regards to the dynamics but also concerning the partition of turbulent energy into ion versus electron heating \citep[cf.][]{kawazura20}. While the properties of isolated waves in collisionless, high-$\beta$ plasmas have now been elucidated, there is clearly much more work to be done.}

%
%
\section*{Acknowledgements}

S.M.~and M.W.K.~were supported in part by NSF CAREER Award No.~1944972. Support for J.S. was provided by Rutherford Discovery Fellowship RDF-U001804, which is managed through the Royal Society Te Ap\=arangi. High-performance computing resources were provided by: the Texas Advanced Computer Center at The University of Texas at Austin under Stampede2 allocation TG-AST160068 and Frontera allocation AST20010; and the PICSciE-OIT TIGRESS High Performance Computing Center and Visualization Laboratory at Princeton University. This work used the Extreme Science and Engineering Discovery Environment (XSEDE), which is supported by National Science Foundation grant number ACI-1548562. The authors thank Archie Bott, Eliot Quataert, Alex Schekochihin, and the participants of the 13th Plasma Kinetics Working Meeting at the Wolfgang Pauli Institute in Vienna for useful discussions, \edit{and the three anonymous referees for insightful comments that sharpened the presentation.} M.W.K.~additionally thanks the Institut de Plan\'etologie et d'Astrophysique de Grenoble (IPAG) for its hospitality and visitor support while this work was being completed.

\appendix

\section{Hermite--Laguerre solution to linear KMHD}\label{app:Delta}

In this appendix, we detail our numerical method for calculating the time-dependent pressure anisotropy generated by a linear NP mode. The task is to integrate the system~\eqref{eqn:linzd} numerically from an appropriate set of initial conditions. Before providing those conditions, we take the time derivative of \eqref{eqn:mom} and use~\eqref{eqn:ind} to obtain the following wave equation for the $\bb{E}{\btimes}\bb{B}$ drift velocity:
\begin{equation}\label{eqn:mom_app}
    \biggl( \DD{t}{} + k^2 v^2_{\rm A} \biggr) u_\perp = - \frac{\imag k_\perp}{m_{\rm i}n_0} \D{t}{} \bigl( \delta p_{\perp\rm i} + T_{\rm e} \delta n \bigr) .
\end{equation}
The right-hand side of this equation is calculated by taking the zeroth and second moments of the linearized Vlasov equation~\eqref{eqn:vlasov}. After assuming an isotropic Maxwellian background, $F_0=F_{\rm M}(v)$, and rewriting the electric and magnetic-mirror forces using \eqref{eqn:ind} and \eqref{eqn:eprl}, equation~\eqref{eqn:vlasov} reduces to
\begin{equation}\label{eqn:vlasov_app}
    \biggl( \pD{t}{} + \imag k_\parallel v_\parallel \biggr) \delta f + \biggl( \imag k_\perp u_\perp \frac{w^2_\perp}{v^2_{\rm th,i}} + \imag k_\parallel v_\parallel \frac{T_{\rm e}}{T_{\rm i0}} \frac{\delta n}{n_0} \biggr) F_{\rm M} = 0 .
\end{equation}
Equations~\eqref{eqn:mom_app} and \eqref{eqn:vlasov_app} are solved numerically as follows.

We express the $v_\parallel$ dependence of $\delta f$ in terms of Hermite polynomials $H_n$ and the $w^2_\perp$ dependence in terms of Laguerre polymonials $L_m$:
\begin{equation}
    \delta f(t,k_\parallel,k_\perp,v_\parallel,w_\perp) = F_{\rm M}(v)  \sum_{m,n=0}^\infty g_{m,n}  H_n\biggl(\frac{v_\parallel}{v_{\rm th,i}}\biggr) L_m\biggl(\frac{w^2_\perp}{v^2_{\rm th,i}}\biggr) .
\end{equation}
This spectral decomposition allows the required moments to be calculated simply as
\begin{equation}
    \frac{\delta n}{n_0} = g_{0,0} , \quad \frac{\delta p_{\perp\rm i}}{p_{\rm i0}} = g_{0,0} - g_{1,0} , \quad \frac{\delta p_{\parallel\rm i}}{p_{\rm i0}} = g_{0,0} + 4g_{0,2} ,
\end{equation}
so that \eqref{eqn:mom_app} becomes
\begin{equation}\label{eqn:mom_app_HL}
    \biggl( \DD{t}{} + k^2 v^2_{\rm A} \biggr) \frac{u_\perp}{v_{\rm th,i}} = - \frac{\imag k_\perp v_{\rm th,i}}{2} \D{t}{} \biggl[ \left( 1 + \frac{T_{\rm e}}{T_{\rm i0}} \right) g_{0,0} - g_{1,0}  \biggr] .
\end{equation}
Because the Hermite and Laguerre polynomials form orthonormal bases with respect to Gaussian and exponential weights, respectively, equation~\eqref{eqn:vlasov_app} may be easily transformed to Hermite--Laguerre space to find
\begin{subequations}\label{eqn:geqn}
\begin{align}
    \D{t}{g_{m,0}} &+ \imag k_\parallel v_{\rm th,i} g_{m,1} + \imag (\delta_{m,0} - \delta_{m,1}) k_\perp u_\perp = 0 ,
    \\*
    \D{t}{g_{m,1}} &+ \imag k_\parallel v_{\rm th,i} \biggl( 2 g_{m,2} + \frac{1}{2} g_{m,0} \biggr) + \imag \frac{T_{\rm e}}{2T_{\rm i0}} \delta_{m,0} g_{0,0} = 0 ,
    \\*
    \D{t}{g_{m,n}} &+ \imag k_\parallel v_{\rm th,i} \biggl[ (n+1) g_{m,n+1} + \frac{1}{2} g_{m,n-1} \biggr] = -\nu n^4 g_{m,n} ,\quad n\ge 2. \label{eqn:gmn}
\end{align}
\end{subequations}
Note that the term $k_\parallel v_\parallel \delta f$ representing the parallel phase mixing of the perturbed distribution function couples together different Hermite moments, representing the generation of fine-scale structure in $v_\parallel$. Because the magnetic field suppresses phase mixing across the magnetic field, there is no cascade to higher $w_{\perp}$ moments and only the first two Laguerre polynomials ($m=0,1$) are needed. To the right-hand side of \eqref{eqn:gmn} we have appended a fourth-order hyper-collision operator; the restriction of the collision operator to $n\ge 2$ guarantees that number and momentum are conserved. The hyper-collisionality is added because only a finite number of Hermite polynomials are usable, so the series must be truncated somewhere. A hard truncation in which the final $v_{\parallel}$ moment is arbitrarily set to zero will result in numerical instability unless a collisionality is employed to ensure the velocity-space cascade (associated with parallel phase mixing of the perturbed distribution function) decays to zero amplitude before the last resolved moment is reached.

A code was written in Fortran~90 to solve~\eqref{eqn:mom_app_HL} and~\eqref{eqn:geqn}. Equation~\eqref{eqn:geqn} is solved and $\delta f$ updated in time using a semi-implicit Crank--Nicholson method; the moments $g_{0,0}$ and $g_{1,0}$ are then used in~\eqref{eqn:mom_app_HL} to update the drift velocity using centered differencing in time. The discrete time axes on which $g_{m,n}$ and $u_\perp$ are stored are staggered to maintain appropriate centering for all derivatives. The matrix inversion needed to update $g_{m,n}$ is performed using the Thomas Tridiagonal Matrix Algorithm (TDMA). 

For the initial conditions, we start from isothermal pressure balance, with $g_{1,0} = g_{0,2} = 0$ and $g_{0,0}\ne 0$ (but arbitrary). The reasoning behind this choice is discussed in~\S\ref{sec:method}. These initial conditions transition rapidly into the NP eigenmode by launching small-amplitude (relative to the amplitude of the NP mode) fast waves that facilitate the adjustment. The linear evolution of the NP mode from this initial condition is shown in figure~\ref{fig:linevo} and discussed in~\S\ref{sec:paniso}.

\section{Magnetosonic modes with arbitrary scattering frequency}\label{app:mslin}

To obtain the linear dispersion relation of kinetic hydromagnetic modes at arbitrary $\nu$, we must use a model that accurately captures the effects of adiabatic invariants, heat fluxes, and collisional isotropization. One such model is given by the \citet{cgl56} equations supplemented, by collisional isotropization and closed by so-called Landau-fluid heat fluxes \citep{shd97}. Assuming isothermal electrons, these equations are:
\begin{subequations}\label{eq:cglmhd}
\begin{gather}
    \bigD{t}{n} = -n \grad\bcdot\bb{u} , \\*
    m_{\rm i} n \bigD{t}{\bb{u}} = -\grad\biggl( p_{\perp\rm i} + n T_{\rm e} + \frac{B^2}{8\upi} \biggr) + \grad\bcdot\biggl[ \eb\eb\biggl( \Delta p_{\rm i} + \frac{B^2}{4\upi}\biggr)\biggr],\label{eq:cglmom} \\*
    \bigD{t}{\bb{B}} = (\bb{B}\bcdot\grad)\bb{u} - \bb{B}\grad\bcdot\bb{u}, \\*
    nB \bigD{t}{} \biggl(\frac{p_{\perp\rm i}}{nB}\biggr) = -\grad\bcdot\bigl(q_{\rm \perp i} \eb\bigr) - q_{\rm \perp i} \grad\bcdot\eb - \frac{1}{3} \nu \Delta p_{\rm i} , \label{eqn:cglpprp}\\*
    \frac{n^3}{B^2} \bigD{t}{} \biggl(\frac{p_{\parallel\rm i} B^2}{n^3}\biggr) = - \grad\bcdot\bigl(q_{\rm\parallel i}\eb\bigr) + 2q_{\rm \perp i} \grad\bcdot\eb + \frac{2}{3} \nu \Delta p_{\rm i} ,\label{eqn:cglpprl}
\end{gather}
\end{subequations}
where ${\rm D}/{\rm D}t \doteq \partial/\partial t + \bb{u}\bcdot\grad$ is the convective derivative for the bulk velocity $\bb{u}$, $\eb\doteq\bb{B}/B$ is the unit vector in the direction of the local magnetic field, $\Delta p_{\rm i} \doteq p_{\perp\rm i} - p_{\parallel\rm i}$ is the dimensional ion pressure anisotropy, $\nu$ is the isotropizing collision frequency, and $q_{\rm \parallel i}$ and $q_{\rm \perp i}$ represent the field-parallel flow of parallel and perpendicular ion heat. For linear perturbations to the ion temperature ($\delta T_{\parallel\rm i}$, $\delta T_{\perp\rm i}$) and magnetic-field strength ($\delta B_\parallel$) having parallel wavenumber $k_\parallel$, the latter may be adopted from equations (48) and (49) of \citet{shd97}:
\begin{gather}\label{eq:qprl}
    q_{{\rm \parallel i},k} = - \frac{4 n v^2_{\rm th\parallel,i}}{2\sqrt{\upi}|k_\parallel|v_{\rm th\parallel,i} + (3\upi-8)\nu}~ \imag k_\parallel \delta T_{\parallel\rm i}, \\*
    q_{{\rm \perp i},k} = - \frac{n v^2_{\rm th\parallel,i}}{\sqrt{2\upi}|k_\parallel|v_{\rm th\parallel,i} + 2\nu} \left( \imag k_\parallel \delta T_{\perp\rm i} +  \imag k_\parallel  T_{\perp i} \Delta_{\rm i} \frac{\delta B_\parallel}{B} \right).\label{eq:qprp}
\end{gather}
These `3+1' heat fluxes accurately reproduce the linear Landau--Barnes damping of the kinetic hydromagnetic modes in the collisionless limit \citep[][\S VIII]{shd97} and take on a form akin to that obtained by \citet{braginskii65} in the collisional limit. Because Braginskii-MHD does not accurately capture the linear heat fluxes when $\nu \lesssim |k_\parallel| v_{\rm th,i}$, the Landau-fluid CGL equations are used to describe the linear propagation of these modes at arbitrary $\nu$, bridging the gap between the fully collisionless ($\nu=0$) and the weakly collisional ($\nu \gg k_\parallel v_{\rm th,i})$. 
Note that, in the absence of heat fluxes and collisionality, equations~\eqref{eqn:cglpprp} and \eqref{eqn:cglpprl} guarantee conservation of the adiabatic invariants $\mu$ and $\mathcal{J}$ associated with Larmor gyrations and bounce motion. One of the advantages of using the Landau-fluid CGL equations over a Vlasov approach is the former's lack of dependence on the plasma dispersion function $\mathcal{Z}(\zeta)$, whose dependence on $\zeta\doteq \omega/|k_\parallel|v_{\rm th,i}$ can only be expressed analytically in the asymptotic limits $\zeta\gg 1$ and $\zeta\ll 1$. Instead, the `3+1' heat fluxes yield polynomial dispersion relations for the modes at all frequencies. As a result, if one wishes to derive an analytic expression for the frequency and damping rate of the oblique IAW, which has $\zeta\sim 1$ when $T_{\rm e}/T_{\rm i0} \sim 1$, they can then do so with ease.

\begin{figure}
\centering
\includegraphics[width=0.95\textwidth]{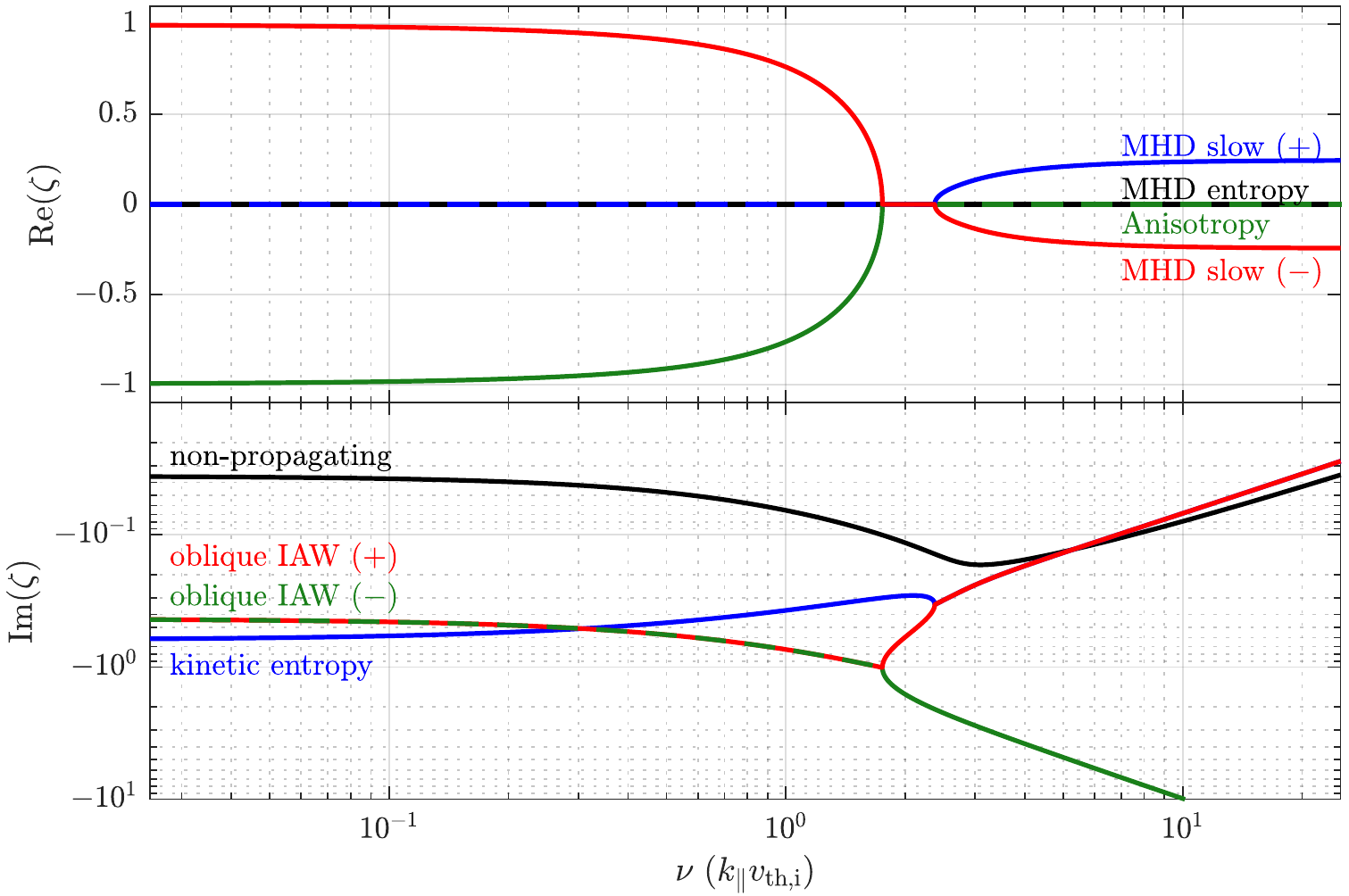}
\caption{Linear dispersion relation of the Landau-fluid CGL-MHD equations~\eqref{eq:cglmhd}. The dimensionless (complex) frequency $\zeta\doteq \omega/|k_\parallel|v_{\rm th,i}$ is computed numerically as a function of collisionality $\nu/|k_\parallel|v_{\rm th,i}$ for $k_\perp = 4|k_\parallel|$, $\beta_{\rm i0} = 16$, and $T_\mathrm{e} = T_\mathrm{i0}$.}\label{fig:cgldisp}
\end{figure}

Proceeding with the linear analysis, we assume zero background pressure anisotropy, neglect all nonlinear terms, and Fourier transform \eqref{eq:cglmhd}--\eqref{eq:qprp} in space and time,  so that $\mathrm{D/D}t \rightarrow -\imag\omega$ and $\grad\rightarrow \imag\bb{k}$. The result is a straightforward algebraic system, some solutions of which are shown in figure~\ref{fig:cgldisp}. In total there are 8 modes associated with 8 unique time derivatives ($\grad\bcdot\bb{B}=0$ fixes one of the components of $\delta\bb{B}_\perp$). The modes not displayed in figure~\ref{fig:cgldisp} are the Alfv\'en waves (which would be lines at $\zeta = \pm\beta^{-1/2}_{\rm i0}$) and both fast waves (which are shown in figure~\ref{fig:fastdisp}). Considering that there exists one additional time derivative in CGL-MHD than in collisional MHD due to the splitting of the thermal pressure into two components, there should be a mode that vanishes in the collisional limit. Indeed, after bifurcation one branch of the oblique IAW becomes non-propagating and is damped at a rate approximately equal to $\nu$ as $\nu \rightarrow \infty$. This strong damping is due to the mode's polarization, having opposing perpendicular and parallel pressure perturbations that satisfy $|\delta p_\perp| \gg |\delta p_\parallel|$ when $k_\perp \gg k_\parallel$. Hence the reason we have termed this mode the ``anisotropy mode'' in figure~\ref{fig:cgldisp}: it remains anisotropic even at arbitrarily large $\nu$, causing it to damp increasingly fast.

The NP mode has in some cases been attributed to the collisionless limit of the MHD slow magnetosonic mode (\textit{e.g.} \citet{verscharen17}), hence its frequently being referred to as the collisionless slow mode. This may be due to the fact that the Braginskii-MHD dispersion relation predicts a non-propagating slow mode at sufficiently low $\nu$, one which remains non-propagating as $\nu \rightarrow 0$. In reality, the slow mode does propagate once again at sufficiently low collisionality, and the NP mode is better identified as the kinetic extension of the MHD entropy mode. In the MHD entropy mode, no pressure perturbation is permitted by the parallel momentum equation, only a density perturbation. However, at lower scattering rates the pressure separates into its field-parallel and perpendicular components, and perpendicular pressure balance becomes achievable (see~\eqref{eq:cglmom}). The assertion that the NP mode is connected to the MHD entropy mode, rather than the slow mode, is likely more desirable as it also avoids degeneracy in different branches of the dispersion relation. Careful inspection of figure~\ref{fig:cgldisp} shows that there exists a band in which both the NP and oblique ion-acoustic modes possess zero real frequency. If it were the case that the MHD slow mode became the NP mode, this branch would have to cross with the kinetic entropy mode and both would have identical decay rates, making them degenerate. Therefore, in our argument for the behaviour of above-threshold NP modes in high-$\beta$ plasmas, we expect that at very large scale separation, and hence large $\nu/|k_\parallel|v_{\rm th,i}$, the NP mode will become more akin to the MHD entropy mode.

The oblique ion-acoustic wave (IAW) also deserves special attention, not least because it possesses a non-propagating band beginning near $\nu \sim k_\parallel v_{\rm th,i}$. Somewhat paradoxically, this is the collisionless extension of the MHD slow mode, never mind the fact that at high~$\beta$ it propagates {\em faster} than the Alfv\'en speed. Even in the collisionless Landau-fluid CGL model, this mode evades a simple general expression for its frequency. However, in the limit of $k_\perp \gg k_\parallel$ and $\beta \gg 1$ with $T_{\rm e}=T_{\rm i0}$, one can obtain the dispersion relation numerically; we find that $\zeta \approx 1-0.43\imag$. This mode therefore has a very similar dispersion relation to its parallel-propagating variant, especially with regards to its rapidly damped nature. Asymptotic analysis for $k_\perp \gg k_\parallel$ reveals that this mode develops a non-propagating band when $\beta \gtrsim 7.1$, occurring in the approximate range of scattering frequencies satisfying $\nu/k_\parallel v_{\rm th,i} \in [2,(3/4)\sqrt{\beta}]$. When $\beta \sim \mathcal{O}(1)$ and smaller, the Braginskii slow mode smoothly transitions into the oblique IAW as $\nu\rightarrow 0$. However, at high~$\beta$, an increasingly large gap forms between the two propagating portions of this mode. This phenomenon is not present in parallel-propagating IAWs at any $\beta$.

\section{Oblique IAWs and micro-instabilities}\label{app:oiaw}

Of the collisionless hydromagnetic modes that do not propagate parallel to the background magnetic field, we have yet to discuss one in the context of high-$\beta$ plasmas and micro-instabilities: the oblique IAW. Given that oblique IAWs share many traits with their parallel propagating counterparts (\S\ref{app:mslin}), generalizing the results of~\citet{kunz20} to the oblique case should not require dramatic changes. Even when propagating across the background magnetic field, at high~$\beta$ these waves are still largely driven by a perturbation to the parallel pressure. As a result, the magnetic tension plays essentially no role, and no interruption-like process can occur as in the case of linearly polarized Alfv\'en waves. Furthermore, the oblique IAW generates equivalent positive and negative pressure anisotropies (there is no pressure balance as in the NP mode). For this reason, both mirror and firehose instabilities can be triggered by this mode. The only notable difference between the oblique and parallel IAWs is the existence of a non-propagating band at certain values of $\nu$ in the dispersion relation of the oblique mode. To see how this difference affects propagation in the presence of instability-induced scattering, we perform an analysis similar to that found in~\S\ref{sec:fastanis}.

Our first task is to determine the amplitude limit above which the anisotropic pressure perturbation in the oblique IAW is unstable to both the mirror and firehose instabilities. Taking the $k_\perp \gg  k_\parallel$ and $\beta \gg 1$ limit, the parallel and perpendicular temperature perturbations in the oblique IAW are
\begin{subequations}\label{eqn:OIAW}
\begin{gather}
    \frac{\delta T_\parallel}{T_{\rm i0}} \approx -\Biggl[2+\biggl(1+\imag\frac{k_\parallel v_{\rm th,i}}{\omega \sqrt{\pi}} \biggr)^{-1}  \Biggr]\biggl(1+2\imag\frac{k_\parallel v_{\rm th,i}}{\omega \sqrt{\pi}} \biggr)^{-1}\frac{\delta B_\parallel}{B_0}, \\*
    \frac{\delta T_\perp}{T_{\rm i0}} \approx \biggl(1+\imag\frac{k_\parallel v_{\rm th,i}}{\omega \sqrt{\pi}} \biggr)^{-1}\frac{\delta B_\parallel}{B_0}.
\end{gather}
\end{subequations}
Substituting in $\omega/k_\parallel v_{\rm th,i} \approx 1-0.43\imag$, equations~\eqref{eqn:OIAW} yield an ion pressure anisotropy $\Delta = (1.88 - 3.03\imag)(\delta B_\parallel/B_0)$. This implies the following amplitude threshold for oblique IAWs to trigger both the firehose and mirror instabilities:
\begin{equation}\label{eq:iawthresh}
    \left|\frac{\delta B_\parallel}{B_0}\right| \gtrsim \frac{1}{\beta_{\rm i}} \quad\textrm{(oblique  IAW  amplitude  threshold)} .
\end{equation}
We argue that, above this threshold, the scattering induced by micro-instabilities will be that required to maintain marginal stability, or $\Delta \sim \beta_{\rm i}^{-1}$. Through the same logic as was applied to the fast mode, this scattering rate is
\begin{equation}
\nu \sim \mathrm{Re} \biggl[ 3\omega \beta_{\rm i} \biggl(\frac{\delta B_\parallel}{B_0} - \frac{2}{3}\frac{\delta n}{n_0}\biggr) \biggr] \approx 3.7k_\parallel v_{\rm th,i}\beta_{\rm i} \biggl|\frac{\delta B_\parallel}{B_0}\biggr|.
\end{equation}
As in the case of the fast wave, the above expression for the limiting collisionality is only valid in the limit that $\nu \gg \omega$. This constraint is nearly satisfied at the amplitude threshold, therefore this scattering rate is likely to be a good approximation even for mode amplitudes of only a few $\beta_{\rm i}^{-1}$. 

With the scaling of the induced scattering rate now known, we may return to the dispersion relation shown in figure~\ref{fig:cgldisp} to surmise how micro-instabilities might modify the propagation of oblique IAWs. Recall from Appendix~\ref{app:mslin} that the oblique IAW becomes non-propagating for $\beta_{\rm i}\gtrsim 7.1$ when $\nu/k_\parallel v_{\rm th,i} \in [2,(3/4)\sqrt{\beta}]$. The form of the effective scattering rate (being dependent on $\delta B_\parallel$) then suggests that the fate of an oblique IAW rests on the amplitude of the initial perturbation. For amplitudes  within the range $\beta^{-1} \lesssim |\delta B_\parallel/B_0| \lesssim \beta^{-1/2}$, the oblique IAW will become a viscously damped mode that does \textit{not} propagate, while above $|\delta B_\parallel/B_0| \gtrsim \beta^{-1/2}$ it will become a Braginskii-like propagating sound wave. The latter of the two regimes is essentially the result obtained by \citet{kunz20} for parallel-propagating IAWs. The former limit of moderate amplitude becomes increasingly important at high~$\beta$ where its range of relevance increases. In plasmas with $\beta \lesssim 10$ however (e.g., the solar wind), this range is either extremely narrow or nonexistent, leading to evolution closer to the parallel IAW. \edit{As in all cases, the action of microinstabilities and their induced scattering can only be expected to last for as long as the wave-associated pressure anisotropy is driven beyond the microinstability thresholds.}

\end{document}